\documentclass[
twocolumn,
aps,prd,
nofootinbib,
superscriptaddress,
tightenlines,
amsmath,
amssymb
]{revtex4}
\usepackage{bm}
\usepackage{color,graphicx}
\usepackage{amsmath}
\usepackage{amssymb}
\usepackage{slashed}
\usepackage{float}
\usepackage[normalem]{ulem}

\newcounter{comment}

\allowdisplaybreaks
\begin{document}
\title{Parametrization of Quark and Gluon Generalized Parton Distributions in a Dynamical Framework} 
%\label{sec-5}

\author{Brandon Kriesten} 
\email{btk8bh@virginia.edu}
\affiliation{Department of Physics, University of Virginia, Charlottesville, VA 22904, USA.}

\author{Philip Velie}
\email{pmv8ev@virginia.edu}
\affiliation{Department of Physics, University of Virginia, Charlottesville, VA 22904, USA.}

\author{Emma Yeats}
\email{ery5ua@virginia.edu}
\affiliation{Department of Physics, University of Virginia, Charlottesville, VA 22904, USA.}

\author{Fernanda Yepez Lopez}
\email{fy2nb@virginia.edu}
\affiliation{Department of Physics, University of Virginia, Charlottesville, VA 22904, USA.}

\author{Simonetta Liuti} 
\email{sl4y@virginia.edu}
\affiliation{Department of Physics, University of Virginia, Charlottesville, VA 22904, USA.}
%\affiliation{Laboratori Nazionali di Frascati, INFN, Frascati, Italy}

\begin{abstract}
We present a parametrization of
the chiral even generalized parton distributions, $H$, $E$, $\widetilde{H}$, $\widetilde{E}$, for the quark, antiquark and gluon, in the perturbative QCD-parton
framework. Parametric analytic forms are given as a function of two equivalent sets of variables $x,\xi,t$ (symmetric frame) and $X,\zeta,t$ (asymmetric frame), at an initial scale, $Q_o^2$.   
In the $X>\zeta$ region a convenient and flexible form is obtained as the product of a Regge term $\propto X^{-\alpha + \alpha' t}$, describing the low $X$ behavior, times a spectator model-based functional form depending on various mass parameters; the behavior at $X<\zeta$, is determined using the generalized parton distributions symmetry and polynomiality properties. 
The parameters are constrained using data on the flavor separated nucleon electromagnetic elastic form factors, the axial and pseudoscalar nucleon form factors,
and the parton distribution functions from both the deep inelastic unpolarized and polarized nucleon structure functions.
For the gluon distributions we use, in particular,  constraints provided by recent lattice QCD moments calculations.
The parametrization's kinematical range of validity is: $0.0001 \leq X \leq 0.85$, $0.01 \leq \zeta \leq 0.85$, $0 \leq -t \leq 1$ GeV$^2$, $2 \leq Q^2  \leq 100$ GeV$^2$. With the simultaneous description of the quark, anti-quark and gluon sectors, this parametrization represents a first tool enabling a global QCD analysis of deeply virtual exclusive experiments.
\end{abstract}
\maketitle

\section{Introduction}
\label{sec:intro}
Deeply virtual exclusive photon and/or meson production processes 
allow us to access Generalized Parton Distributions (GPDs) \cite{Ji:1996ek,Radyushkin:1997ki}, the universal quantities that lie at the heart of all studies of the 3D structure  of  the  proton  \cite{Burkardt:2005hp}.  GPDs can also give access to the mechanical properties of angular  momentum \cite{Ji:1996ek,Ji:1996nm},  pressure,  and shear forces \cite{Polyakov:2002wz,Polyakov:2002yz,Polyakov:2018zvc} defining the internal structure and dynamics of hadrons.  Analogous to the  parton distributions functions (PDFs) obtained from inclusive deep inelastic scattering  (DIS) processes,  GPDs  parametrize  the  quark, antiquark  and gluon correlation functions involving matrix elements between proton states of operators at a light-like  separation  between  the respective parton fields. 

An important difference with inclusive scattering is that GPDs enter the cross section for deeply virtual exclusive experiments such as Deeply Virtual Compton Scattering (DVCS) \cite{Ji:1996nm}, deeply virtual meson production (DVMP) and related cross channel reactions, at the amplitude level, multiplied by the Wilson coefficient functions and integrated over the longitudinal momentum fraction, $x$. 
Quantum Chromodynamics (QCD) factorization theorems similar to the inclusive DIS case have been proven for DVCS in Refs. \cite{Ji:1997nk,Ji:1998xh} and for DVMP  (see Ref.\cite{Collins:2011zzd}). Because the proton states have different momenta, GPDs depend on two additional kinematic variables: the momentum transfer squared between the initial and final proton which is proportional to the invariant, $t$, and the light cone (LC) momentum transfer fraction, $\xi$, or $\zeta$ (see {\it e.g.} Refs.\cite{Diehl:2001pm,Belitsky:2001ns} for extensive reviews, definitions and notations) .  
The phenomenology of perturbative QCD evolution is therefore similar to the one extensively developed for inclusive scattering. The observables, the Compton Form Factors (CFFs), are complex quantities obtained as convolutions of GPDs with kernels governed by perturbative QCD. 

Notwithstanding this additional complication, the quark, antiquark and gluon components of CFFs and consequently of GPDs, can be extracted from deeply virtual exclusive experiments with the same logic behind DIS, {\it i.e.}  merging information from a combination of electron and neutrino probes, including meson production, {\it e.g.} $J/\psi$ production which is sensitive to the gluon content (see reviews in \cite{Kumericki:2016ehc,Lin:2020rut}), and crossed channel experiments such as deeply virtual exclusive pion-proton Drell Yan scattering \cite{Sawada:2016mao,Chang:2020rdy}.
While present available data sets cover somewhat limited kinematic ranges which are neither sufficient to separate out the various components, nor to gauge their relative importance in the various regions, the exclusive program at Jefferson Lab@12 GeV, as well as upcoming measurements at COMPASS and JPARC will provide, in the upcoming years, a large amount of precise data. A wide range of diverse experiments from various targets will be performed, from DVCS, to timelike Compton scattering (TCS), and various meson production processes. The future planned Electron Ion Colliders (EIC, EIcC), will further these exploration at both higher four momentum transfer squared $Q^2$ and low Bjorken $x$. 

It is therefore timely that a flexible parametrization including valence, sea quarks and gluon components which can be perturbatively evolved to the scale of the data, is made available. Our parametric forms build on the previously determined valence distributions which are modeled at a low initial scale, $Q_o^2 \approx 0.1$ GeV$^2$.
At this scale only valence quarks are present. Gluons and sea quarks (quark-antiquark pairs) are resolved as independent degrees of freedom at a larger scale, $Q_o^2 \approx 0.58$ GeV$^2$. These components subsequently undergo perturbative evolution and generate additional gluon and sea quarks dynamically through gluon bremmstrahlung.

The GPDs dynamical framework uses for the initial scale a  parametrization based on the reggeized spectator model \cite{Ahmad:2006gn,Ahmad:2007vw,Goldstein:2010gu,GonzalezHernandez:2012jv}. In this model we envisage scattering from either a valence quark, a sea quark, or a gluon; leaving behind, respectively, a spectator diquark, tetraquark, or color octet proton. The proton-parton-spectator vertex is modeled with a form factor which provides a cut-off in the parton's $k_T$ integration. Finally, Regge behavior is obtained by allowing the spectator mass to vary modulated by a spectral function and using the relation, $x \approx [M_X^{q,g}]^{-1}$, $M_X^{q,g}$ being the spectator's variable mass.  As we explain in detail later on, the model's parameters are constrained recursively by first fitting the GPD in the forward limit, $H_{q,\bar{q},g}(x,0,0)$, for $q=u,d$, to the corresponding PDFs, and in a subsequent step, fitting the $t$ dependence to the form factor. The GPD property of polynomiality is therefore obtained by definition for the leading Mellin moment. Since this is in essence an overlap model, polynomiality does not hold directly for higher order Mellin moments by construction, whereas it can be imposed with a measurable uncertainty. 
The parametrization describes all chiral-even GPDs, $H, E, \widetilde{H}, \widetilde{E}$ in the quark sector, similar to Refs.\cite{Goldstein:2010gu,GonzalezHernandez:2012jv}. We introduce a new parametrization for $H_{g,\bar{q}}$ and $E_{g,\bar{q}}$ for the gluon and antiquarks. Perturbative QCD evolution is performed at leading order (LO). We also study an extension to next to leading order (NLO) which provides the basis for quantitatively determining the parameters of our next version to be presented in future work.

The paper is organized as follows: in Section \ref{sec:sec2} we give the relevant definitions including various symmetry and integral properties of the quark and gluon GPDs; in Section \ref{sec:param} we present the analytic form of the parametrization for the various quark and gluon GPDs at the initial scale $Q_o^2$; in Section \ref{sec:param_1} the numerical values of the different quark flavor and gluon GPDs parameters are displayed in tabulated form and details of perturbative QCD evolution of GPDs are illustrated; numerical results for the various GPD components are shown and discussed in Section \ref{sec:numerical}. 
%illustrating the role of LO and NLO evolution for the observables; in Section \ref{sec:Fou} we show spatial images of the various components; 
 %in Section \ref{sec:Fou} we discuss the impact of our parametrization on the Fourier transformation to impact parameter space.
 Finally, in Section \ref{sec:conclusions} we present our conclusions and outlook. Many supplementary formulae explaining the details of the parametrization along with its summarized, ready to use, version are presented in the appendices.

%%%%%%%
%%%%%%%
\section{Definitions}
\label{sec:sec2}
\begin{center}
\begin{figure}
\includegraphics[width=7cm]{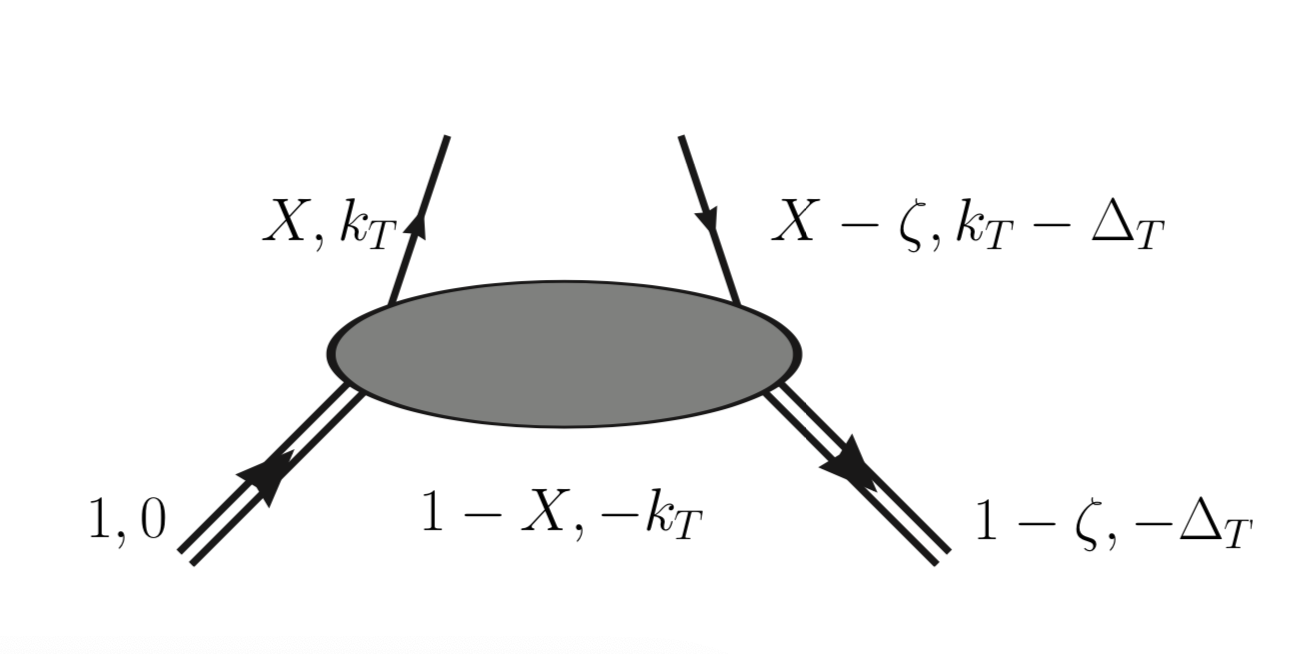}
\includegraphics[width=4cm]{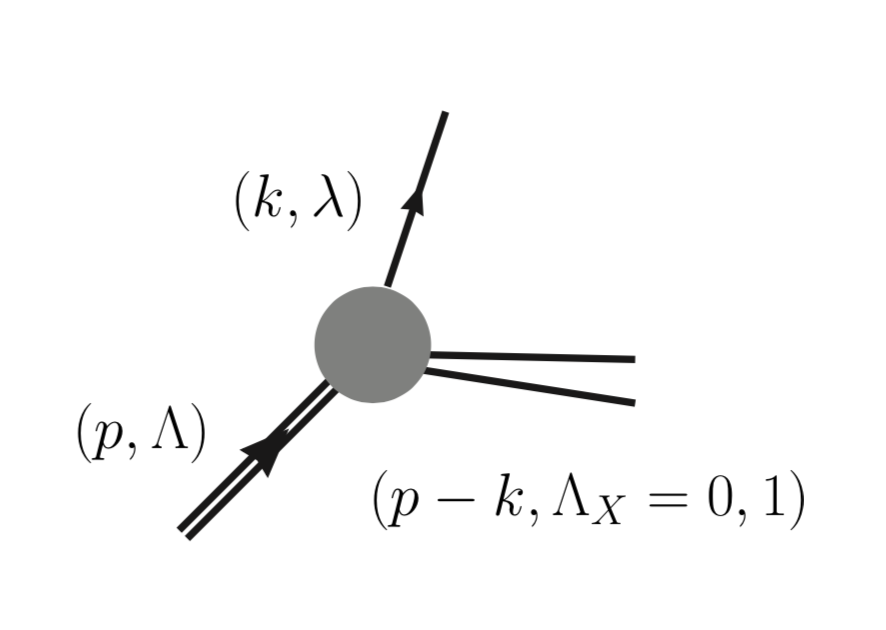}
\includegraphics[width=4cm]{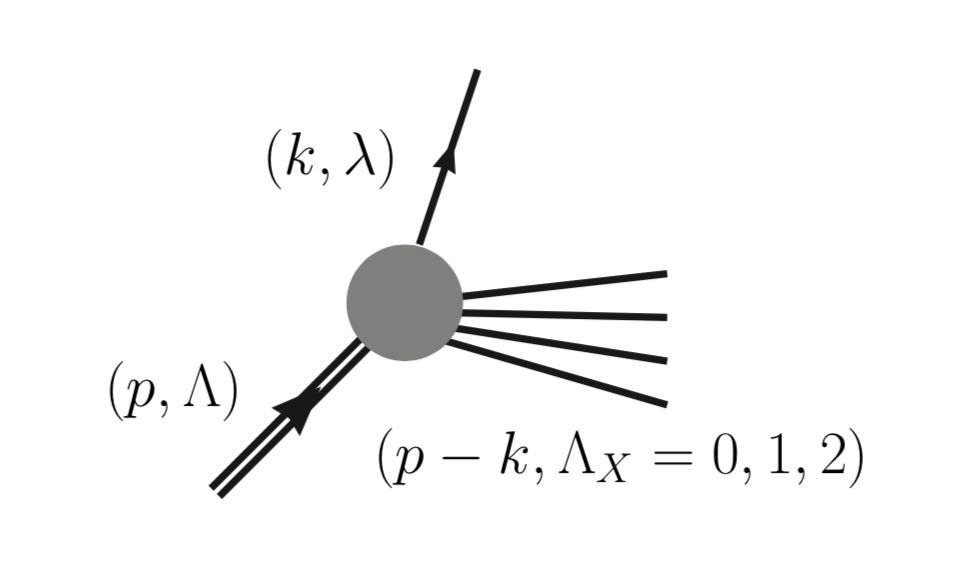}
\includegraphics[width=4cm]{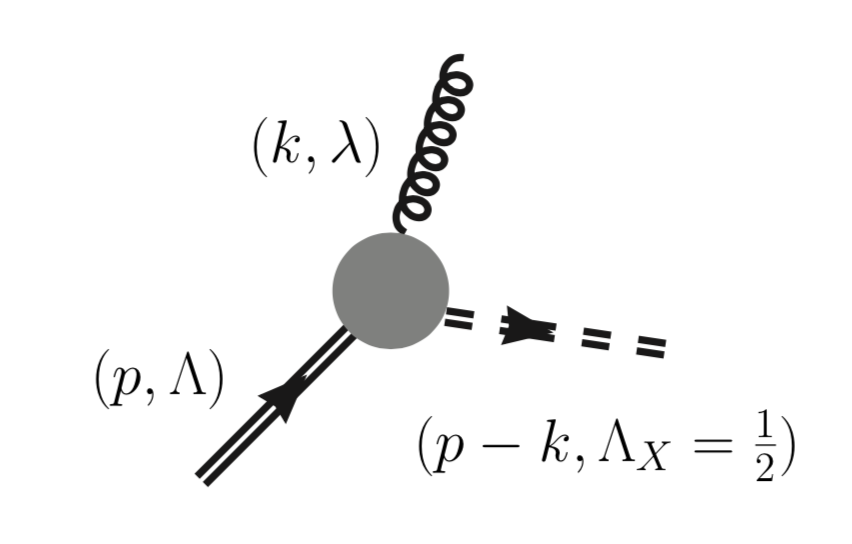}
\caption{(top) Cut diagram for the calculation of GPDs in the spectator model with labels for both the longitudinal momentum fractions and transverse momentum of the parton, proton and spectator system with respect to the initial proton. In the symmetric system of variables the momentum fractions are evaluated using Eqs.(\ref{eq:symm1},\ref{eq:symm2}); (bottom) vertices for the spectator system: scalar ($\Lambda_X=0$) or axial vector diquark ($\Lambda_X=1$) for valence quarks; tetraquark ($\Lambda_X=0,1, 2$)for sea quarks, and color octet proton ($\Lambda_X=1/2$) for gluons. }
\label{fig:spectator}
\end{figure}
\end{center}

The parametrization for all twist-two chiral-even GPDs in the quark and gluon sectors is given in terms of a set of two light cone (LC) momentum fractions, and the Mandelstam invariant, $t$. The LC variables represent the quark/gluon longitudinal momentum fraction, $X$, and the difference between the longitudinal momentum fractions of the outgoing and incoming quark, $\zeta$, respectively (Figure \ref{fig:spectator}, for reviews see Refs.\cite{Diehl:2003ny,Belitsky:2005qn,Kumericki:2016ehc}). 

The support in $X$ is expressed in the following form,
\begin{equation}
\label{eq:F_DGLAPERBL}
F^{q,\bar{q},g}(X,\zeta,t; Q^2)  =  \left\{ 
\begin{array}{ll}
 F^{DGLAP}_{q,g} & \mbox{ $\zeta \leq X \leq1$} \\ 
 & \\
 F^{ERBL}_{q,\bar{q},g} & \mbox{$0 \leq X  < \zeta $} \\
 & \\
  F^{DGLAP}_{\bar{q},g} & \mbox{$-1 + \zeta \leq X  < 0 $}
\end{array} 
\right.
\end{equation}
%%%%%%%
where $F^{q,\bar{q},g} \equiv H^{q,\bar{q},g}, E^{q,\bar{q},g}, \widetilde{H}^{q,\bar{q},g}, \widetilde{E}^{q,\bar{q},g}$. 
The acronyms ``DGLAP" and ``ERBL" designating specific $X$ ranges in Eq.\eqref{eq:F_DGLAPERBL} refer to the two different modes of perturbative QCD evolution in these regions. 
%The labels "DGLAP" and "ERBL" refer to how perturbative QCD evolution is described in the different kinematic domains.

The kinematic variables are defined using deeply virtual exclusive photo-production, $e p \rightarrow e' p' \gamma$, as a testing ground experiment for GPDs,
\begin{itemize}
    \item $X$, the LC momentum fraction; $X=k^+/p^+$, where $k$ and $p$ are the parton/proton four-momenta (see Appendix \ref{sec:appa} for detailed kinematics definitions).
    \item $t$, the four-momentum transfer squared between the initial and final proton; $t = \Delta^2 = (p-p')^2= \Delta_\mu\Delta^\mu = \Delta_0^2-\Delta_\bot^2-\Delta_3^2$. $t$ ultimately gives access to the transverse spatial distribution, and is always negative.
    \item $\zeta$ the skewness parameter; $\zeta = \Delta^+/p^+ > 0$
    \item $Q^2$, the virtuality of the initial photon exchanged between the initial and final electrons; $Q^2 = -(k_e - k_e')^2$. 
%    $Q_0^2$ corresponded to a smaller subsection of the proton (less mass, more momentum), whereas approaching $Q^2$ would bring you to the whole proton (more mass, less momentum).
\end{itemize}

We use the so-called asymmetric frame
where the initial (final) proton, $p$ ($p'$), and initial (final) parton, $k$ ($k'$), four-momentum components are given in the form, $v \equiv(v^+,v^-,{\bf v}_\perp)$,  (see Figure \ref{fig:spectator} and appendix \ref{sec:appa}).

The asymmetric system of LC variables was introduced to better describe the dynamics of the spectator model including perturbative QCD evolution \cite{Goldstein:2010gu,GonzalezHernandez:2012jv}. In this case,  the initial proton is set along the $z$-axis. A more commonly used system uses a symmetric set, ($x,\xi$). The conversion between symmetric ($x, \xi$) and asymmetric variables ($X,\zeta$) is given by,
\begin{eqnarray}
\label{eq:symm1}
x & = & \frac{k^+ + k^{\prime\,+}}{P^+ + P^{\prime\,+}} = \frac{X-\zeta/2}{1-\zeta/2} \quad \Rightarrow X= \frac{x+\xi}{1+\xi} \\
\label{eq:symm2}
 \xi & = & \frac{2\Delta^+}{P^+ + P^{\prime\,+}} =\frac{\zeta}{2-\zeta}  \quad 
 \Rightarrow \zeta =\frac{2 \xi}{1+\xi} .
\end{eqnarray}
For the GPDs we have,
\begin{eqnarray}
F^{q,\bar{q},g}(x,\xi) = 
\begin{cases}
F_{q}^{\text{DGLAP}} &\text{for   } x > \xi\\
\\
F_{q,\overline{q}}^{\text{ERBL}} &\text{for   } -\xi < x < \xi \\
\\
F_{\overline{q}}^{\text{DGLAP}} &\text{for   } -1 < x < -\xi ,
\end{cases}
\end{eqnarray}
where similar definitions hold for the helicity and gluon distributions, $\widetilde{F}$.
Other variables used to define GPDs in the LC frame are (see also \cite{Goldstein:2010gu}),
\begin{eqnarray}
\label{eq:Xprime}
 X' & = &\displaystyle\frac{X-\zeta}{1-\zeta},  \quad \quad
1- X'  = \displaystyle\frac{1-X}{1-\zeta} 
\\
\label{eq:ktilde}
 {\bf \tilde{k}} & = & {\bf k}_\perp - \frac{1-X}{1-\zeta} {\bf \Delta}_\perp 
 \\
 \label{eq:tvszeta}
 t & = & \Delta^2 = -\frac{M^2 \zeta^2}{1-\zeta} - \frac{{\bf \Delta}_\perp^2}{1-\zeta} = -\frac{4M^2  \xi^2}{1-\xi^2} - {\bf \Delta}_\perp^2 \frac{1-\xi}{1+\xi} \nonumber \\
 \end{eqnarray}
 where it should be underlined that the expression of the invariant, $t$, in terms of the longitudinal and transverse variables, $\zeta (\xi)$ and $\Delta_T$ is specific to the chosen LC frame.   
 
 The minimum kinematically allowed value of $t$, obtained for ${\bf \Delta}_T=0$ is,
 \begin{equation}
 t_0 = - \frac{4 \xi^2 M^2}{1-\xi^2}.
 \label{eq:tmin}
\end{equation}

%%%%%%%%%%%%%
%%%%%%%%%%%%%
%%%%%%%%%%%%% LIMITS AND CONSTRAINTS
\subsection{Limits and constraints}
GPDs are subject to constraints in the forward limit ({\it i.e.} for $\zeta,t \rightarrow 0$) and in their Mellin moments structure (polynomiality). Furthermore they satisfy positivity bounds written in terms of PDFs from DIS. Although these limits were written in several reviews, {\it e.g.} Ref.\cite{Diehl:2003ny}, we provide an essential list below. 

%%%%%%
%%%%%% FORWARD LIMIT
%%%%%%
\subsubsection{Forward limit}
\label{sec:forward}
In the forward limit the quark GPDs $H$ and $\widetilde{H}$ define the PDFs,
\begin{eqnarray}
H^{q}(X,0,0;Q^2) &\equiv& f_1^{q,}(X, Q^2) \\  \widetilde{H}^{q}(X,0,0;Q^2) &\equiv & g_1^{q}(X, Q^2) 
\end{eqnarray}
where $f_1^{q}$, $g_1^{q}$ are the unpolarized and helicity PDFs, respectively. 
In the gluon sector,
\begin{eqnarray}
H^{g}(X,0,0;Q^2) &\equiv& X g(X, Q^2) \\  \widetilde{H}^{g}(X,0,0;Q^2) &\equiv & X \Delta g(X, Q^2) ,
\end{eqnarray}
$g(X)$ and $
\Delta g(X)$ being the unpolarized and helicity PDFs, respectively.

\subsubsection{Polynomiality}
\label{sec:polinom}
Stemming from the property of polynomiality (see discussion in \cite{Diehl:2003ny}), in the symmetric frame notation,
%%%
\footnote{The same integral properties can be written in the asymmetric frame with a switch of variables, and inserting the Jacobian, $\frac{1}{1-\frac{\zeta}{2}}$ \cite{Goldstein:2010gu}. }
%%%
the integrals in $x$ of the quark GPDs are independent of $\xi$ and give the various proton elastic form factors,
\begin{eqnarray}
\label{eq:Dirac}
 \int_{-1}^1 dx  H^{q}(x,\xi,t;Q^2) & = F_1^{q}(t) \\
\label{eq:Pauli}
 \int_{-1}^1 dx E^{q}(x,\xi,t;Q^2) &  = F_2^{q}(t) \\
\label{eq:axial}
 \int_{-1}^1 dx \widetilde{H}^{q}(x,\xi,t;Q^2) &  =  G_A^{q}(t) \\
\label{eq:pseudo}
 \int_{-1}^1 dx \widetilde{E}^{q}(x,\xi,t;Q^2) &  =  G_P^{q}(t) \; . 
\end{eqnarray}
$F_1^q$ and $F_2^q$ are the quark $q$ contribution to the proton Dirac and Pauli form factors; similarly, $G_A^q$ and $G_P^q$ are the quark $q$ axial and pseudoscalar form factors.

The second moments of the quark GPDs, $H$ and $E$ read,
\begin{eqnarray}
\label{eq:Dirac2}
 \int_{-1}^1 dx  x H^{q}(x,\xi,t;Q^2) & = A_q(t) + (2\xi)^2 C_q(t)\\
\label{eq:Pauli2}
 \int_{-1}^1 dx x E^q(x,\xi,t;Q^2) & = B_q(t) - (2\xi)^2 C_q(t) 
\end{eqnarray}
(similar relations are found in the axial vector sector).

In the gluon sector we consider constraints given by the moments, 
\begin{eqnarray}
\label{eq:Diracg}
 \int_{0}^1 dx  H^{g}(x,\xi,t;Q^2) & =   A_g(t) + (2 \xi)^2 C_g(t) \\
\label{eq:Paulig}
\int_{0}^1 dx E^g(x,\xi,t;Q^2) & = B_g(t) - (2 \xi)^2 C_g(t)
\end{eqnarray}
$A_g$, $B_g$, $C_g$ have been recently calculated in lattice QCD \cite{Shanahan:2018pib}. All of the form factors presented above have been either measured or calculated in lattice QCD and, therefore,  provide essential constraints to the parametrization. 

%\begin{eqnarray}
%\int_{-1}^{1} dx x H_{\overline{q}}(x,\xi,t) = A_{2,0}^{\overline{q}}(t) + (2\xi)^{2} D_{2,0}^{\overline{q}}(t) \\
%\int_{-1}^{1} dx x E_{\overline{q}}(x,\xi,t) = B_{2,0}^{\overline{q}}(t) - (2\xi)^{2} D_{2,0}^{\overline{q}}(t) 
%\end{eqnarray}

%\begin{eqnarray}
%\int_{0}^{1}dx H_{g}(x,\xi,t) &=& A_{2,0}^{g}(t) + (2\xi)^{2} D^{g}_{2,0}(t) \\
%\int_{0}^{1}dx E_{g}(x,\xi,t) &=& B_{2,0}^{g}(t) - (2\xi)^{2} D^{g}_{2,0}(t)
%\end{eqnarray}
%\subsubsection{Form Factors}
%Notice that the region of support for the anti-quarks is twice as large as the valence distribution, therefore when integrating the anti-quark distribution for the form factor one must include a factor of 2.
The form factors $A_{q,g}$, $B_{q,g}$, $C_{q,g}$
depend on the scale, $Q^2$, and are also scheme dependent at Next to Leading order (NLO) in perturbative QCD.  

Summing Eqs.(\ref{eq:Dirac2},\ref{eq:Pauli2},\ref{eq:Diracg},\ref{eq:Paulig}) at $t=0$, one finds the (scale independent) angular momentum sum rule \cite{Ji:1996ek},  
\begin{equation}
\label{eq:JiSR}
  \frac{1}{2}\left[  \int_{-1}^{1} x (H_q+E_q) +  \int_{-1}^{1}  (H_g+E_g) \right] = J_q + J_g = \frac{1}{2} , 
\end{equation}
whereas momentum conservation of the nucleon constituents is expressed by,
\begin{equation}
   \int_{-1}^{1} x H_q +   H_g = M_q + M_g = 1 .
\end{equation}
%Similarly one can write out the first moment of the gluon helicity distribution where the $\xi^{2}$ term is present in even moments.
%
%\begin{eqnarray}
%\int_{0}^{1}dx \widetilde{H}_{g}(x,\xi,t) &=& \widetilde{A}_{2,0}^{g}(t) \\
%\int_{0}^{1}dx \widetilde{E}_{g}(x,\xi,t) &=& \widetilde{B}_{2,0}^{g}(t)
%\end{eqnarray}
Equations for the full polynomiality structure involving the Mellin moments for any value of integer value $n$  
are reviewed in \cite{Diehl:2001pm,Belitsky:2005qn}.

%%
%%%%% FIGURE 2
%%%%%
\begin{figure*}
\includegraphics[scale=0.35]{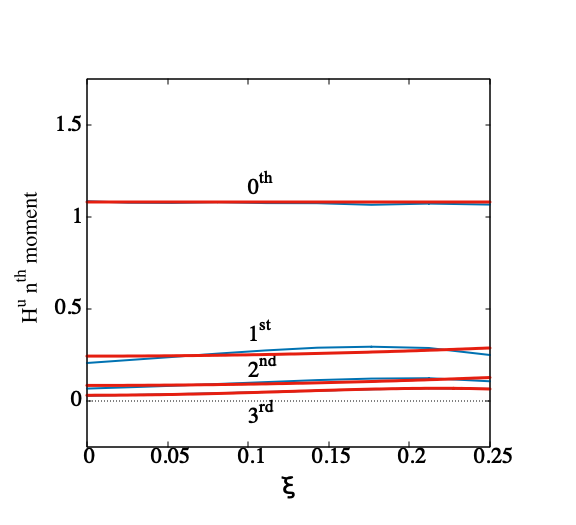}
\includegraphics[scale=0.35]{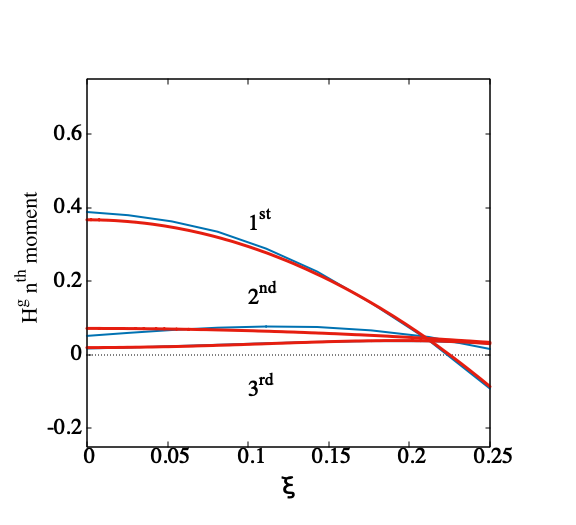}
\caption{Polynomiality property for $H_{u/g}$ in our parametrization calculated for a typical JLab kinematic bin $t = -0.3$ GeV$^{2}$ and $Q^{2} = 4$ GeV$^{2}$. The blue line is our parametrization results and the red lines are correspond to a polynomial fit in powers of $\xi^{2}$. We demonstrate that polynomiality is satisfied in our parametrization.}
\label{fig:poly}
\end{figure*}

Polynomiality is imposed numerically, and not an ab initio property in parton-like models such as the reggeized spectator model. In our approach we fit the first moments, Eqs.(\ref{eq:Dirac},\ref{eq:Pauli},\ref{eq:axial},\ref{eq:pseudo}) to the measured form factors; the $n=1$ moments are fitted to moments of PDFs at $t=0$ and at a given $Q^2$ value. Their value at $t<0$ can be only constrained from lattice QCD results since no measurements of these form factors are available.
For illustration, in Figure \ref{fig:poly} we show the first few Mellin moments, calculated with our parametrization, for the GPDs $H_u$ (left panel) and $H_g$ (right panel) compared to polynomial forms in $\xi^2$ at $t=-0.3$ GeV$^2$, and $Q^2= 4$ GeV$^2$. Notice that the range in $\xi$ is reduced because of the kinematic limit obtained imposing $\Delta_\perp^2 \geq 0$ in Eq.(\ref{eq:tvszeta}).
Although polynomiality is a fitted property, we find that the first few Mellin moments, which are most important to determine the GPDs behavior, follow this property well within the given error from the fit. We ascribe this behavior to the Lorentz invariance of the model. 
%Future work will be directed towards improving these small discrepancies. 
To further address this issue one could explore ansazte similar to the one devised for pion GPDs in Ref.\cite{Chouika:2017itz,Chouika:2017dhe}.

\begin{figure*}
\includegraphics[scale=0.35]{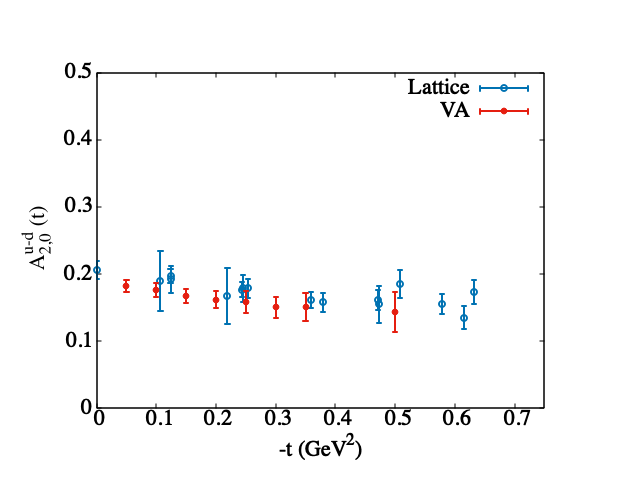}
\includegraphics[scale=0.35]{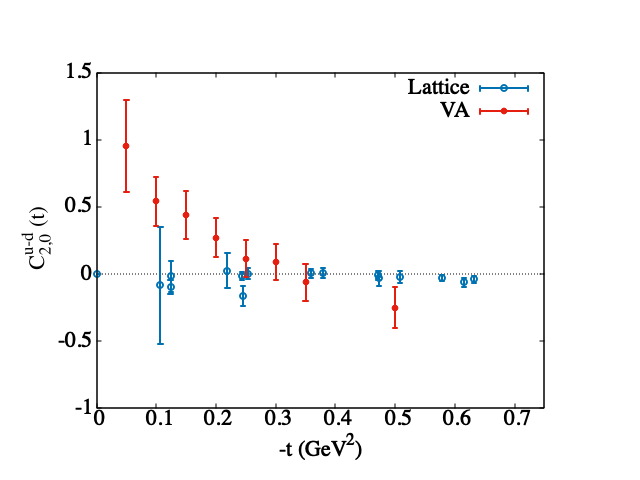}
\includegraphics[scale=0.35]{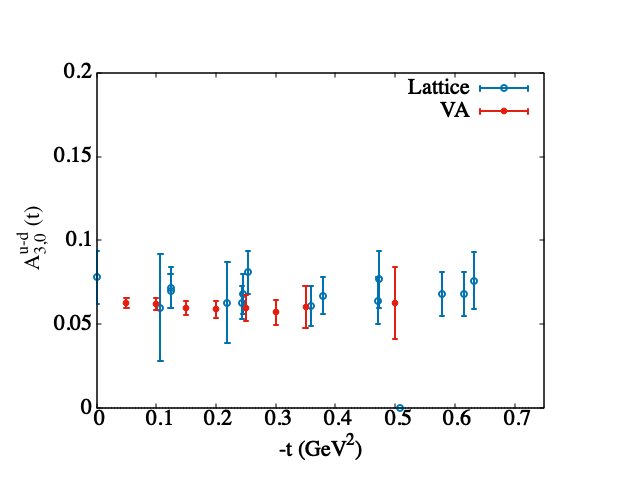}
\caption{Moments of GPDs as calculated by the polynomial fits in $\xi^{2}$ in red according to our parametrization evolved to a final scale of $Q^{2} = 4$ GeV$^{2}$. These are shown  compared to moments calculated in lattice QCD \cite{Hagler:2007xi}. The errors of ``VA" are the propagated errors of the fit parameters.}
\label{fig:polyn_lattice}
\end{figure*}

In Figure \ref{fig:polyn_lattice} we show results from our fit compared to lattice QCD calculations for the flavor non-singlet, $n=2$ $u-d$ moments, namely $A_{2,0}^{u-d} \equiv A_u-A_d$, $C_{2,0}^{u-d} \equiv C_u-C_d$, Eq.(\ref{eq:Dirac2}), and the $n=3$ moment,  $A_{3,0}^{u-d} = \int dx x^2 (H_u - H_d)$  \cite{Hagler:2007xi,Alexandrou:2015rja}. Our fit was constrained using data at $t=0$ only. It shows an excellent agreement with lattice calculations for the $A$ form factors, whereas a discrepancy with the $C$ form factor seems to emerge at small $t$ values.  
%%%%%%
%%%%%%
%%%%%% FIGURE 3

\vspace{0.5cm}
%%%%%%
%%%%%%
\subsubsection{Positivity}
\label{sec:positivity}
Generalized parton distributions are bounded by the forward parton distribution functions at two different momentum fraction values. This bound has been studied in Refs. \cite{Martin:1997wy,Pire:1998nw,Diehl:2000xz,Boffi:2007yc}, for an essential review see Ref.\cite{Diehl:2003ny}.   

The GPDs in the DGLAP region limit to the PDFs in the forward limit where $\xi$ and $t$ are equal to 0. Therefore, one would expect relations between the two distributions. Using the wave function description of the GPDs, one can work out the Schwartz inequality constraining the GPDs with an upper limit in terms of the unpolarized PDFs. Stronger constraints can be made when using all of the polarized PDFs in this constraint. 

The momentum fractions for an incoming quark with respect to the incoming proton, and an outgoing quark with respect to the outgoing proton read,
\begin{eqnarray}
x_{\text{in}} &=& \frac{x+\xi}{1+\xi} \implies \, X_{\text{in}} = X \\
x_{\text{out}} &=& \frac{x-\xi}{1-\xi} \implies \, X_{\text{out}} = \frac{X-\zeta}{1-\zeta}
\end{eqnarray}
The positivity constraints are given by,
\begin{widetext}
\begin{eqnarray}
\label{eq:hpos}
(1-\xi^{2})\Big(H^{q}(x,\xi,t) - \frac{\xi^{2}}{1-\xi^{2}}E^{q}(x,\xi,t) \Big)^{2} + \Big(\frac{\sqrt{t_{0}-t}}{2M\sqrt{1-\xi^{2}}}E^{q}(x,\xi,t) \Big)^{2} \le \frac{q(x_{\text{in}})q(x_{\text{out}})}{1-\xi^{2}}
\end{eqnarray}
\begin{eqnarray}
E^{q}(x,\xi,t) \le \frac{2M}{\sqrt{t_{0}-t}}\sqrt{q(x_{\text{in}})q(x_{\text{out}})}
\end{eqnarray}
\begin{eqnarray}
(1-\xi^{2})\Big(\widetilde{H}^{q}(x,\xi,t) - \frac{\xi^{2}}{1-\xi^{2}}\xi\widetilde{E}^{q}(x,\xi,t) \Big)^{2} + \Big(\frac{ \sqrt{t_{0}-t}}{2M\sqrt{1-\xi^{2}}}\xi\widetilde{E}^{q}(x,\xi,t) \Big)^{2} \le \frac{q(x_{\text{in}})q(x_{\text{out}})}{1-\xi^{2}}
\end{eqnarray}
\begin{eqnarray}
\widetilde{E}^{q}(x,\xi,t) \le \frac{2M}{\xi \sqrt{t_{0}-t}}\sqrt{q(x_{\text{in}})q(x_{\text{out}})}
\end{eqnarray}
\end{widetext}

%%%%%%
%%%%%% FIGURE 4 POSITIVITY
\begin{figure}[ht]
\begin{center}
\includegraphics[scale=0.375]{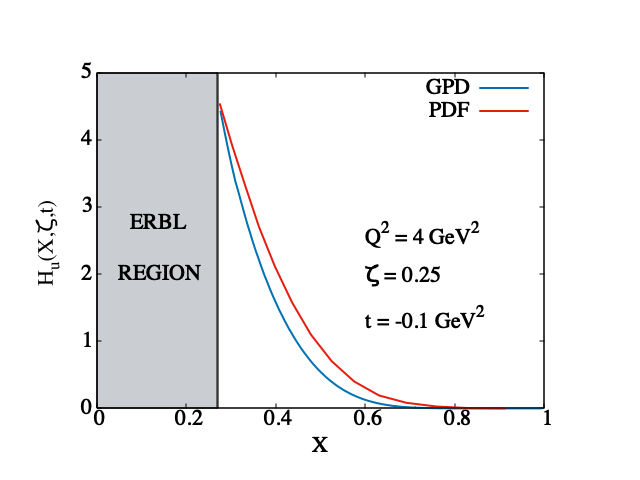}
\caption{Positivity constraints for the GPD $H_{u}$ distribution at the kinematics $\zeta = 0.25, \,\, Q^{2} = 4$ GeV$^{2}$, and $t = -0.1$ GeV$^{2}$. We use the LO PDF parametrization \cite{Alekhin:2002fv} for comparison. The red curve is the right hand side of (\ref{eq:hpos}) and the blue curve is the left hand side of (\ref{eq:hpos}). %No positivity constraints have been worked out for the ERBL region.
}
\label{fig:positivity}
\end{center}
\end{figure}
An illustration of how our parametrization satisfies the positivity constraints is shown in Figure \ref{fig:positivity}, for the GPD $H_u$ plotted vs. $X$ at the kinematic point $\zeta =0.25$, $Q^2= 4$ GeV$^2$, and $t=-0.1$ GeV$^2$.

%%%%%%%%%%%%%
%%%%%%%%%%%%%
%%%%%%%%%%%%%
%
% SYMMETRIES 
%
%%%%%%%%%%%%%
%%%%%%%%%%%%%
%%%%%%%%%%%%%
\subsection{Symmetries} 
\label{sec:symm}
%An important role in GPD modeling is played by their specific symmetry relations in the longitudinal variable $X $ or $x$ in the asymmetric or symmetric systems, respectively
Symmetry relations in the longitudinal momentum fraction $X$, or $x$ in the symmetric system of variables, play an important role in GPD modeling
\cite{GolecBiernat:1998ja,Diehl:2003ny,Kirchner:2003wt}). 

In the symmetric system of variables,  the support of $F(x,\xi)$ ranges from $x \in [-1,1]$, where, in particular, 
the quark distribution is defined in the range $x \in [-\xi,1]$, and the anti-quark distribution is defined in the range $x \in [-1,\xi]$. The two regions overlap  in the range $x \in [-\xi,\xi]$, called the ERBL region.
For the gluons, we have $0 < x < 1$, and since the gluons are their own anti-particle they are equivalently described through symmetries about the $x = 0$ axis. 

The antiquark distributions are defined for the unpolarized, $F=H,E$, and the helicity, $\widetilde{F} = \widetilde{H}, \widetilde{E}$, GPDs as,
\begin{subequations}
\label{eq:simm_def}
\begin{eqnarray}
F_{\bar{q}}(x,\xi) & = & - F_q(-x,\xi), \\ \widetilde{F}_{\bar{q}}(x,\xi) & =  & \widetilde{F}_q(-x,\xi)
\end{eqnarray}
\end{subequations}
From these definitions we obtain the contributions to the  ($-$) (flavor non-singlet, NS) and ($+$) distributions for the unpolarized case,
\begin{subequations}
\begin{eqnarray}
\label{eq:Fminus}
F^{NS} = F_{q_V}\equiv F^-_q(x,\xi) &= &F_{q}(x,\xi) - F_{\overline{q}}(x,\xi)  \\
\label{eq:Fplus}
F^+_q(x,\xi) & = & F_{q}(x,\xi) + F_{\overline{q}}(x,\xi) ,
\end{eqnarray}
\end{subequations}
and, similarly, for the helicity dependent GPDs we have,
\begin{subequations}
\begin{eqnarray}
\widetilde{F}^{NS}  = \widetilde{F}_q^-(x,\xi) &=& \widetilde{F}_{q}(x,\xi) - \widetilde{F}_{\overline{q}}(x,\xi)  
\\
\widetilde{F}^+_q(x,\xi) &=&  \widetilde{F}_{q}(x,\xi) + \widetilde{F}_{\overline{q}}(x,\xi) 
\end{eqnarray}
\end{subequations}
For perturbative evolution (see Section \ref{sec:evol}) we introduce the flavor singlet distributions given by the combinations,
\begin{eqnarray}
F^\Sigma \equiv \sum_{q} F^+_q(x,\xi)  &= &\sum_{q}  \left[ F_{q}(x,\xi) + F_{\overline{q}}(x,\xi) \right] , \\
\widetilde{F}^\Sigma \equiv \sum_q \widetilde{F}^+_q(x,\xi) &= &\sum_{q} \left[ \widetilde{F}_{q}(x,\xi) + \widetilde{F}_{\overline{q}}(x,\xi) \right] 
\end{eqnarray}
From Eq.(\ref{eq:simm_def}) it follows that the symmetries of these distributions around $x=0$ are,
\begin{subequations}
\begin{eqnarray}
F^{NS}(x,\xi) &=& F^{NS}(-x,\xi), \quad \\
F^\Sigma(x,\xi) &= &- F^\Sigma(-x,\xi), \quad \\ \nonumber \\
\widetilde{F}^{NS}(x,\xi) &=& - \widetilde{F}^{NS}(-x,\xi)
\label{eq:simmF-}
\\
\widetilde{F}^\Sigma(x,\xi) & = & \widetilde{F}_{q}^S(-x,\xi) \label{eq:simmF+}
\end{eqnarray}
\end{subequations}
For the gluon distributions we have that the unpolarized distributions are symmetric around $x=0$, while the helicity distributions are antisymmetric,
\begin{subequations}
\begin{eqnarray}
F_{g}(x,\xi) &=& F_{g}(-x,\xi), \quad \quad \\
\widetilde{F}_{g}(x,\xi) &=& - \widetilde{F}_{g}(-x,\xi) 
\end{eqnarray}
\end{subequations}
We must also acknowledge a second symmetry about the off-diagonal direction $\xi$. Along with symmetry or asymmetry under $x \rightarrow -x$ these off-diagonal distributions all have a symmetry under $\xi \rightarrow -\xi$ meaning that these distributions are all time reversal even. 
%This is due to the fact that the parton distributions studied in this paper are leading twist, meaning that there are no final state interactions to time order the diagram.

In the asymmetric system, by changing sets of variables from $(x,\xi)$ to ($X,\zeta)$ using Eqs.(\ref{eq:symm1},\ref{eq:symm2}) one has a similar set of symmetries where now the support region is, 
\begin{eqnarray}
X\in [-1+\zeta,1] 
\end{eqnarray}
while the symmetry axis changes from $x= 0$ to $X ={\zeta/2}$, whereby the quark distribution is defined in the range $X \in [0,1]$, and the anti-quark distribution is defined in the range $X \in [-1 +\zeta,\zeta]$, the two regions overlapping in $X \in [0,\zeta]$, the ERBL region.
The ($+$) and ($-$) distributions are defined as,
\begin{eqnarray}
\label{eq:symm-}
F_{q}^-(\zeta-X,\zeta) &=& F_{q}(\zeta-X,\zeta) - F_{\overline{q}}(\zeta-X,\zeta) \nonumber \\ &=&
-F_{\overline{q}}(X,\zeta) + F_{q}(X,\zeta) \nonumber \\ &=&F_{q}^-(X,\zeta)
\end{eqnarray}
Similarly for the flavor singlet, plus, distribution one finds.
\begin{eqnarray}
\label{eq:symm+}
F^+(\zeta-X,\zeta) &=& \sum_{q} F_{q}(\zeta-X,\zeta) + F_{\overline{q}}(\zeta-X,\zeta) \nonumber \\
&=&\sum_{q} -F_{\overline{q}}(X,\zeta) - F_{q}(X,\zeta) \\
&=& -F^+(X,\zeta)
\end{eqnarray}
The same argument can be used for the gluon distribution in which we find
\begin{eqnarray}
F_{g}(\zeta-X) = F_{g}(X)
\end{eqnarray}

For the helicity GPDs one has,
\begin{eqnarray}
\widetilde{F}_{q}^-(\zeta - X ,\zeta) &=& - \widetilde{F}_{q}^-(X ,\zeta) \\
\widetilde{F}^+(\zeta - X , \zeta) &=& \widetilde{F}^+(X , \zeta)
\end{eqnarray}
Lastly the gluon helicity distribution symmetry in the ERBL region can similarly be found.
\begin{eqnarray}
\widetilde{F}_{g}(\zeta - X) &=& - \widetilde{F}_{g}(X)
\end{eqnarray}

%%%%%%
%%%%%%
%%%%%% FIGURE 5 SYMMETRIES 
\begin{figure}
\includegraphics[scale=0.35]{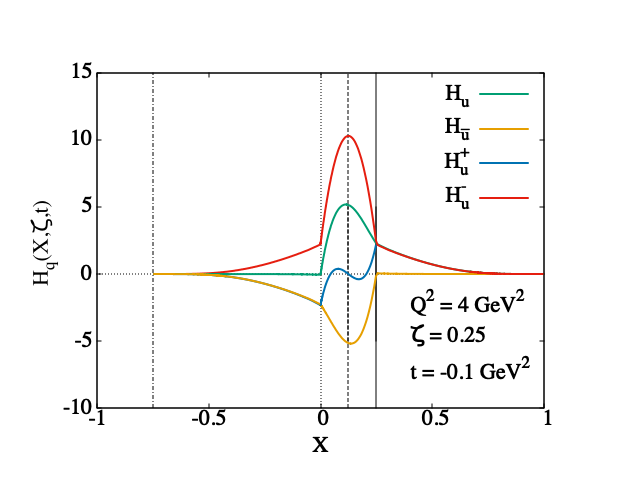}
\includegraphics[scale=0.35]{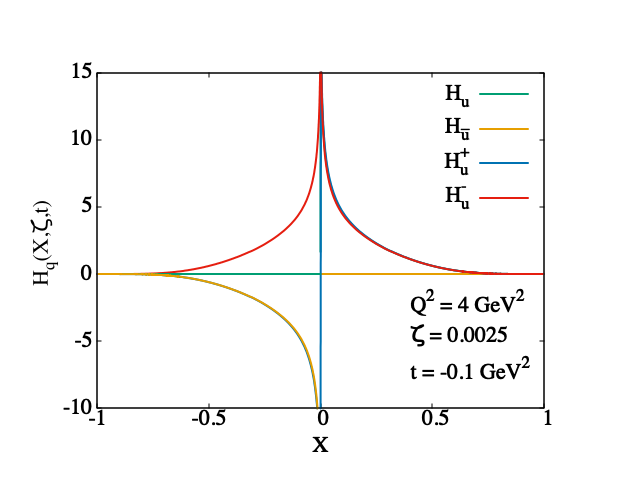}
\includegraphics[scale=0.35]{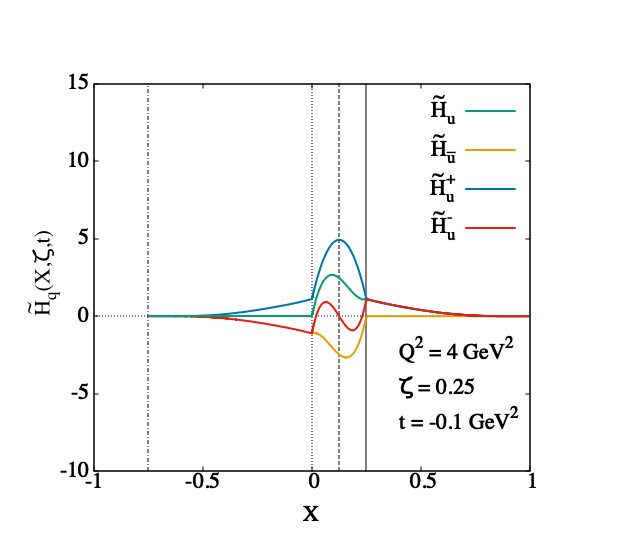}
\caption{Quark symmetries. Thick dotted line is at $\zeta / 2$,the solid line is at $\zeta$, thin dotted line is at 0, and the dot dashed line is at $-1 + \zeta$. The $\overline{u}$ quark distribution has been added here. We can see that the symmetries are made explicit here. In the case of the helicity GPD, the $-$ distribution (representing the valence helicity distribution) is now anti-symmetric due to symmetry constraints and the $+$ distribution (representing the quark sea helicity distribution) is symmetric.}
\label{fig:symmetries}
\end{figure}

%%%%%%%
%%%%%%%
%%%%%%% FIGURE 6
\begin{figure}
\includegraphics[scale=0.35]{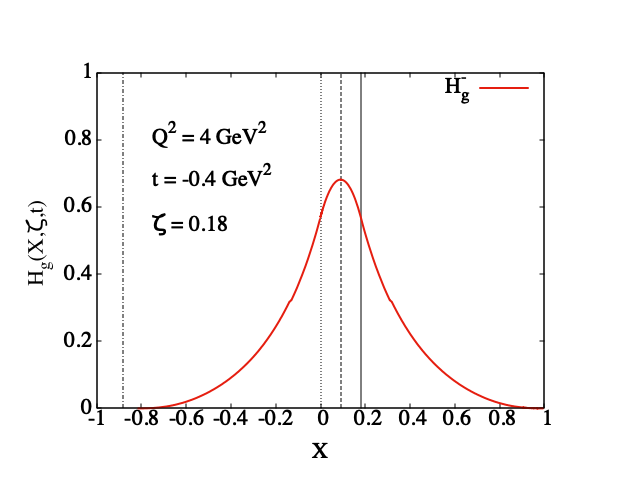}
\caption{The bosonic nature of the gluon means that it is its own anti-particle, therefore the gluon is symmetric about $X = \zeta/2$ or $x = 0$ in the symmetric system of variables.}
\label{fig:gluonsymm}
\end{figure}

The behavior of the valence quark and sea quark distributions around $X=\zeta/2$ is illustrated in Figure \ref{fig:symmetries}, where in the upper panel we show the GPDs $H_u^-$ (red curve) and $H_u^+$ (blue curve), which are respectively, symmetric and antisymmetric with respect to $X=\zeta/2$ (for illustration purposes we take $H_{\bar{u}}=0$ for $X>\zeta$, and $H_u=0$ for $X<0$, in the DGLAP region). From the figure it appears clearly that $H_u$ and $H_{\bar{u}}$ are not symmetric. The middle panel illustrates the symmetries for a low value of $\zeta$, where the ERBL region is suppressed. Finally, the lower panel shows the symmetries for the GPD $\widetilde{H}_u$. 

In Figure \ref{fig:gluonsymm} we show the symmetry of the gluon distribution with respect to $X=\zeta/2$.

%%%%%%%
%%%%%%%
%%%%%%%
\subsection{Valence quark GPDs: SU(4) wave function}
\label{sec:SU4}
For valence quarks the proton-quark-diquark vertex function, Fig.\ref{fig:spectator} and Appendix \ref{sec:appb}, can have two possible couplings depending on whether the outgoing diquark is a scalar ($S=0$), or an axial vector ($S=1$). 
Using the SU(4) symmetry of the proton wave function one has \cite{Jakob:1997wg},
\begin{eqnarray}
\mid p \uparrow \rangle &=& \sqrt{\frac{2}{1+a_S^2}}  
\Big[ \frac{a_S}{\sqrt{2}} \mid u\uparrow S_0^0 \rangle + \frac{1}{3 \sqrt{2}} \mid u\uparrow T_0^0 \rangle \nonumber 
\\
&-& \frac{1}{3} \mid u\downarrow T_0^{1} \rangle
-\frac{1}{3} \mid d\uparrow T_{1}^0 \rangle + \frac{\sqrt{2}}{3} \mid d\downarrow T_{1}^{1} \rangle \Big]
\end{eqnarray}
where $S_0^0 \equiv S_{I_3}^{S_3}$ is the scalar diquark with isospin 0 and spin component 0; $T_{0,1}^{0,1} \equiv T_{I_3}^{S_3}$ is the axial vector (triplet) diquark with indicated isospin and spin components, and the parameter $a_S=1$ for SU(4) symmetry and can differ from 1 to allow for symmetry breaking \cite{Bacchetta:2008af}. 
When matrix elements are formed with this state and the corresponding spin down proton, the sum over the spin states leaves  purely flavor or isospin couplings. This feature of the model allows us to separate out the $u$ and $d$ quark flavors.
 The GPDs, $F=H,E$, decompose as, 
\begin{eqnarray}
F_u &=& \frac{2}{1+a_S^2} \left( \frac{3}{2}a_S^2 F^{(0)} + \frac{1}{2} F^{(1)} \right) \nonumber \\
\label{Fu}
F_d &=& \frac{2}{1+a_S^2} F^{(1)},
\label{Fd}
\end{eqnarray}
For the helicity dependent GPDs, $\widetilde{F}_{q} = \widetilde{H}_{q}, \widetilde{E}_{q}$, only the quark spin state $\mid 0,\, \uparrow \rangle$ contributes and one has,
\begin{eqnarray}
\label{FTu}
\widetilde{F}_u & = &  \frac{2}{1+a_S^2} \left(\frac{3}{2} a_S^2 \widetilde{F}^{(0)} - \frac{1}{6} \widetilde{F}^{(1)}\right)   \nonumber   \\
\label{FTd}
\widetilde{F}_d & = & -  \frac{2}{1+a_S^2} \frac{1}{3} \widetilde{F}^{(1)}, 
\end{eqnarray}
If the outgoing spectactor is a tetraquark, {\it i.e.} in the case of a proton-antiquark-tetraquark coupling (Fig.\ref{fig:spectator}, {\it rhs}), one can also have an $S=2$ outgoing system. However,  we consider only $S=0,1$, and model the $\bar{u}$, $\bar{d}$ distributions similarly to the quark case.

%%%%%%%
%%%%%%%
%%%%%%%
\subsection{Parametrization form}
\label{sec:dglap}
We present our parametric forms  separately for the valence quark ($F^-_q, \widetilde{F}^-_q$), antiquark ($F_{\overline{q}}, \widetilde{F}_{\overline{q}}$)
%and consequently singlet distributions ($F^{+}_{q}, \widetilde{F}^{+}_{q}$)
, and gluon ($F_g, \widetilde{F}_g$) components. 
%In each case we present our formulation for the DGLAP and ERBL regions. 
%
These expressions are valid at an initial scale, $Q_o^2$, therefore the scale does not appear among the arguments.

For all components the functional form in the DGLAP region is given as,
\begin{equation}
F_{DGLAP}(X,\zeta,t)  = F_{M_X,m}^{M_\Lambda}(X,\zeta,t) \,  
R^{\alpha,\alpha^\prime}_p(X,\zeta,t) 
% X^{-(\alpha + \alpha^\prime(X)  t )}   \]
%
\label{fit_form}
\end{equation}
where the functions $F_{M_X,m}^{M_\Lambda} \equiv$ $H_{M_X,m}^{M_\Lambda}$, $E_{M_X,m}^{M_\Lambda}$, $\tilde{H}_{M_X,m}^{M_\Lambda}$, $\tilde{E}_{M_X,m}^{M_\Lambda}$  are obtained as the product of proton-parton-spectator vertices (Fig. \ref{fig:spectator}, Appendix \ref{sec:appb}).
These functions depend on mass parameters: $M_X$, the minimum spectator mass, $m$, the struck parton mass, and $M_\Lambda$, the dipole form factor cut-off mass value.   
$R^{\alpha,\alpha^\prime}_p$ ensures the proper low $X$ Regge behavior resulting from a generalization of the spectator model picture in which 
the mass of the spectator, $M_X$ varies according to a spectral distribution \cite{Brodsky:1973hm}. The spectral function produces a smearing in $M_X \propto 1/X$ such that it reproduces the experimentally observed slope in $X$ for $X\rightarrow 0$, or equivalently at large values of the spectator mass. The role of the spectral function for GPDs was studied in detail in Ref.\cite{GonzalezHernandez:2012jv}.

The parametrization in the ERBL region is obtained by introducing polynomial forms in $X$ that are either symmetric or antisymmetric with respect to the point $X=\zeta/2$, and by imposing the continuity condition at the cross over points, $X=0$, $X=\zeta$, and the polynomiality condition (Section \ref{sec:polinom}). 

All parametric forms are evaluated at an initial scale, $Q_o^2$,  and evolved to the scale where constraints from either experimental data or lattice QCD calculations can be imposed. The value of $Q_o^2$ is, therefore, also a parameter in our fit forms. Its impact on evolution for the various components is presented and discussed in Section \ref{sec:evol}.

The expressions for all GPDs at the initial scale $Q_o^2$, to be readily used in numerical calculations, are summarized in Appendix \ref{sec:appc}. 

%%%%%%
%%%%%%% SECTION PARAMETRIZATION
%%%%%
\section{Parametrization description}
\label{sec:param}
We now present expressions for the parametrization of the valence, antiquark, and gluon distributions evaluated at the initial scale, $Q_o^2$. The parametric forms are subsequently evolved numerically to the scale of current experimental data and can be used directly in the cross section and asymmetry evaluations, including MonteCarlo simulations. The detailed calculations in the spectator model leading to the expressions for the various GPDs are shown in Appendix \ref{sec:appb}. The numerical values of the parameters are listed in the tables in Section \ref{sec:param_1}, where a description of the fitting procedure is also given. 

The current parametrization represents an extension of the one presented in Refs.\cite{Ahmad:2006gn,Ahmad:2007vw,Goldstein:2010gu,GonzalezHernandez:2012jv} in the valence quark sector, to the antiquark and gluon sectors.
The parametrization includes now the GPDs $H$, and $E$ for the following flavors, $u_v$, $d_v$, $\bar{u}$, $\bar{d}$, and $g$; $\widetilde{H}$, and $\widetilde{E}$ for $u_v$, $d_v$.
%, $\bar{u}$, $\bar{d}$. 
The extension to strange and charm quarks can be considered as soon as more stringent constraints from data and lattice QCD will be available.  

An important benchmark for GPD parametrizations is given by the ability to reproduce the behavior of the nucleon form factors when integrated in $X$. In the valence sector, in particular, one can benefit from the flavor separated nucleon Dirac and Pauli form factors obtained in the accurate analysis of Ref.\cite{Cates:2011pz}. For the gluon  GPDs we rely on lattice QCD calculations recently made available in Ref.\cite{Shanahan:2018pib}. On the other hand, we used an approximated method to normalize the antiquark GPDs since, while there exist lattice computations of the second Mellin moments of flavor singlet PDFs \cite{Hagler:2007xi,Alexandrou:2015rja},  a clearcut analysis of flavor separation in the antiquark sector is still lacking (see however Refs.\cite{Deka:2008xr,Deka:2013zha}). 

\begin{figure*}[ht]
\includegraphics[scale=0.375]{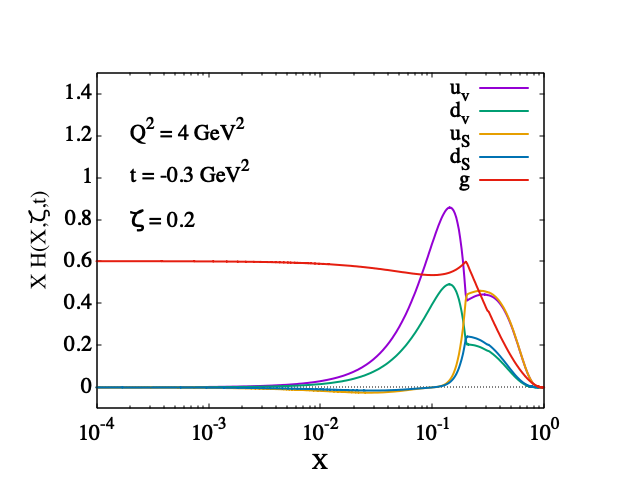}
\includegraphics[scale=0.375]{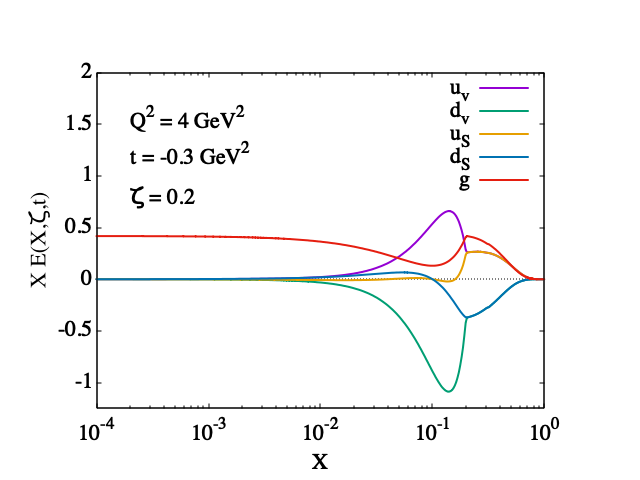}
\caption{GPDs $H$ (left) and $E$ (right) with all flavors for the kinematics $Q^{2} = 4$ GeV$^{2}$, $t = -0.3$GeV$^{2}$, and $\zeta = 0.2$. }
\label{fig:allgpds}
\end{figure*}

An example of the $u_v,d_v,u+\bar{u},d+\bar{d}, g$ GPDs $H$ and $E$, generated with our parametrization is shown in Figure \ref{fig:allgpds} at the kinematic point $t=-0.3$ GeV$^2$, $x_{Bj} \approx \zeta = 0.2$, and $Q^2= 4$ GeV$^2$. 

The range of validity of our parametrization is:
\begin{itemize}
    \item $0.0001 \leq X \leq 0.85$
    \item $0.01 \leq \zeta \leq 0.85$, 
    \item $0 \leq -t \leq 2$ GeV$^2$
    \item $1 \leq Q^2  \leq 100$ GeV$^2$ .
\end{itemize}

%%%%%%
%%%%%% FIGURE 7

%%%%
\subsection{Valence quarks}
\label{sec:valence}
Our model uses two different descriptions of the valence quark distribution in the DGLAP and ERBL regions.
The DGLAP region can be considered a direct extension of
the parton model in the forward region, where the struck quark with initial
longitudinal momentum fraction $X$ is reinserted in the
proton target after reducing it to
$X-\zeta$, $\zeta$ being the fraction transferred in the exclusive scattering process. In the DGLAP region, the initial and final quarks
are both off-shell, while the diquark intermediate state is
on mass shell. The ERBL region is described through a minimal mathematical form that is consistent with the properties of continuity at $X=\zeta$, polynomiality, and $X$ symmetry. This form is sufficiently flexible to describe the data where GPDs appear integrated over in the CFFs, while avoiding ambiguities due to semi-disconnected diagrams which are inherent to a partonic formulation \cite{Goldstein:2010gu}.

\subsubsection{DGLAP region: $\zeta < X <1 $ ($\xi<x <1$)}
For the valence quark distributions the spectator is a system with diquark quantum numbers and variable mass, $M_X$, with spin $S=0,1$. 
The analytic expressions of our model are given directly as a function of the mass parameters for the quark, $m$, diquark, $M_X$, and dipole mass parameter, $M_\Lambda$. We set $a_s=1$, but allow the mass parameters to vary in the axial-vector sector ($\widetilde{H}$, $\widetilde{E}$) with respect to the same parameters for $H$ and $E$. 
%All parameters along with the fitting procedure are described in Section \ref{sec:param}. 

The parametric forms read,
%\vspace{-0.1cm}
%%%%%
%%%%% H and E and H tilde
%%%%%
\begin{widetext}
\begin{eqnarray}
\label{GPDHpar} 
H_{M_X,m}^{M_\Lambda} & = &  2\pi  \mathcal{N} \left(1-\displaystyle\frac{\zeta}{2}\right) %\frac{(1-X)^3}{(1-\zeta)^2} 
 \int_0^\infty  \frac{d k_\perp k_\perp}{1-X} 
\frac{% 
 a \,  \left[(m+M X)  \left(m + M X' \right) +  k_\perp^2 \right]   - b \,  (1-X') \, k_\perp \Delta_\perp }%
{D^2 \,  \left(a^2 - b^2\right)^{3/2}  }       +   \frac{\zeta^2}{4(1-\zeta)} E_{M_X,m}^{M_\Lambda}, \nonumber  \\ \\
\label{GPDEpar}
E_{M_X,m}^{M_\Lambda}& = &  2\pi  \mathcal{N}  \left(1-\displaystyle\frac{\zeta}{2}\right) \int_0^\infty    \frac{d k_\perp k_\perp}{1-X}
%%%
 \frac{-4M k_\perp^2 [-(m + M X)+ (m+ M X')] + a \, [2 M(-m + M X) ] (1-X') }%
{(1-\zeta) \,  D^2 \,  \left(a^2 - b^2\right)^{3/2}  }  \\
%%%
\widetilde{H}_{M_X,m}^{M_\Lambda} & = &  2\pi  \mathcal{N}  \left(1-\displaystyle\frac{\zeta}{2}\right) 
\int_0^\infty \frac{d k_\perp k_\perp}{1-X} 
\frac{% 
 a \,  \left[(m+M X)  \left(m + M X' \right) -  k_\perp^2 \right]   + b \,  (1-X')  \, k_\perp \Delta_\perp }%
{ D^2 \,  \left(a^2 - b^2\right)^{3/2} } 
 +   \frac{\zeta^2}{4(1-\zeta)}\widetilde{E}_{M_X,m}^{M_\Lambda}
\label{GPDHTILDEpar} \nonumber \\ 
\\
\widetilde{E}_{M_X,m}^{M_\Lambda} & = &  2\pi  \mathcal{N}  \left(1-\displaystyle\frac{\zeta}{2}\right) \, \frac{1-\zeta}\zeta \int_0^\infty    \frac{d k_\perp k_\perp}{1-X}
%%%
 \frac{-4M k_\perp^2 [(m + M X)+ (m+ M X')] - a \, [4 M(-m + M X) ] (1-X') }%
{ D^2 \,  \left(a^2 - b^2\right)^{3/2}  } \nonumber \\
\label{GPDETILDEpar}
\end{eqnarray} 
\end{widetext}
%%%%%
where $M$ is the proton mass, $X'$ given in Eq.(\ref{eq:Xprime}) and,
\begin{subequations}
\label{eq:abD}
\begin{eqnarray}
& a  =  \mathcal{M}(X') - \displaystyle\frac{k_\perp^2}{1-X'}  - \Delta_\perp^2  (1-X'), \quad\quad 
b  =  2 k_\perp \Delta_\perp, \nonumber \\ 
\\
&D  =  \mathcal{M}(X)- \displaystyle\frac{k_\perp^2}{1-X}, \\
\label{eq:calliM}
&\mathcal{M}(Y)  =   Y M^2 - M_{\Lambda}^{2} - M_{X}^{2} \displaystyle\frac{Y}{1-Y} ,
\end{eqnarray}
\end{subequations}
where $Y=X,X'$.
%\begin{eqnarray}
%D_0 & = & \left[(1-X){\cal M}(X)^2 - k^2_\perp\right] \\
%D_1 & = & \left[(1-X'){\cal M}(X')^2 - k^2_\perp - \Delta_\perp^2 (1-X')^2 \right] \\
%D_2 & = & (1-X')  k_\perp  \Delta_\perp  \\
%\mathcal{M}(X) =  x M^2 - (M_\Lambda)^2 - (M_X)^2 \frac{X}{1-X}  \, .
%\end{eqnarray} 
%

%%%%
%%%% REGGE
%%%%
\noindent We parametrize the Regge term as follows,
\begin{equation}
R^{\alpha,\alpha^\prime}_{p} =  X^{-[\alpha + \alpha^\prime(1-X)^p t  ]},
\label{eq:regge}
\end{equation}
where the parameters: $\alpha, \alpha', p$ take on different values depending on the GPD.
%Only the term $X^{-\alpha}$ can be considered a proper Regge contribution.
Notice that, although our parametric form is given in terms of the asymmetric set of variables, $(X,\zeta,t)$, these can readily be transformed into the symmetric set $(x,\xi,t)$ using Eqs.\eqref{eq:symm1},\eqref{eq:symm2}.

Summarizing, the expressions for the valence quarks in the DGLAP region are given by,
\begin{eqnarray}
\label{eq:Hqv}
H^-_q &= H_{q_v}(X,\zeta,t)  = H_{M_X,m}^{M_\Lambda}(X,\zeta,t) \,  
R^{\alpha,\alpha^\prime}_{p}(X,t) \\ \nonumber \\
\label{eq:Eqv}
E^-_q &= E_{q_v}(X,\zeta,t)  =  E_{M_X,m}^{M_\Lambda}(X,\zeta,t) \,  
R^{\alpha,\alpha^\prime}_{p}(X,t) \\ \nonumber \\
\label{eq:Htilqv}
\widetilde{H}^-_q &= \widetilde{H}_{q_v}(X,\zeta,t)  = \widetilde{H}_{M_X,m}^{M_\Lambda}(X,\zeta,t) \,  
R^{\alpha,\alpha^\prime}_{p}(X,t) \\ \nonumber \\
\label{eq:Etilqv}
\widetilde{E}^-_q &= \widetilde{E}_{q_v}(X,\zeta,t) =  \widetilde{E}_{M_X,m}^{M_\Lambda}(X,\zeta,t) \,  
R^{\alpha,\alpha^\prime}_{p}(X,t) \;,
\end{eqnarray}
where $H_{M_X,m}^{M_\Lambda}$ is given in Eq.\eqref{GPDHpar}, $E_{M_X,m}^{M_\Lambda}$ in Eq.\eqref{GPDEpar}, $\widetilde{H}_{M_X,m}^{M_\Lambda}$ in Eq.\eqref{GPDHTILDEpar} and $\widetilde{E}_{M_X,m}^{M_\Lambda}$ in Eq.\eqref{GPDETILDEpar}. While these are different functional forms, all GPDs have the same form of the Regge term. The parameters values specific to each GPD 
 are listed in Section \ref{sec:param_1}.

%%%%%%%
%%%%%%% VALENCE, ERBL
\subsubsection{ERBL region: $0<X<\zeta$ ($\xi<x<\xi$)}
\label{sec:ERBL_val}
To parametrize the valence component of the GPDs $H$ and $E$ in the ERBL region we use the symmetry around the point $X=\zeta/2$ ($x=0$) for
the $F^-$, flavor non-singlet distributions, given in Eq.\eqref{eq:Fminus}.
By choosing a quadratic form for the $X$ dependence, we can fix the three unknown parameters,
\begin{equation}
\label{eq:valenceerbl}
 F_{q_v}(X,\zeta,t) = F_q^- = a_F X^2 +b_F X + c_F
 \end{equation}
by imposing the following conditions on the symmetric component, $F_{ERBL}^-=H^-,E^-$:
%\widetilde{H}^-,\widetilde{E}^-$,

\vspace{0.5cm}
\noindent 
(1) symmetry around $X=\zeta/2$ $\Rightarrow$  $F^-(\zeta,\zeta,t) = F^-(0,\zeta,t)$,

\vspace{0.3cm} 
\noindent 
(2) continuity condition at $X= \zeta$, $F^-_{ERBL}(\zeta,\zeta,t) = F^-_{DGLAP}(\zeta,\zeta,t)$,

\vspace{0.3cm} \noindent
(3) polynomiality at leading order,  Eqs.(\ref{eq:Dirac}-\ref{eq:pseudo}), taking $H$, for instance, 
\begin{eqnarray}
    F_1^q(t) &=& \int_{-1+\zeta}^1 \frac{d X}{1 -\zeta/2} H^q(X,\zeta,t) \nonumber \\
 && \hspace{-0.8cm} = \frac{1}{2} \int_{\zeta/2}^\zeta \frac{d X}{1 -\zeta/2} H^-_q(X,\zeta,t) + \int_\zeta^1 \frac{d X}{1 -\zeta/2} H_q^-(X,\zeta,t) \nonumber \\
\end{eqnarray}
By using the constraints $(1)$ and $(2)$, one finds,
\[ b_F=-\zeta a_F \quad\quad c_F=F_{DGLAP}(\zeta,\zeta,t). \] 
The parameter $a$ is determined imposing constraints $(3)$, giving,
\begin{eqnarray}
\label{eq:aminusH}
a_F & = & \frac{6}{\zeta^3} \Big[\zeta F(\zeta,t) - 2 S_F(\zeta, t) \Big], 
\end{eqnarray}
where, 
\begin{eqnarray}
H({\zeta,t}) = H_{DGLAP}(\zeta,\zeta,t), \quad E({\zeta,t}) = E_{DGLAP}(\zeta,\zeta,t) \nonumber 
\end{eqnarray}
 are the GPD values at the crossover point between the ERBL and DGLAP regions calculated using Eqs.(\ref{GPDHpar}-\ref{GPDETILDEpar}). 
$S_F$ is the area subtended by $F^- \equiv H^-, E^-$, $\widetilde{H}^+, \widetilde{E}^+$, respectively Eqs.(\ref{eq:Fminus}, \ref{eq:Fplus}, \ref{eq:simmF-}, \ref{eq:simmF+}), in the ERBL $(X<\zeta)$ region. 
This is obtained by subtraction from the various form factors, Eqs.(\ref{eq:Dirac}-\ref{eq:pseudo}) as, 
%%%%
%%%%
\begin{subequations}
\begin{eqnarray}
&&S_H  =  \int_0^\zeta  dX \, H^-(X,\zeta,t) = \nonumber \\
 && \hspace{0.4cm} =\left(1-\frac{\zeta}{2} \right) \left(F_1 - \int_\zeta^1 \frac{H(X,\zeta,t)}{1-\zeta/2} dX \right) 
\\
&&S_E  =  \int_0^\zeta  dX \, E^-(X,\zeta,t) = \nonumber \\
&& \hspace{0.4cm} = \left(1-\frac{\zeta}{2} \right) \left(F_2 - \int_\zeta^1 \frac{H(X,\zeta,t)}{1-\zeta/2} dX \right) 
%\\
%&&S_{\widetilde{H}}  =  \int_0^\zeta  dX \, \widetilde{H}^+(X,\zeta,t) = \nonumber \\
%&& \hspace{0.4cm} = \left(1-\frac{\zeta}{2} \right) \left(G_A - \int_\zeta^1 \frac{\widetilde{H}(X,\zeta,t)}{1-\zeta/2} dX \right) 
%\\
%&&S_{\widetilde{E}}  =  \int_0^\zeta  dX \, \widetilde{E}^-(X,\zeta,t) = \nonumber \\
%&& \hspace{0.4cm} = \left(1-\frac{\zeta}{2} \right) \left(G_P - \int_\zeta^1 \frac{\widetilde{E}(X,\zeta,t)}{1-\zeta/2} dX \right)  
\end{eqnarray}
\end{subequations}
%%
%%%%
Notice that $S_F$ appears in the definition of $a$, Eq.\eqref{eq:aminusH}, multiplied by a factor of $2$ because of the crossing symmetry property for the
areas subtended by $F^-$ and $F$ (see Section \ref{sec:symm}). 

The final analytic expressions are given by,
\begin{eqnarray}
\label{eq:hqverbl}
H_{q_v}(X,\zeta,t) & = & a_H X^2 -a_H \, \zeta X + H({\zeta,t}) \nonumber \\ \\
\label{eq:eqverbl}
E_{q_v}(X,\zeta,t) & = & a_E X^2 -a_E \, \zeta X + E({\zeta,t}) \nonumber \\ 
%\\
%\widetilde{H}_{ERBL}(X,\zeta,t) & = &  {a}_{\widetilde{H}} X^2 -a_{\widetilde{H}} \, \zeta X + \widetilde{H}({\zeta,t}) \nonumber \\ \\
%\widetilde{E}_{ERBL}(X,\zeta,t) & = & {a}_{\widetilde{E}} X^2 -a_{\widetilde{E}} \, \zeta X + \widetilde{E}({\zeta,t}) \nonumber \\
\end{eqnarray}
where $a_H$ and $a_E$ are calculated from Eq.(\ref{eq:aminusH}).

To conclude, our parametric form in the ERBL region introduces no free parameters. As more data from DVCS and related experiments become available, thus allowing a larger number of  parameters, more flexibility could be introduced by {\it e.g.} including higher powers in $X$.

For the valence components of the GPDs $\widetilde{H}$ and $\widetilde{E}$ the symmetry is opposite, {\it i.e.} $\widetilde{F}^-$ is antisymmetric around $X=\zeta/2$ ($x=0$). We consider the following form,
\begin{eqnarray}
\label{h+}
\widetilde{F}^-_q =\widetilde{F}_{q_v}(X,\zeta,t) & = & a_{\tilde{F}} X^3 + b_{\tilde{F}} \zeta X^2 + c_{\tilde{F}} X + d_{\tilde{F}}   . \nonumber \\
\end{eqnarray}
Similar to $H, E$, the parameters can be fixed by considering the symmetry conditions for anti-quarks, 

\vspace{0.5cm}
\noindent (1) antisymmetry around $X=\zeta/2$ $\Rightarrow$ $\widetilde{F}(\zeta,\zeta,t) = - \widetilde{F}(0,\zeta,t)$. 

\noindent (2) continuity condition at $X=\zeta$, $\widetilde{F}_{ERBL}(\zeta,\zeta,t) = \widetilde{F}_{DGLAP}(\zeta,\zeta,t)$,

\noindent (3) $\widetilde{F}_{ERBL}=0$ at $X=\zeta/2$ (the integral in $X$ is zero).

One can therefore determine three of the parameters as,
\begin{eqnarray}
b_{\tilde{F}} & = & - \frac{3}{2} a_{\tilde{F}} \zeta, \quad \quad 
c_{\tilde{F}}  =  \frac{1}{\zeta} \left[2  \widetilde{F}(\zeta,t) + \frac{1}{2} a_{\tilde{F}} \zeta^3\right]  \quad \quad \nonumber \\
{d}_{\tilde{F}} & = &- F(\zeta,t) 
\end{eqnarray}
with $\widetilde{F}(\zeta,t) = \widetilde{F}_{DGLAP}(\zeta,\zeta,t)$,
while $a_{\tilde{F}}$ is a free parameter which was determined numerically (see Table \ref{tab:H}). 

The analytic expressions for the valence helicity GPDs are given by,
\begin{eqnarray}
\label{eq:h-}
\widetilde{H}_{q_v}(X,\zeta,t) & = &  {a}_{\widetilde{H}} X^3 - \frac{3}{2} a_{\widetilde{H}} \, \zeta X^2 + {c}_{\widetilde{H}} X + \widetilde{H}({\zeta,t}) \nonumber \\ \\
\label{eq:e-}
\widetilde{E}_{q_v}(X,\zeta,t) & = & {a}_{\widetilde{E}} X^3 - \frac{3}{2} a_{\widetilde{E}} \, \zeta X^2 + {c}_{\widetilde{E}} X + \widetilde{E}({\zeta,t}) \nonumber \\
\end{eqnarray}
The values of the parameters $a_{\widetilde{H}}$, $a_{\widetilde{E}}$ are given in the tables in Section \ref{sec:param_1}. All other parameters are constrained. 
%%%%%%%%%%%%%%%%%
%%%%%%%%%%%%%%%%% SEA QUARKS
%%%%%%%%%%%%%%%%%
\subsection{Antiquarks}
\label{sec:antiqDGLAP}
Similar to the valence quarks, we describe the antiquark GPDs in a spectator model in the DGLAP region, and in a symmetric parametric form in the ERBL region.   

%%%%%
%%%%%
\subsubsection{DGLAP region: $-1 +\zeta<X<0$, \, ($-1 < x <-\xi$)}
\label{sec:seaDGLAP}
In the spectator model, if the struck parton is an antiquark, the spectator is a tetraquark (Figure \ref{fig:spectator}). Because the tetraquark can have spin $S=0,1,2$, the wave function has, in principle, a more complicated form than the SU(4) form described in Section \ref{sec:SU4}, which would allow for more quark flavors than just the $u$ and $d$ quarks. We, however, consider a simplified version and we adopt the same mathematical expressions given for the valence quarks in Eqs.(\ref{GPDHpar}, \ref{GPDEpar}, \ref{GPDHTILDEpar}, \ref{GPDETILDEpar}), with different values of the mass parameters. 
The parametrization forms in the antiquark sector are
\begin{eqnarray}
\label{eq:Hqbar}
H^+_q &=  H_{M_X,m}^{M_\Lambda}(X,\zeta,t) \,  
R^{\alpha,\alpha^\prime}_{p}(X,t) \\ \nonumber \\
\label{eq:Eqbar}
E^+_q &=   E_{M_X,m}^{M_\Lambda}(X,\zeta,t) \,  
R^{\alpha,\alpha^\prime}_{p}(X,t) \\ \nonumber \\
\label{eq:Htilqbar}
\widetilde{H}^+_q &=  \widetilde{H}_{M_X,m}^{M_\Lambda}(X,\zeta,t) \,  
R^{\alpha,\alpha^\prime}_{p}(X,t) \\ \nonumber \\
\label{eq:Etilqbar}
\widetilde{E}^+_q &=  \widetilde{E}_{M_X,m}^{M_\Lambda}(X,\zeta,t) \,  
R^{\alpha,\alpha^\prime}_{p}(X,t) \;,
\end{eqnarray}
The parameter values are listed in Table \ref{tab:E}. Notice that the Regge term also has the same form as for the valence contribution, Eq.(\ref{eq:regge}).

\subsubsection{ERBL region}
\label{sec:ERBL_sea}
In the ERBL region the GPDs $H$ and $E$ for antiquarks are antisymmetric with respect to $X=\zeta/2$, while $\widetilde{H}$ and $\widetilde{E}$ are symmetric. We choose, therefore the following form for $H$ and $E$ (analogous to the axial vector sector in Section \ref{sec:ERBL_val}) where $a^+$ is a free parameter,
\begin{eqnarray}
\label{eq:h+}
F^+_{ERBL}(X,\zeta,t) & = & a^+ X^3 - \frac{3}{2} a^+ \zeta X^2 + c X + d   , \nonumber \\
\end{eqnarray}
The coefficients $b$, $c$, $d$ are constrained similarly to Eqs.(\ref{eq:h-},\ref{eq:e-}). 

For the GPDs $\widetilde{H}$ and $\widetilde{E}$ we take a symmetric form analogous to Eq.(\ref{eq:valenceerbl}), in the vector sector. We have,
\begin{eqnarray}
\label{eq:Htilqbarerbl}
\widetilde{H}_{ERBL}(X,\zeta,t) & = &  {a}_{\widetilde{H}} X^2 -a_{\widetilde{H}} \, \zeta X + \widetilde{H}({\zeta,t}) \nonumber \\ \\
\label{eq:Etilqbarerbl}
\widetilde{E}_{ERBL}(X,\zeta,t) & = & {a}_{\widetilde{E}} X^2 -a_{\widetilde{E}} \, \zeta X + \widetilde{E}({\zeta,t}) \nonumber \\
\end{eqnarray}
Therefore, we have no free parameters for the antiquark  axial-vector GPDs in the ERBL region.

%%%%%%%%%%%%%%%%%
%%%%%%%%%%%%%%%%% GLUONS
%%%%%%%%%%%%%%%%%
\subsection{Gluons}
\label{sec:gluons}
A well known issue to PDF fitters is that a non negligible gluon density needs to be present already at a low scale in order to ensure that perturbative QCD evolution of the parton distributions produces a steep enough slope to reproduce the data at low $X$. If, on the contrary, gluon distributions are initially set to zero, and only generated perturbatively, the resulting quark/antiquark distributions become too soft.
We model the gluon distribution at the initial scale  
%a strong gluon is emitted at the proton vertex, we assume 
in a spectator model with a strong gluon emitted at the proton vertex leaving behind an octet color state with proton quantum numbers. The vertex is described by (Figure \ref{fig:spectator}, Appendix \ref{sec:appb}),
\[ \Gamma(k) \bar{u}(p-k)\gamma_\mu U(p)\, \varepsilon^\mu(k) \]
where $u(p-k)$ is the outgoing color octet proton,  $U(p)$ is the incoming proton, $\varepsilon^\mu$ is the struck gluon wave function; $\Gamma(k) \gamma_\mu$ describes the coupling at the proton-octet proton-gluon vertex in a similar way to the proton-quark-diquark vertex (details are given in Appendix \ref{sec:appb}).  This model allows us to evaluate the gluon GPDs in the DGLAP region. We extend our calculation to the ERBL region using the symmetry properties of gluon distributions described in Section \ref{sec:symm}. An important part of our calculation is given by the fact that we can model the $t$ dependence of gluon GPDs by ensuring that our model follows the normalization provided by the Mellin moments evaluations for $H_g$ andf $E_g$ given in Ref.\cite{Shanahan:2018pib}.

\subsubsection{DGLAP region: $\zeta <X<1$, \ ($\xi < x < 1$)}
The  gluon-proton amplitudes for the GPDs, $H_g$, $E_g$, constructed from the tree level vertex for the process,
\[ p \rightarrow g + p_8 ,\]
where $p_8$ is a color octet spetactor baryon with spin $1/2$ and momentum $k_X= p-k = p^\prime -k^\prime$,
are given by the following expressions,
\begin{widetext}
\begin{eqnarray}
\label{eq:Hg} 
H_{M_X^g}^{M_\Lambda^g} & = & 2 \pi \mathcal{N} \int d k_{\perp} \frac{k_{\perp}}{1-X} \,  
 \frac{1}{D^2 \,  \left(a^2 - b^2\right)^{3/2}  } 
 \nonumber \\
 & \times& \left\{ a\left[ X X' \Big((1-X) M - M_X \Big) \Big((1-X') M - M_X \Big) + \left(\displaystyle\frac{1}{1-X'} - (1-X) \right)k_\perp^2 \right]  - b \, \left(\displaystyle\frac{1}{1-X} + (1-X') \right) k_\perp \Delta_\perp \right\} 
 \nonumber \\
 & + & \frac{\zeta^2}{4(1-\zeta)} \, E_g  
\\
E_{M_X^g}^{M_\Lambda^g} & = & 2 \pi \mathcal{N} \int d k_{\perp} \frac{k_\perp}{1-X}  
\frac{1}{D^2 \,  \left(a^2 - b^2\right)^{3/2}  } 
\frac{-2M(1-\zeta)}{1-\frac{\zeta}{2}}
 \nonumber \\ 
 &\times & 
 \left\{ 2 k_T^2 \left[ X((1-X)M-M_X) - X'(1-\zeta)((1-X') M - M_X)  \right] - a (1-X') X\left[(1-X)M - M_X  \right]\right\} ,\nonumber \\ 
\label{Eg}
\end{eqnarray}
%%%%%
%%%%%
%%%
where $a$, $b$, $D$ are given by  the same definitions as in Eqs.\eqref{eq:abD}.

For the gluon helicity dependent GPDs we find,
\begin{eqnarray}
\tilde{H}_{M_X^g}^{M_\Lambda^g} &= &\mathcal{N} \int d^2k_{\perp}  \frac{1}{(1-X)^2} 
\frac{\left[X(X-\zeta)((1-X)M-M_X)\left(\displaystyle\frac{1-X}{1-\zeta}M - M_X \right)+\left(1-\zeta - (1-X)^2 \right) {\bf k}_T \cdot \tilde{\bf k}_T \right]}{(k^2-M_{\Lambda}^2)^2(k^{\prime2}-M_{\Lambda}^2)^2}  \nonumber \\
&+& \frac{\zeta^2}{4(1-\zeta)}\tilde{E}_g 
\label{Htildeg}
\\
\tilde{E}_{M_X^g}^{M_\Lambda^g} &= &\mathcal{N} \int d^2k_{\perp}  \mathrm{} \, \frac{2}{\zeta} \,\, \frac{(-2M)(1-\zeta)}{(1-X)}  
\times \frac{\left[X((1-X)M-M_X) \displaystyle\frac{\tilde{\bf k}_T \cdot {\bf \Delta}_T}{\Delta_T^2} + (X-\zeta)(\frac{1-X}{1-\zeta})M-M_X) \displaystyle\frac{\tilde{\bf k}_T \cdot {\bf \Delta}_T}{\Delta_T^2} \right]}{(k^2-M_{\Lambda}^2)^2(k^{\prime2}-M_{\Lambda}^2)^2}
\label{Etildeg}
\end{eqnarray} 
\end{widetext}
%The appearance of $\Delta_\perp E_g, \Delta_\perp {\tilde E}_g$ integrands and  $1/\zeta$ in the ${\tilde E}_g$ integrand requires that for both of those GPDs the rest of the integrand must  vanish as $\Delta_\perp \, \& \, \zeta\rightarrow 0$. This is the case as we can see by evaluating the ${\vec k}_\perp$ angular integration exactly. 
%The ${\vec k}_\perp$ angular integration of the partially integrated helicity amplitudes in Eq.~\ref{A01s}-\ref{fliphelicity} can be written in the form
%\begin{eqnarray}
%A_{\Lambda^{\prime},\Lambda_{g^{\prime}};\Lambda,\Lambda_g}(X, \zeta, t) &=& \int dk_\perp k_\perp \int_0^{2\pi} d\phi \mathcal{F}(X,\zeta) 
%\frac{(1, \cos\phi, \cos^2\phi)}{(k^2-m_g^2)^2(k^{\prime \, 2} - m_g^2)^2},
%\end{eqnarray}
%with the $k^{\prime \, 2}$ containing $k_\perp \, \Delta_\perp \, \cos\phi$.
The gluon Regge term is the same as Eq.(\ref{eq:regge}) where the $\alpha_g$ parameter is obtained from fitting to the power of $H_g(X,0,0) \equiv X g(X)$, the gluon PDF.

The expressions for the gluon distributions for unpolarized gluons in the DGLAP region are given by,
\begin{eqnarray}
\label{eq:Hgspect}
H^g(X,\zeta,t)  = H_{M_X^g}^{M_\Lambda^g}(X,\zeta,t) \,  R^{\alpha_g,\alpha^\prime_g}_p(X,t) \\ \nonumber \\
\label{eq:Eg}
E^g(X,\zeta,t)  = E_{M_X^g}^{M_\Lambda^g}(X,\zeta,t) \, R^{\alpha_g,\alpha^\prime_g}_p(X,t) 
\end{eqnarray}
We limit ourselves to listing the expressions for the helicity gluon distributions, since their parameters cannot be sensibly constrained at present, as neither experimental results nor lattice QCD calculations are either accurate enough or available. 
%e present a model calculation based on current phenomenological information.  

%%%%%%
%%%%%% GLUONS ERBL
\subsubsection{ERBL region}
%The gluon form factor is defined through integration from 0 to 1 similar to the valence quark distribution. 
Gluons share the same symmetry properties as the valence quark distributions, that is they are symmetric around $x=0$. Therefore the property of their form factor integration over $x$ is used in a similar way to obtain the behavior in the ERBL region. Notice that for the gluon GPDs  are $H_g(x,0,0) = x g(x)$, therefore the form factor integrals of the gluon GPDs correspond to the second Mellin moments of the energy momentum tensor form factors.

The analytic expressions in the unpolarized gluon sector are given by,
\begin{eqnarray}
\label{eq:Hgerbl}
H_g(X,\zeta,t) & = & a_H X^2 -a_H \, \zeta X + H({\zeta,t}) \nonumber \\ \\
\label{eq:Egerbl}
E_g(X,\zeta,t) & = & a_E X^2 -a_E \, \zeta X + E({\zeta,t}) \nonumber \\ 
\end{eqnarray}
The values of the parameters are listed in Section \ref{sec:param_1}.

%%%%%
%%%%%% FORWARD
\subsection{Forward limit, $\zeta = 0$ and $t = 0$}
\label{sec:teq0}
The limits: (1) $\zeta=0$, $t \neq 0$; (2) $\zeta=0$, $t=0$, represent important physical cases (note that in this case, $X=x$). The former is needed to perform Fourier transforms in the transverse plane \cite{Burkardt:2000za}, while (2) gives the connection to PDFs, namely $H_{q}(x,0,0) \equiv f_1^q(x)$, $H_{g}(x,0,0) \equiv x g(x)$ $\widetilde{H}_q \equiv g_1^q(x)$, $\widetilde{H}_g \equiv \Delta x \Delta g(x)$ (Section \ref{sec:forward}. Moreover, the GPDs $H_{q,g}(X,0,0)$, $E_{q,g}(X,0,0)$ define the angular momentum sum rule, Eq.(\ref{eq:JiSR}) \cite{Ji:1996ek}.

For valence and sea quarks the parametric expressions simplify to,
%%%%%
%%%%% H and E
%%%%%
\begin{eqnarray}
f_1^q(x) & = & \displaystyle \mathcal{N}  x^{-\alpha} \int d^2k_\perp 
\frac{% 
 \left[ (m+M x)^2  + {\bf k}_\perp^2 \right](1-x)^3}%
{\left[(1-x) {\cal M}(x) - {\bf k}_\perp^2 \right]^4}     \\
\label{GPDH00}
E_q(x) & = & \displaystyle \mathcal{N} x^{\alpha} \int  d^2k_\perp 
%%%
\frac{ 2M %
 \left(m + M x \right)  (1-x)^4}%
 {\left[(1-x) {\cal M}(x) - {\bf k}_\perp^2 \right]^4 }
 \label{GPDE00}  \\
%%%
%%%%%
%%%%% H tilde and and E tilde
%%%%%
g_1^q(x) & = &  \displaystyle \mathcal{N} x^{\alpha} \int d^2k_\perp 
\frac{% 
 \left[ (m+M x)^2  - {\bf k}_\perp^2 \right](1-x)^3}{\left[(1-x) {\cal M}(x) - {\bf k}_\perp^2 \right]^4  }    
\label{GPDHTILDE00}
\end{eqnarray} 
\begin{widetext}
\begin{eqnarray}
\widetilde{E}_q(x)  =  \displaystyle  \frac{ \mathcal{N} \pi M^2}{\mathcal{M}(x) } \, x^{-\alpha} \, (1-x)^3 \Big[ \: \frac{1}{3} 
 - \frac{4}{5} \frac{(M+mx)(M^2 (1-2x) - M_X^2 + M_\Lambda^2) }{M \mathcal{M}(x)(1-x)} \Big]
\label{GPDETILDE00}
\end{eqnarray} 
 \end{widetext}
where $\mathcal{M}$ defined in Eq.(\ref{eq:calliM}), having dimensions of $M^2$, contains the parameters, $m, M_X, M_\Lambda$. The total number of parameters per GPD flavor is therefore, five, namely,  $m, M_X, M_\Lambda, \cal{N}, \alpha$.
    
For the gluon GPDs we have,
\begin{widetext}
\begin{eqnarray}
\label{eq:Hgt0}
x g(x) &=& 
x^{-\alpha}\mathcal{N} \int d^2k_{\perp}  (1-x)^2 
\frac{[x^2((1-x)M-M_X)^2+(1+(1-x)^2)k_{\perp}^2]}{(x M_X^2+(1-x)M^2_{\Lambda}-x(1-x)M^2+k_{\perp}^2)^4}   
\\
%%%
E_g &=& 
x^{-\alpha}\mathcal{N} \int d^2k_{\perp}  (1-x)^4
\frac{-2M x((1-x)M-M_X)}{(x M_X^2+(1-x)M^2_{\Lambda}-x(1-x)M^2+k_{\perp}^2)^4} 
\\
%%%
x \Delta g(x) &=& 
x^{-\alpha}\mathcal{N} \int d^2k_{\perp}  (1-x )^2 
\frac{[x^2((1-x)M-M_X)^2+(1-(1-X)^2)k_{\perp}^2]}{(x M_X^2+(1-x)M^2_{\Lambda}-X(1-X)M^2+k_{\perp}^2)^4}  
\\
%%%
\tilde{E}_g&=& x^{-\alpha}\mathcal{N} \int d^2k_{\perp} (1-x)^3\frac{[8M((1-x)^2M-M_X) k_\perp^2]}{(xM_X^2+(1-x)M^2_{\Lambda}-x(1-x)M^2+k_{\perp}^2)^3} ,
\end{eqnarray}
\end{widetext}
Notice that the integrals in $d^2 k_\perp$ defining $\widetilde{E}$ in Eqs.(\ref{eq:Etilqv},\ref{Etildeg}) do not diverge for $\zeta \rightarrow 0$, since the terms in the numerator also go to zero, thus canceling the divergence. This can be seen by inspecting the helicity amplitude substructure shown in Appendix \ref{sec:appb}, where one has,
 \[A_{++,-+} + A_{-+,++}\mid_{\zeta=0} = 0,  \] 
leading to $\lim_{\zeta \to 0} \tilde{E}_{q,g}  \rightarrow \; {\rm constant}, $.

%%%%%%%%%%%%%
%%%%%%%%%%%%%
%%%%%%%%%% PARAMETERS
\section{PDF fit parameters}
\label{sec:param_1}
We present our fit parameters for: $H$ ($u_v$, $d_v$, $g$, $\bar{u}$, $\bar{d}$) in Table \ref{tab:H}; $E$ ($u_v$, $d_v$, $g$), in  \ref{tab:E}; $\tilde{H}$, and $\tilde{E}$ ($u_v$, $d_v$) in Table \ref{tab:tilde}. 
A fully quantitative fit, constrained by either data or lattice QCD calculations is presently not  attainable for the GPD $E$ in the antiquark sector, and for the helicity GPDs for antiquarks and gluons. 

\newpage
\begin{widetext}
%%%%%
%%%%%%%%%%%%%%%%%%%%%%%
%%%%%%%%%%%%%%%%%%%%%%%
%%%%%%%%%%%%%%%%%%%%%%% Table H
%%%%%%%%%%%%%%%%%%%%%%%
%%%%%%%%%%%%%%%%%%%%%%%
%%%%%%%%%%%%%%%%%%%%%%% Table H and H tilde

\begin{table}[ht]
\centering
\begin{tabular}{|c|c|c|c|c|c|}
\hline
\hline

Parameters             &  $H_{u_{v}}$        &  $H_{d_{v}}$        & $H_{g}$            & $H_{\overline{u}}$ & $H_{\overline{d}}$ \\ 
\hline
\hline
$m $ (GeV)           & 0.420               &  0.275              &  -               &  0.380             &  0.300      \\
$M_X$ (GeV)          & 0.604               &  0.913              &  0.726             &  3.250             &  2.105      \\
$M_\Lambda$ (GeV)    & 1.018               &  0.860              &  0.979             &  1.372             &  1.495      \\
$\alpha$             & 0.210               &  0.0317             &  -0.622             &  1.144             &  1.125      \\
$\alpha^\prime$      & 2.448  $\pm$ 0.0885 &  2.209 $\pm$ 0.156  &  2.000  $\pm$ 0.10  &  0.100 $\pm$ 0.060   &  0.125 $\pm$ 0.023     \\
$p$                  & 0.620  $\pm$ 0.0725 &  0.658 $\pm$ 0.257  &  2.000  $\pm$ 0.05  &  0.100 $\pm$ 0.025   &  0.120 $\pm$ 0.05     \\
${\cal N}$           & 2.043               &  1.570              &  1.467  $\pm$ 0.228  &  1.206 $\pm$ 0.008   &  1.230 $\pm$  0.082  \\
$a $           & 2000               &  1000              &  1000   &  2000    &  1000 
\\ \hline
%%%%%%
%
%$m_2 $ (GeV)           & -               & -               &  -               &  1.211         &        \\
%$M_X^2$ (GeV)          & -               & -               &  0.726             &  0.699         &        \\
%$M_\Lambda^2$ (GeV)    & -               & -               &  0.980             &  0.836         &        \\
%$\alpha_2$             & -              & -              &  -0.622            &  0.0417        &        \\
%$\alpha^\prime_2$      & -    & -    &  1.859  $\pm$ TBD  &  3.302  $\pm$ TBD   & \\
%$p_2$                  & -   & -    &  0.237  $\pm$ TBD  &  2.804  $\pm$ TBD   & \\
%${\cal N}_2$           & -               & -               &  1.467  $\pm$ TBD  &  -1.156 $\pm$ TBD   & \\ \hline

\hline
\hline
\end{tabular}
\caption{Parameters for $H_{u_v}$, $H_{d_v}$, $H_g$, $H_{\bar{u}}$, $H_{\bar{d}}$. The valence quark GPD parameters are determined at an initial scale of $Q_o^2= 0.1 $ GeV$^2$. The gluon and antiquark GPDs are determined at the scale $Q_o^2= 0.58$ GeV$^2$. All parameters, as well as the fitting procedure are described in the text.  }
\label{tab:H} 

\end{table}
%%%%

%%%%%%%%%%%%%%%%%%%%%%%
%%%%%%%%%%%%%%%%%%%%%%%
%%%%%%%%%%%%%%%%%%%%%%% Table E
%%%%%%%%%%%%%%%%%%%%%%%
%%%%%%%%%%%%%%%%%%%%%%%
\begin{table}[htp]
\center
\begin{tabular}{|c|c|c|c|}
\hline
\hline

Parameters             &  $E_{u_{v}}$        &  $E_{d_{v}}$        & $E_{g}$               \\ 
\hline
\hline
%$m_1 $ (GeV)           & 0.420               &  0.275              &  n/a                 \\
%$M_X^1$ (GeV)          & 0.604               &  0.913              &  1.120                     \\
%$M_\Lambda^1$ (GeV)    & 1.018               &  0.860              &  1.100                      \\
%$\alpha_1$             & 0.210               &  0.0317             &  0.053                    \\
%$\alpha^\prime_1$      & 2.835  $\pm$ TBD &  1.281 $\pm$ TBD  &  0.000  $\pm$ TBD  \\
%$p_1$                  & 0.969  $\pm$ TBD &  0.726 $\pm$ TBD  &  0.000  $\pm$ TBD \\
%${\cal N}_1$           & 1.803               &  -2.780             &  3.990  $\pm$ TBD  \\ \hline
%%%%%%
%$m_2 $ (GeV)           & 0.420               & 0.275               &  n/a                       \\
%$M_X^2$ (GeV)          & 0.604               & 0.913               &  0.490                        \\
%$M_\Lambda^2$ (GeV)    & 1.018               & 0.860               &  0.485                          \\
%$\alpha_2$             & 0.210               & 0.0317              &  -0.622                          \\
%$\alpha^\prime_2$      & 2.835  $\pm$ 0.146  & 1.281  $\pm$ 3.176  &  0.000 $\pm$ 0.629     \\
%$p_2$                  & 0.969 $\pm$ 0.3355  & 0.726  $\pm$ 1.543  &  0.000  $\pm$ 0.552    \\
%${\cal N}_2$           & 1.803               & -2.780              &  0.02  $\pm$ 0.0273   \\ \hline

$m $ (GeV)           & 0.420               & 0.275               &  n/a                       \\
$M_X$ (GeV)          & 0.604               & 0.913               &  0.490                        \\
$M_\Lambda$ (GeV)    & 1.018               & 0.860               &  0.485                          \\
$\alpha$             & 0.210               & 0.0317              &  -0.622                          \\
$\alpha^\prime$      & 2.835  $\pm$ 0.146  & 1.281  $\pm$ 3.176  &  0.000 $\pm$ 1.212    \\
$p$                  & 0.969 $\pm$ 0.3355  & 0.726  $\pm$ 1.543  &  0.000  $\pm$ 1.197    \\
${\cal N}$           & 1.803               & -2.780              &  0.034  $\pm$ 0.05   \\ 
\hline

\hline
\hline
\end{tabular}
\caption{Parameters for $E_{u_v}$, $E_{d_v}$, $E_g$, $H_{\bar{u}}$, $H_{\bar{d}}$. The valence quark GPD parameters are determined at an initial scale of $Q_o^2= 0.1 $ GeV$^2$. The gluon and antiquark GPDs are determined at the scale $Q_o^2= 0.58$ GeV$^2$. All parameters, as well as the fitting procedure are described in the text. }
\label{tab:E} 
\end{table}
%%%%%%%%%%%%%%%%%%%%%%%
%%%%%%%%%%%%%%%%%%%%%%%%
\begin{table}[htp]
\centering
\begin{tabular}{|c|c|c|c|c|}
\hline
\hline

Parameters &  $\widetilde{H}_{u_{v}}$ &  $\widetilde{H}_{d_{v}}$ &  $\widetilde{E}_{u_{v}}$ &  $\widetilde{E}_{d_{v}}$ \\ 
\hline
\hline
$m $ (GeV)           & 2.624       & 2.603       & 2.624       & 2.603       \\
$M_X$ (GeV)          & 0.474       & 0.704       & 0.474       & 0.704       \\
$M_\Lambda$ (GeV)    & 0.971       & 0.878       & 0.971       & 0.878       \\
$\alpha$             & 0.219       & 0.0348      & 0.219       & 0.0348.     \\
$\alpha^\prime$      & 1.543 $\pm$ 0.296 & 1.298 $\pm$  0.245 & 5.130 $\pm$  0.101 & 3.385 $\pm$ 0.145 \\
$p$                  & 0.346 $\pm$ 0.248 & 0.974 $\pm$ 0.358 & 3.507 $\pm$ 0.054 & 2.326 $\pm$ 0.137\\
${\cal N}$           & 0.0504      & -0.0262     & 1.074       & -0.966      
\\ 
$a $           & 2000     & 1000    & 2000       & 1000 \\
\hline

\hline
\end{tabular}
\caption{Parameters for $\widetilde{H}_{q}$ and $\widetilde{E}_{q}$ where $q \in \{u_{v}, d_{v} \}$. All parameters are described in the text.  }
\label{tab:tilde} 

\end{table}
\end{widetext}
%%%%%%%%%%%%%%%%%%%%%%%%%%%%%%%%
%%%%%%%%%%%%%%%%%%%%%%%%%%%%%%%%
All parameters are used to evaluate directly the forms given in Section \ref{sec:param}, where the $k_T$ integration limits are taken as $[0, 5]$ GeV.

Our analysis is valid in the kinematic region of $10 ^{-4} < (X, \zeta) < 0.85$, $1<Q^2<100$ GeV$^2$, {\it i.e.} in the multi-GeV region accessible at present and currently planned facilities, and $-t<Q^2$. 
\vspace{0.2cm}

\noindent A few comments are in order: 

\vspace{0.2cm}
\noindent -- The initial scale, $Q_o^2$, is a fitted parameter. 
Antiquarks and gluons are fitted at a higher scale than valence quarks according to the physical picture where
at an initial low scale only valence quarks are present, while gluons and sea quarks (quark-antiquark pairs) are resolved as independent degrees of freedom as the scale increases. Sea quarks and gluons undergo perturbative evolution beyond their initial scale and generate additional gluon and sea quarks dynamically through gluon bremmstrahlung, Section \ref{sec:evol}.  

\noindent -- DGLAP region: our fit is recursive in that we first fitted the mass parameters $m$ (quark), $M_X$ (spectator), $M_\Lambda$ (dipole), the Regge parameter $\alpha$, as well as the normalization ${\cal N}$,  to the forward limit, {\it i.e.} setting $t=\zeta=0$, and using the definitions from Section \ref{sec:forward}. The error on these parameters can be evaluated relative to the PDF parametric forms for the valence, antiquark and gluon distributions in \cite{Alekhin:2002fv}, thus not directly using experimental data. Because of this, in Refs.\cite{Goldstein:2012az,GonzalezHernandez:2012jv} we did not quote these errors. The error on the valence quarks parametrization is determined entirely by the form factor fit. For sea quarks and gluons it is given by the error from the fit to the gluon form factors, in addition to the error on the normalization $\mathcal{N}$, as we explain in what follows.

\vspace{0.2cm}
\noindent -- DGLAP region: the parameters $\alpha'$, $p$,  were fitted subsequently, by keeping the previous set of parameters fixed, switching on the $t$ dependence, and calculating the integrals for the various form factors (Eqs.(\ref{eq:Dirac},\ref{eq:Pauli},\ref{eq:axial}\ref{eq:pseudo},\ref{eq:Diracg},\ref{eq:Paulig}). For the quark sector we used flavor separated nucleon form factor data \cite{Cates:2011pz}, and  lattice results  from \cite{Hagler:2007xi}  (see also \cite{GonzalezHernandez:2012jv}). In the gluon sector we used the results from \cite{Shanahan:2018pib}. 

\vspace{0.2cm}
\noindent -- DGLAP region: the sum of the quark and spectator masses obeys the constraint, $m + M_X > M$ guaranteeing that the system is bound.

\noindent -- DGLAP region: the values of $\alpha$ are not directly related to the Regge predictions for PDFs because the spectator functional form also contributes to the slope at low $X$
(this point is described in detail in Ref.\cite{GonzalezHernandez:2012jv}.

\noindent -- ERBL region: the parameter $a$ is the only free-varying one in our present parametric form for the ERBL region. The choice of having only one fixable parameter is motivated by the present scarcity of data. Our parametrization can be easily extended to a more flexible form than the one presented here, including an enlarged set of parameters for the ERBL region, as more abundant and precise data from deeply virtual exclusive processes become available in the future.

%%%%%%%%%%%%%%%%%%%%%%%

\subsection{Fitting procedure}
\label{sec:fittingprocedure}
Following the method introduced in \cite{Ahmad:2006gn,Ahmad:2007vw,Goldstein:2010gu,GonzalezHernandez:2012jv}, we adopt the flexible parameteric forms given in Section \ref{sec:param} and let the experimental data on deep inelastic scattering reactions and on the nucleon elastic form factors guide the shape
of the parameterization as closely as possible, consistently with the various constraints using elastic scattering and DIS experimental data, and information from lattice QCD calculations whenever applicable:
\begin{itemize}
\item Forward limit (Section \ref{sec:forward}) $\Rightarrow$ DIS data 
\item Polynomiality  (Section \ref{sec:polinom} and Fig.\ref{fig:poly}) $\Rightarrow$ elastic scattering data \cite{Cates:2011pz}, lattice QCD \cite{Hagler:2007xi,Shanahan:2018pib}.
\item Positivity (Section \ref{sec:positivity})
\item Symmetry for $x \rightarrow -x$ (Section \ref{sec:symm}).
\end{itemize}
The fit results in the valence quark sector, namely for $H_{u_v,d_v}$, $E_{u_v,d_v}$, $\widetilde{H}_{u_v,d_v}$, $\widetilde{E}_{u_v,d_v}$ were already performed in Ref.\cite{GonzalezHernandez:2012jv}. Nevertheless, for completeness, we present the values of the parameters in the first two columns of Tables \ref{tab:H}, \ref{tab:E}, and \ref{tab:tilde} alongside the new results derived in this paper for gluons and antiquarks. 

\begin{figure}
    \centering
    \includegraphics[width=8cm]{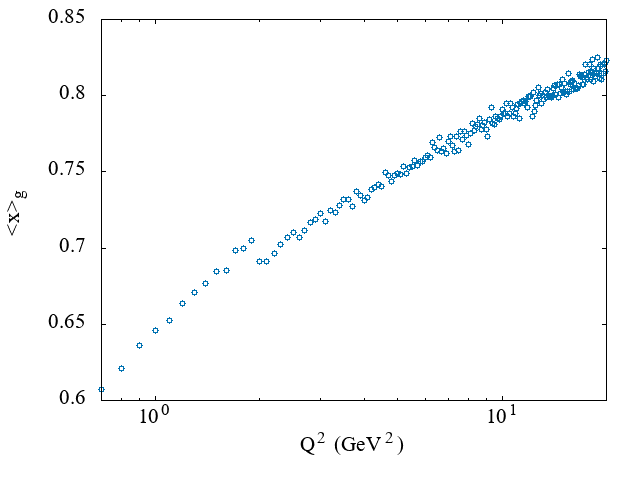}
    \caption{The first Mellin moment of the gluon GPD $H_{g}$ as a function of $Q^{2}$ evolved using LO pQCD evolution tools.}
    \label{fig:gluonQ2dep}
\end{figure}

In the gluon sector we first perform a fit to the gluon PDF at $t=0$. Notice, however, that the expression given in Eq.(\ref{eq:Hgt0}) has to be evolved in perturbative QCD to the $Q^2$ of the data/lattice results. Standard fitting procedures are, therefore, cumbersome.   
%implementing the various constraints, 
%two different methods were developed. 
To overcome this issue, for a practical fit, we devised an algorithm that produces a root mean square error (RMSE) based on a given number of combinations of parameter values 
%The combination of parameters was created by the user, giving the number of combinations and 
varying within specific ranges for each parameter.
The latter form an ``envelope" of gluon GPDs. 
After each iteration of the algorithm, the distance between each subsequent parameter combination decreases according to,
%%%
$$\frac{2 \times {\rm initial\; value\; of\;  parameter}}{({\rm \#\; of\; combinations\;) / (\#\;  of\;  parameters})}$$ 
%%%
%%%%%%% FIGURE 8
\begin{figure}
    \includegraphics[width=8cm]{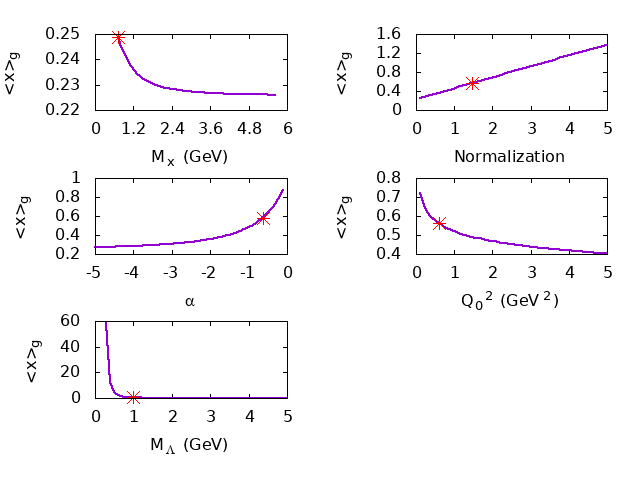}
    \caption{First moment of the gluon GPD $H_{g}$ versus the various forward limit parameters whose functional form is given in Eq.(\ref{eq:Hg}) where the distribution is evolved to a final $Q^{2}$ of 4 GeV$^{2}$. The stars on the plots correspond to the parameters values given in Table \ref{tab:H}.}
    \label{fig:gluonparam}
\end{figure}
whereby the algorithm is iterated using the combination of parameter values that yields the lowest RMSE from the previous step.
By implementing a multi-linear interpolation, one can %test orders of magnitude of more cases 
largely increase the number of trials to better constrain the range of parameters that result in a favorable fit to the data.
 An example of the spread of the various parameter values for the moment $\int dx H_g(x,0,0) = \langle x_g\rangle$, is given in Figure \ref{fig:gluonparam}, while the spread in the $Q^2$ dependence of $\langle x_g \rangle$, obtained using the envelope GPDs is shown in Figure \ref{fig:gluonQ2dep}. 
 The error on the normalization parameter is defined such that the width of the envelope is the size of the error given in \cite{Alekhin:2002fv} (we choose this parametrization because our current fit is done at Leading Order, LO in perturbative QCD). Therefore, the errors on the parametrization in \cite{Alekhin:2002fv} are used as a constraint on the errors of our distribution.

%%%%% DISCUSSION
%%%%%
%%%%%%%
Once the parameters defining the $x$ dependence, $ M_X, M_\Lambda, \alpha, \mathcal{N}$ are determined, we find the $t$-dependent parameters of the gluons, $\alpha'$ and $p$, by recursively fitting the integral of $H_g$, Eq.\eqref{eq:Diracg}, to lattice QCD data at the scale $Q^{2} = 4 \,\text{GeV}^{2}$.   
%however, since we currently only have lattice data at $Q^{2} = 4$ GeV$^{2}$ we fit the parameters at this kinematic point and evolve the gluon distributions through pQCD evolution equations.

%We also note that the gluon form factor data from the Lattice is at an unphysical pion mass of $m_{\pi} = 496$ GeV; therefore, our calculation of the form factor does not represent a physical gluon until there is lattice data at the physical pion mass.

%\begin{figure}[ht]
%    \centering
%    \includegraphics[width=8.5cm]{new_graphs/comparison058to097tobrandon.png}
%    \caption{Gluon GPD, $H_g$, }
%    \label{fig:my_label}
%\end{figure}
%\begin{figure}
%    \centering
%    \includegraphics[width=8cm]{new_graphs/integralofgpdvsqsupdated.png}
%    \caption{Integral of GPDs vs log scale}
 %   \label{fig:gluonQ2dep}
%\end{figure}
%%
%%
%% FIGURE 9

%%%%
%%%% FIGURE 10
\begin{figure}
    \centering
    \includegraphics[width=8cm]{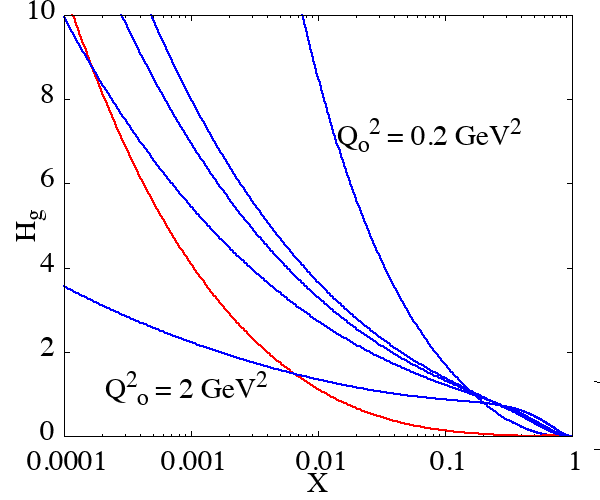}
    \caption{Study of the effect of the initial scale of the gluon GPD $H_{g}$. We keep the forward limit parameters $M_{X}, M_\Lambda$ and $\alpha$ fixed while varying the initial $Q_{o}^{2}$. The red line corresponds to the physical value of the parametrization $Q_{o}^{2} = 0.58 $ GeV$^{2}$ and the blue lines are values of the initial $Q_{o}^{2}$ in a range of $0.2 - 2$ GeV$^{2}$.}
    \label{fig:gluonQ0}
\end{figure}
%%%%%%%%%%%%%%%%%%%%%
To fit the antiquark sector we would need flavor separated lattice QCD results which are not directly available at present.
Nevertheless, we used our valence quark model as a means to estimate the values of the $\bar{u}$ and $\bar{d}$ contribution to the form factors. An improved version of the fit could be readily obtained once flavor separated lattice results will be available. 

We conclude this Section by noting that the initial scale, $Q_o^2$ is also a parameter, to be determined from fits to the data. In Refs.\cite{Ahmad:2006gn,Ahmad:2007vw} it was found that
in the valence sector, $Q_o^2 \approx 0.1$ GeV$^2$. This value is consistent with the more recent fits from Ref.\cite{Goldstein:2012az,GonzalezHernandez:2012jv}, and from the present paper. The fit to the gluon and sea quarks distributions, however,  yields as expected, a larger value of $Q_o^2$. Samples from the envelope for different $Q_o^2$ values, keeping the rest of the parameters fixed, are shown in Figure \ref{fig:gluonQ0}. We found that equivalently viable GPD parametrizations can be obtained for two distinct values of $Q_o^2$. In Tables \ref{tab:Hgluon} and \ref{tab:Egluon} we show the parameters for $Q_o^2 = 0.58$ GeV$^2$. In Section \ref{sec:gluresults} we show results for the GPD $E_g$ obtained for a higher value of $Q_o^2$.   
%%%%%%%%
%%%%%%%%
%%%%%%%% FIGURE 11
\begin{figure*}
%\begin{center}
\includegraphics[scale=0.37]{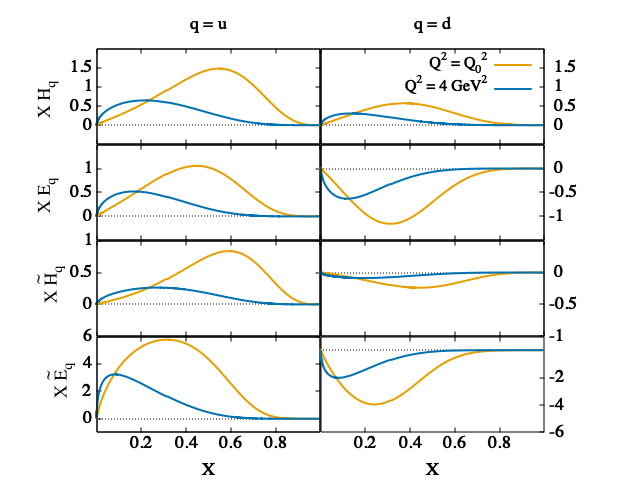}
\includegraphics[scale=0.37]{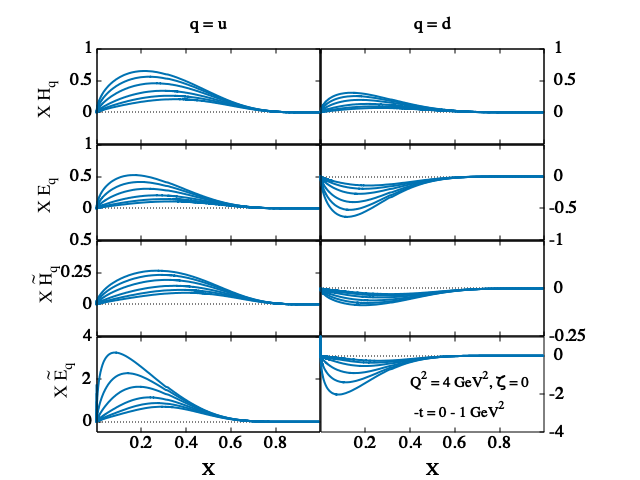}
%\end{center}
\caption{({\it Left}) GPDs $F_{q}(X,0,0)$ where $ F \in (H,E,\widetilde{H},\widetilde{E})$ and $q \in (u_v,d_v)$. We show the initial scale valence quark GPD at $Q_{o}^{2} = 0.1$ GeV$^{2}$, and the evolved GPD at a final scale of $Q^{2} = 4$ GeV$^{2}$.
({\it Right}) GPDs $F_{q}(X,0,t)$ where $ F \in (H,E,\widetilde{H},\widetilde{E})$ and $q \in (u_v,d_v)$. All GPDs are shown at the scale $Q^2 = 4$ GeV$^{2}$, for a range of momentum transfer values from $-t = 0$ GeV$^{2}$ (upper curves) to $-t = 1$ GeV$^{2}$ (lowest curves), and $\zeta = 0$.}
%\label{fig:val2evol_xi0}
%%%%%%
\label{fig:val_tdep_xi0}
\end{figure*}
%%%%%%%

%%%%%%%
%%%%%%%
%%%%%%% FIGURE 12
\begin{figure}[ht]
\includegraphics[scale=0.35]{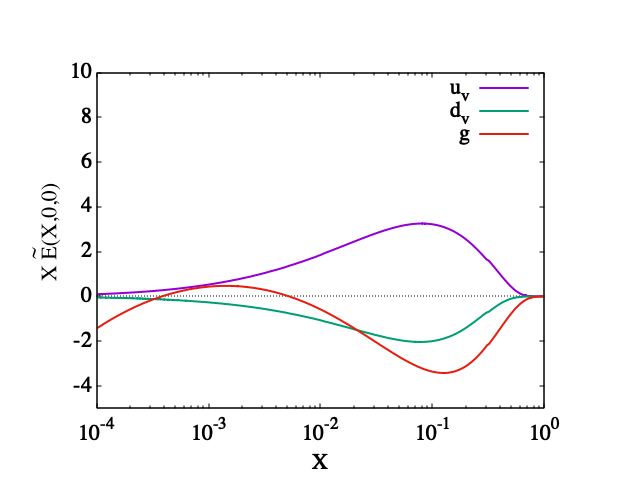}
\caption{We show the GPD $\widetilde{E}$ in the forward limit for the valence quark distributions and for the gluon distribution evolved to $Q^2=4$ GeV$^2$. The parameters at the initial scale are described in Table \ref{tab:tilde} for the valence quarks and in Table \ref{tab:Egluon} for the gluon. }
\label{fig:val_Etilde}
\end{figure}

%%%%%%%%%%%%% SECTION IV B 
%%%%%%%%%%%%%
\subsection{QCD Evolution}
\label{sec:evol}
The QCD anomalous dimensions and Wilson coefficient functions for the off-forward case have been derived and tested at LO in Refs.\cite{Ji:1996nm,Musatov:1999xp,Blumlein:1999sc,GolecBiernat:1998ja}. Calculations of the coefficient functions up to Next-to-Leading-Order (NLO) can be found in Refs.\cite{Ji:1997nk,Ji:1998xh,Belitsky:1999hf,Freund:2001hd}. Correspondingly one can, in principle, evaluate both the LO and NLO kernels of the perturbative QCD evolution equations \cite{Belitsky:1999hf}. The main issue for evolution beyond LO is the non holomorphism of GPDs at the crossover point between the ERBL and DGLAP regions. Numerical calculations have been, therefore, performed computing moments in the conformal partial wave expansion, the $\overline{CS}$ scheme (\cite{Kumericki:2007sa} and references therein). The latter is, however, not directly connected to the $\overline{MS}$ scheme  which is preferable to unanbiguosly constrain the GPDs with forward limit PDFs and to compare with experiment.  

For a practical study, the parametrization presented here implements pQCD evolution of GPDs using LO kernels. This gives a sufficiently accurate description of the data since the effects from LO vs. NLO GPDs could not be possibly observed within the current level of experimental precision. For future studies, NLO evolution will be presented in a separate, dedicated publication. 

Our procedure is as follows: (1) GPDs are evolved in the DGLAP region from $Q_0^2$ to the final $Q^2$, at any given kinematic bin ($\zeta$, $t$); (2) the parameters defining the $X$-dependent curve in the ERBL region are determined at the given scale, $Q^2$, to match the GPDs at the crossover point, $X=\zeta$, while preserving polynomiality. This step in our procedure implies that evolution in the ERBL region is smooth, and that it preserves both` the symmetry properties around $X=\zeta/2$, and the shape of the GPDs. Results illustrating this behavior are plotted in Section \ref{sec:numerical}.
The 

%%%%%%
%\subsubsection{Compton Form Factors}
The structure functions to compare with experiment are the Compton Form Factors (CFFs), which correspond to convolutions of GPDs with the Wilson coefficient functions (or the hard scattering functions). 

This represents, perhaps, the most important difference with the forward case, where one starts from the pQCD evolved PDFs  depending on $x_{Bj}$ and $Q^2$, and considers the convolution in the longitudinal variable with the Wilson coefficient functions. The latter yields structure functions which still depend on $x_{Bj}$ and $Q^2$ \cite{Ellis:1991qj}. For GPDs
%%%% CFFs
the CFFs are defined by the following convolutions for each quark flavor, $q$,
${\cal F}_q$ = (${\cal H}_q$, ${\cal E}_q$), and $\widetilde{\cal F }_q$= ($\widetilde{\cal H}_q$ $\widetilde{\cal E}_q$), and for the gluon, $\mathcal{F}_g$, respectively, as,
\begin{eqnarray}
\label{eq:CFFq}
\mathcal{F}_q(\zeta,Q^2)  & = &    C^+(X,\zeta,Q^2) \otimes F_q(X,\zeta,Q^2), 
\\
\label{eq:CFFqtil}
\mathcal{\widetilde{F}}_q(\zeta,Q^2) & = &  C^-(X,\zeta,Q^2) \otimes \widetilde{F}_q(X,\zeta,Q^2) ,
\end{eqnarray}
while for the gluon,
\begin{eqnarray}
\label{eq:CFFg}
\mathcal{F}_g(\zeta,Q^2) & = & \frac{\alpha_S(Q^2)}{2\pi} 
 C_g^{\overline{MS}}(X,\zeta) \otimes F_g(X,\zeta,Q^2) , 
\nonumber \\
\label{CFF_g}
\end{eqnarray}
where the convolutions are given by the integral:
\[ \otimes \quad \rightarrow \quad  \int_{-1+\zeta}^{1} \frac{dX}{(1-\zeta/2)} ,\]
and we omitted the $t$ dependence for ease of presentation.
The coefficients functions in $\overline{MS}$ scheme are given by \cite{Ji:1997nk,Ji:1998xh,Belitsky:1999hf,Freund:2001hd},
\begin{eqnarray}
\label{eq:CvecLO} 
 C^{\pm}(X,\zeta,Q^2)
& = &  C^{\pm}_0(X,\zeta) + \frac{\alpha_S}{2 \pi} C_q^{\overline{MS}}(X,\zeta) 
\\
C^{\pm}_0(X,\zeta) & = &  \frac{1}{X-\zeta-i\epsilon} \mp \frac{1}{X-i\epsilon} 
\end{eqnarray}
where the expression for $C_q^{\overline{MS}}$ and $C_g^{\overline{MS}}$ were given in Ref.\cite{Ji:1998xh}.
At variance with PDFs, for GPDs it appears clearly that even at leading order, the variable $X$ is integrated over, and the observable longitudinal variables is $\zeta$. 

The GPDs entering Eqs.\eqref{eq:CFFq},\eqref{eq:CFFqtil} and \eqref{eq:CFFg}, are obtained by solving the pQCD DGLAP evolution equations,
\begin{widetext}
\begin{eqnarray}
\frac{\partial}{\partial \ln Q^2} F_{q_v}(X,\zeta,Q^2) &=& \frac{\alpha_S}{2 \pi} P_{qq}\left(\frac{X}{Z},\frac{X-\zeta}{Z-\zeta},\alpha_S\right) \otimes F_{q_v}(Z,\zeta,Q^2) \\
%% singlet
\frac{\partial}{\partial \ln Q^2} F^{\Sigma}(X,\zeta,Q^2) \! &=& \!\! \frac{\alpha_S}{2 \pi} \Big[ P_{qq}\left(\frac{X}{Z},\frac{X-\zeta}{Z-\zeta},\alpha_S \right) \otimes F^{\Sigma}(Z,\zeta,Q^2) + 2 N_f P_{qg}\left(\frac{X}{Z},\frac{X-\zeta}{Z-\zeta},\alpha_S\right) \otimes F_{g}(Z,\zeta,Q^2) \Big]  \\
%% gluon
\frac{\partial}{\partial \ln Q^2} F_{g}(X,\zeta,Q^2) &=& \frac{\alpha_S}{2 \pi} \Big[ P_{gq}\left(\frac{X}{Z},\frac{X-\zeta}{Z-\zeta},\alpha_S\right) \otimes F^{\Sigma}(Z,\zeta,Q^2) +  P_{gg}\left(\frac{X}{Z},\frac{X-\zeta}{Z-\zeta},\alpha_S\right) \otimes F_{g}(Z,\zeta,Q^2) \Big]
\end{eqnarray}
\end{widetext}
where the LO kernels were first derived in Ref.\cite{Ji:1997nk,GolecBiernat:1998ja}, and we defined,
\[ \otimes \quad \rightarrow \quad  \int_X^1 \frac{d Z}{Z} \quad\quad {\rm and} \quad\quad \frac{X'}{Z'}=\frac{X-\zeta}{Z-\zeta}.\]
The definitions and symmetry properties for the flavor non singlet (NS), $F_{q_v}$ and $+$ distributions, for the flavor singlet $F^\Sigma$, and for the gluon, $F_g$,  described in Section \ref{sec:symm}, are conserved under evolution. 

In summary, putting all together, we find the proton and neutron CFFs, ${\cal F}^N$, $N=p,n$, which can be determined up to NLO in the coefficient functions by summing over the $N_f$ active light quark flavors,
\begin{equation}
\mathcal{F}^N(\zeta,t,Q^2) =  \sum_{q = u, d} \left( e_{q}^2 \mathcal{F}_q + e_{\bar{q}}^2 \mathcal{F}_{\bar{q}} \right) + \mathcal {F}_g \, .
  \label{eq:defQs}
\end{equation}
Numerical results for evolution are given in Section \ref{sec:numerical}.

%%%%
%%%% FIGURE 13
%\begin{widetext}
\begin{figure*}
\includegraphics[scale=0.4]{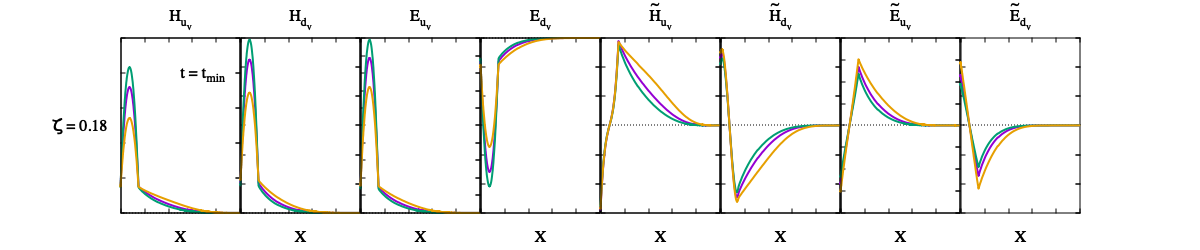}
\caption{GPDs, $H_{q_v}$, $E_{q_v}$, $\widetilde{H}_{q_v}$, $\widetilde{E}_{q_v}$, ($q=u,d$) evaluated at $\zeta = 0.18$, $-t = -t_{\text{min}}= 0.03 \, \, \text{GeV}^{2}$, evolved to $Q^{2}$ values: $1$ (yellow curve), $4$ (purple curve), $50$ (green curve). The latter cover a range of values from Jlab current kinematic settings to the EIC.}
\label{fig:valGPD}
\end{figure*}
%%%%%%
%%%%%%%%%
%%%%%%%%. NUMERICAL
%%%%%%%%%
%%%%%%%%%%
\section{Numerical results}
\label{sec:numerical}
This Section highlights the effect of various parameters
parameters which control the behavior of GPDs in different regions of $X$, $\zeta \approx x_{Bj}$, $t$, and $Q^2$. Several plots were generated to visualize the changes in the different quark flavor and gluon GPDs, plotted as a function of $X$, varying their $t$, $\zeta$ and $Q^2$ dependences. The $H$ and $E$ GPDs for all quark and gluon components are also summarized in Fig.\ref{fig:allgpds} in Section \ref{sec:sec2}.

%%%%%
%%%%% VALENCE
\subsection{Valence Quark Distributions}
\label{sec:valresults}
The valence quark fits confirm the results of Refs.\cite{Goldstein:2010gu,GonzalezHernandez:2012jv} for the GPDs $H_{q_v}$, $E_{q_v}$, and $\widetilde{H}_{q_v}$ $(q=u,d)$. We added new fit results for the GPD $\widetilde{E}_{q_v}$.
All results are summarized in Figures \ref{fig:val_tdep_xi0},  \ref{fig:val_Etilde} and \ref{fig:valGPD}. 

Fig. \ref{fig:val_tdep_xi0} shows the behavior of all four GPDs, $H_{q_v}$, $E_{q_v}$, $\widetilde{H}_{q_v}$ and $\widetilde{E}_{q_v}$ as a function of $Q^2$ at $(\zeta,t)=0$, on the {\it l.h.s.} panels, and as a function of $t$ at $\zeta=0$ and at $Q^2= 4$ GeV$^2$,  on the {\it r.h.s.} panels, respectively. The {\it l.h.s.} panels show a dramatic effect of $Q^2$ evolution from the initial scale of $Q_o^2 = 0.1$ GeV$^2$ to $Q^2= 4$ GeV$^2$, albeit evolution slows down in the multi-GeV region, as we show later on. The values of $t$ on the {\it r.h.s.} panels range from $t=0$ (highest peaked curves), to $t=-1$ GeV$^2$ (lowest curves).

In Fig. \ref{fig:val_Etilde} we focus on the GPD $\widetilde{E}$, calculated for $\zeta$, $t=0$, and $Q^2=4 $ GeV$^2$, both in the quark sector and for the gluon (the latter is discussed below).   

Fig. \ref{fig:valGPD} shows the effect of pQCD evolution for the GPDs  $H_{q_v}$, $E_{q_v}$, $\widetilde{H}_{q_v}$, $\widetilde{E}_{q_v}$ ($q=u,d$), plotted  as a function of $X$ ($0<X<1$), for $\zeta=0.18$, $t_0 \equiv t_{min} = -0.03$ GeV$^{2}$ (Eq.\ref{eq:tmin}). The different curves in each panel are the GPD values evolved  to $Q^{2}$ between $1$ and $50$ GeV$^{2}$. The yellow curve which is highest in the DGLAP region and lowest in the ERBL region corresponds to the lowest value of $Q^2 = 1$ GeV$^2$; the green curve, which is lowest in DGLAP and highest in ERBL, corresponds to $Q^2= 50$ GeV$^2$. We explain this behavior as follows: (1) Perturbative QCD for $X \geq 0.2$ shifts ``strength" from higher $X$ to lower $X$, resulting in the depletion shown in the figure from the low to high values of $Q^2$; (2) because of polynomiality, or the normalization to the nucleon form factors, the curves at higher $Q^2$ must peak higher in the ERBL region (the difference between the yellow and green peaks is noticeable in the figure). Notice how in the pQCD evolution framework defined in Section \ref{sec:evol}, the symmetry around $X=\zeta/2$ is conserved. Overall, the effects of pQCD evolution of GPDs in the range of current and future experiments, are logarithmic. We, therefore, expect the $Q^2$ dependence of DVCS type experiments to be more substantially influenced by the behavior of the NLO Wilson coefficient function. 
%%%%%%%%%% TABLE IV
%%%%%%%%%%%%%%%%%%%%%%%
\begin{table}[htp]
\center
\begin{tabular}{|c|c|c|}
\hline
\hline

Parameters             &  $Q_o^2 = 0.58$ GeV$^2$       &  $Q_o^2 = 0.97$ GeV$^2$                      \\ 
\hline
\hline
$M_X$ (GeV)          & 0.726             &       1.12                            \\
$M_\Lambda$ (GeV)    & 0.979               &    1.05                                \\
$\alpha$             & -0.622             &   0.005                             \\
$\alpha^\prime$      & 2 $\pm$ 0.10 &        0.28 $\pm$ 0.10  
\\
$p$                  & 2 $\pm$ 0.05  &   0.17  $\pm$  0.05
\\
${\cal N}$           & 1.4672 $\pm$ 0.228             &   1.525  $\pm$ 0.228       \\ 
%%%%%%

\hline
\hline
\end{tabular}
\caption{Parameters for the gluon GPD, $H_{g}$ for  $Q_0^2 = 0.58$ GeV$^2$, and $Q_0^2 = 0.97$ GeV$^2$. }
\label{tab:Hgluon} 
\end{table}
%%%%%%%%%%%%%%%%%%%%%%%%%%%%%%
%%%%%%%
%%%%%%%
%%%%%%%
%%%%%%%
%%%%%%%
%%%%%%%%% TABLE EG
\begin{table}[htp]
\scriptsize
\center
\begin{tabular}{|c|c|c|c|}
\hline
\hline

Parameters             &   $E_g^{(1)}$    &   $E_g^{(2)}$  &     $E_g^{(3)}$               \\ 
\hline
\hline
$Q_o^2$  (GeV$^2$) & $0.97$  & 0.97 & $0.58$ \\ 
$M_X$ (GeV)          & 1.120             &       0.490                     &   0.490    
\\
$M_\Lambda$ (GeV) & 1.100  
& 0.485               &    0.485                  
\\
$\alpha$   &  0.053          & -0.622             &   -0.622            
\\
$\alpha^\prime$   & 0.45 $\pm 0.30$   & 0.000 $\pm$ 1.221 &        0.000 $\pm$ 1.212 
\\
$p$        & -0.20 $\pm 0.30$      & 0.000 $\pm$ 1.205 &   0.000  $\pm$ 1.197 
\\
${\cal N}$ &  3.970 $\pm$      1.950                &   0.020  $\pm$ 0.0273   & 0.034  $\pm$ 0.050  \\ 
%%%%%%
\hline
\hline
\end{tabular}
\caption{Parameters for the gluon GPD, $E_{g}$. The first column show parameters for $E_{g}^{(1)}$, obtained with the initial scale $Q_0^2 = 0.97$ GeV$^2$, and fitted to a dipole form (Eq.\eqref{eq:dipoleEg}). The second and third columns show parameters  evolved from $Q_0^2 = 0.97$ GeV$^2$ and $Q_0^2 = 0.58$ GeV$^2$, respectively labeled $E_{g}^{(2)}$, $E_{g}^{(3)}$, and fitted to a constant value. In this case, the GPD functional form displays a node (see Figure \ref{fig:eg_q2}).}
\label{tab:Egluon} 
\end{table}
%%%%%%%%%%%%%
%%%%%%%%%%%%%
%%%%%%
%%%%%% FIGURE 14
\begin{figure*}
\includegraphics[scale=0.4]{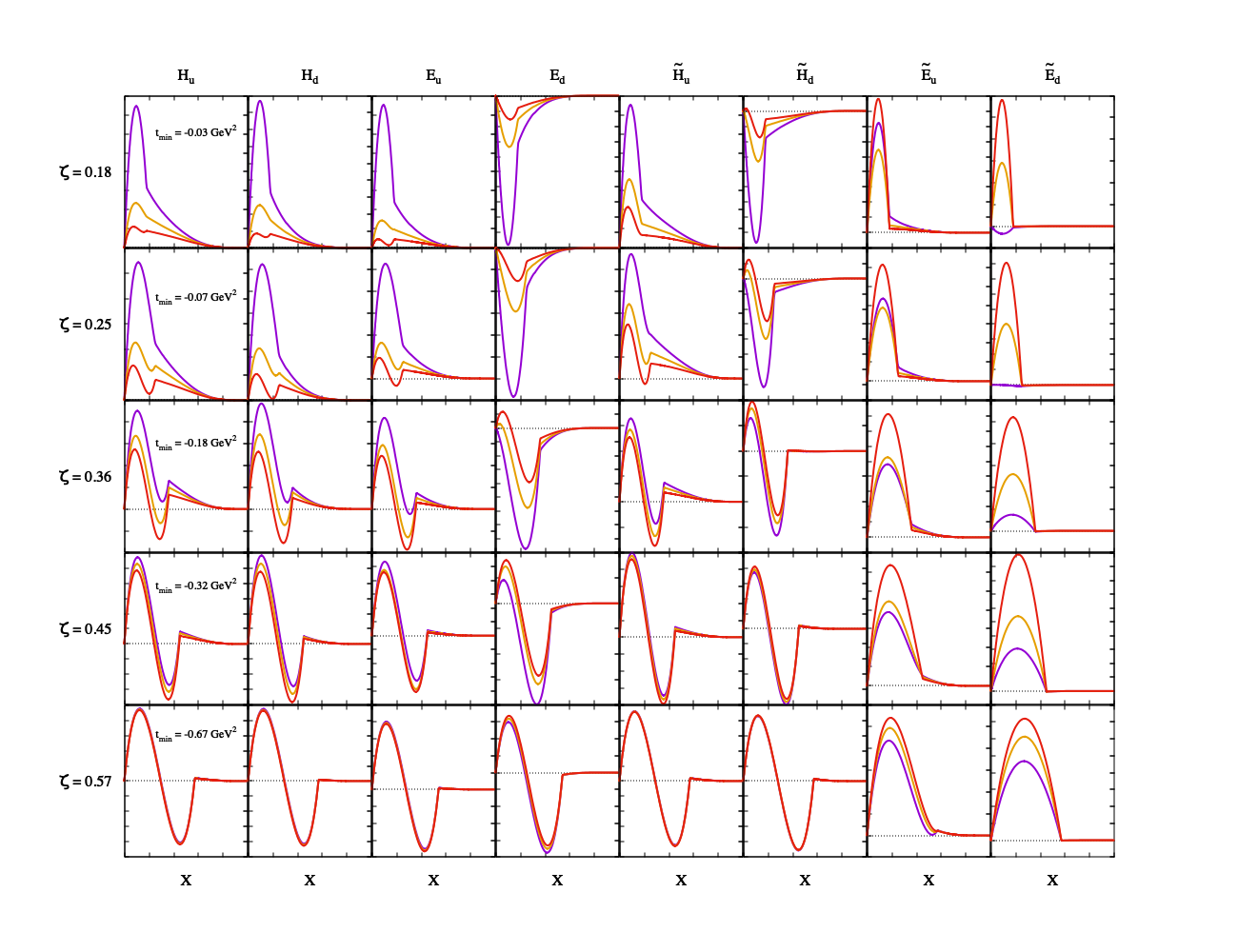}
\caption{GPDs, $H$, $E$, $\widetilde{H}$, and $\widetilde{E}$ plotted vs. $X$, separated into quark contributions $u$ and $d$ (columns), evaluated at different $\zeta \approx x_{Bj}$ values, $\zeta=0.18, 0.25, 0.36, 0.45, 0.57$ (rows). In each panel we show momentum transfer, $-t$, values: $t_0 \equiv t_{min}$, Eq.\eqref{eq:tmin} (purple lines), $1$ GeV$^{2}$ (red line), and an intermediate  value for each $\zeta$, in the interval $[t_0,1$ GeV$^2]$ (yellow line).  All panels correspond to $Q^2= 4$ GeV$^2$. }
\label{fig:val_stamps}
\end{figure*}
%%%%%%%%%%%
%%%%%%%%%%%
%%%%%%%%%%%%
%%%%%%%%%%%% FIGURE 15 EVOLUTION
\begin{figure}
    \centering
    \includegraphics[width=7cm]{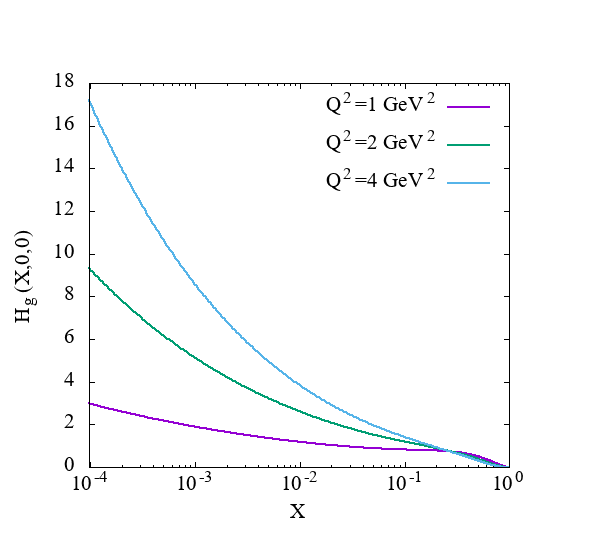}
    \caption{$Q^2$ dependence of the GPD $H_g(X,0,0)$.}
    \label{fig:gluonH_evol}
\end{figure}
%%%%%%%%
%%%%%%%% FIGURE 16
\begin{figure}
\includegraphics[scale=0.35]{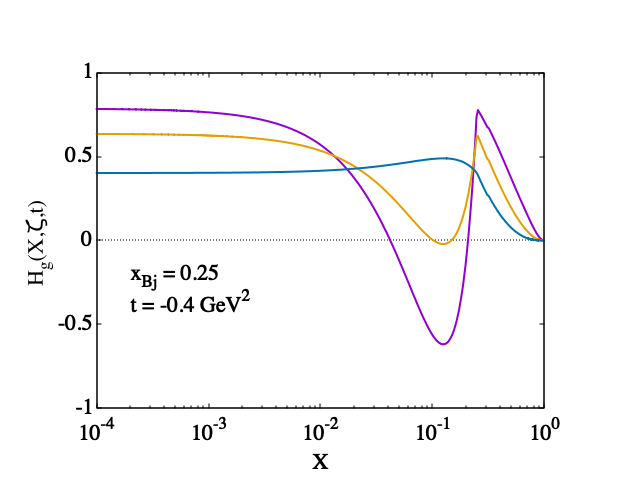}
\caption{$Q^2$ dependence of the gluon GPD, $H_g$ from pQCD evolution equations at LO in the kinematic bin: $x_{Bj} = 0.25$, $-t = 0.4$ GeV$^{2}$,  for $Q^{2} = 1$ GeV$^{2}$ (purple), $Q^{2} = 4$ GeV$^{2}$ (yellow), and $Q^{2} = 50$ GeV$^{2}$ (blue). The range in $Q^2$ covers the kinematics from the present JLab kinematic setting to the EIC. }
\label{fig:gluon_q2}
\end{figure}
%%%%%%%%%%%%%%%%%%%%%%%%%%%%%%
%%%%%%%%%%%%%%%%%%%%%%%%%%%%%%

%%%%%%%%% 
%%%%%%%%% FIGURE 17
\begin{figure}
\includegraphics[scale = 0.35]{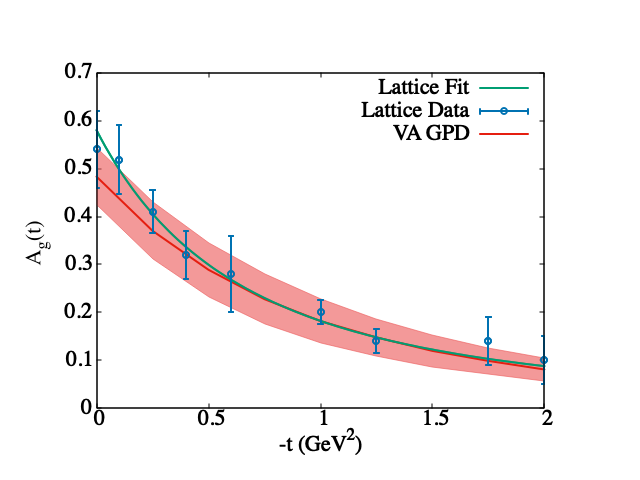}
\includegraphics[scale = 0.35]{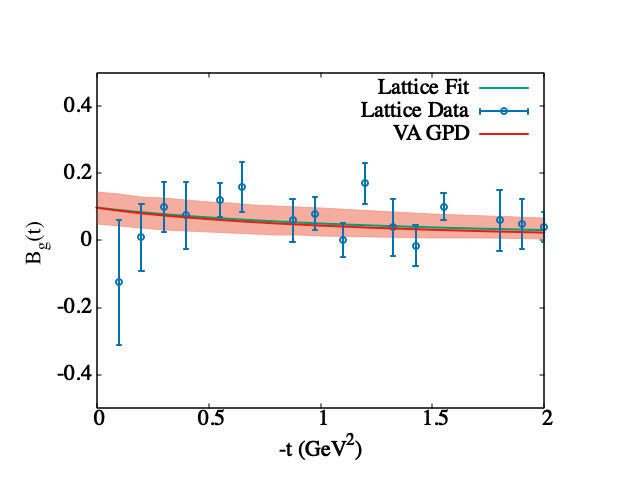}
\caption{({\it Upper Panel}) GPD $H_{g}(X,0,t)$ integrated over $X$ at $\zeta = 0$, fitted to lattice QCD results from Ref.\cite{Shanahan:2018pib} at $Q^{2} = 4$ GeV$^{2}$. Parameters are given in Table \ref{tab:Hgluon} and described in the text.
({\it Lower Panel}) GPD $E_{g}(X,0,t)$ integrated over $X$ at $\zeta = 0$, fitted to a dipole form (lattice QCD resulkts from Ref.\cite{Shanahan:2018pib}). Parameters are given in Table \ref{tab:Egluon} (first column). }
\label{fig:HgEg_ff}
\end{figure}

%%%%%%%%%%%
%%%%%%%%%%%
%%%%%%%%%%% gluon and antiquark
\subsection{Antiquark and Gluon Distributions}
\label{sec:gluresults}
In Fig.\ref{fig:val_stamps} the GPDs $F_q = F_{q_V}+ F_{\bar{q}}$, ($q=u,d$), are plotted as a function of $X$, at $Q^2 = 4 $ GeV$^2$,  for different values of $\zeta$ and corresponding ranges in $t$. Similar to Fig.\ref{fig:valGPD}, the GPDs with the smallest values of $t$ ($t=t_0\equiv t_{min}$, Eq.(\ref{eq:tmin}), are the largest. One can clearly see how the relative values of the GPDs in the  DGLAP region compared to the ERBL region become increasingly important as $-t$ increases. 
Notice that in this case there is no symmetry constraint around $X=\zeta/2$. This impacts the ERBL region where we notice that all GPDs, with the exception of $\widetilde{E}$,  the symmetric, or $-$ component Eq.(\ref{eq:symm-}), dominates at the smallest value of $\zeta=0.18$. As $\zeta$ increases, the DGLAP region shrinks and, in order to preserve polynomiality, the ERBL gradually becomes dominated by the $+$, anti-symmetric distribution, Eq.(\ref{eq:symm+}). The transition can be visualized in the figure, proceeding from top to bottom. This behavior will be altered as $Q^2$ increases (cf. Fig.\ref{fig:valGPD}).

%%%%%%%%%
%%%%%%%%% FIGURE 17
%\begin{figure}
%\includegraphics[scale=0.55]{FinalGraphfit.png}
%\caption{Integrated GPDs $E_g^{(2)}(X,t)$ and %$E_g^{(3)}(X,t)$, with their error, shown for values of %$t$ ranging from $0$ to $-2$ GeV$^2$. This plot therefore %shows the fit of these integral values to the constant %$B(t) = 0.075$, and also compares them with the lattice data.}
%\label{fig:g1}
%\end{figure}

%%%%%%%%%
%%%%%%%%% FIGURE 18
\begin{figure*}
\includegraphics[scale=0.62]{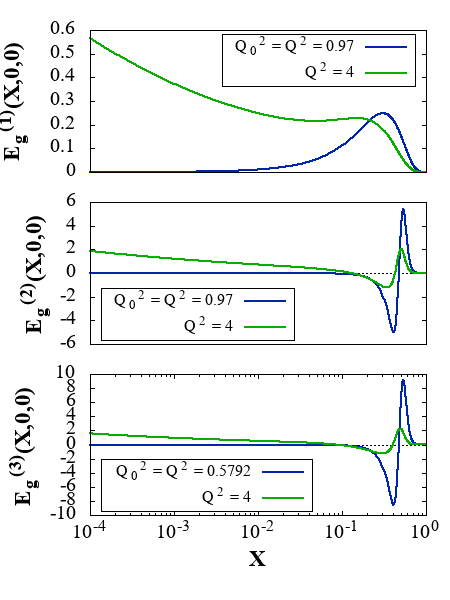}
\hspace{0.5cm}
\includegraphics[scale=0.65]{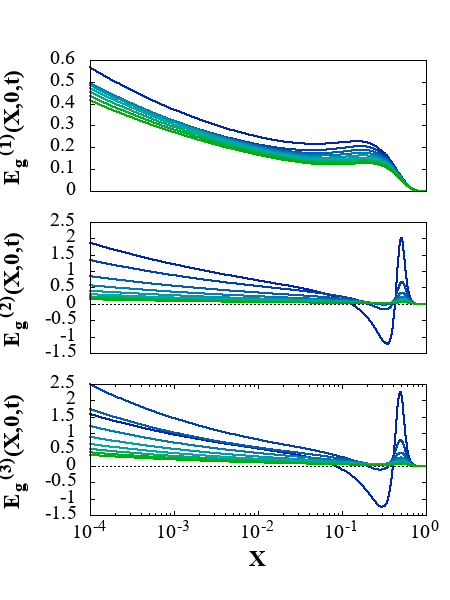}
\caption{{\it Left} The three parametric forms for $E_g(X,0,0)$, described in the text, namely $E_g^{(1)}$ (top), $E_g^{(2)}$(middle) and $E_g^{(3)}$ (bottom). $E_g^{(1)}$ is fitted to a dipole form, whereas $E_g^{(2,3)}$ are fitted to a constant value in $t$ for the gluon form factor from lattice QCD using different values of the initial scale $Q_o^2$. Notice the node in $E_g^{(2,3)}$. The green lines are evolved starting from the blue lines evaluated at $Q_o^2$, to $Q^2 = 4 $ GeV$^2$ in order to match the lattice QCD values \cite{Shanahan:2018pib}; {\it Right}: the $t$-dependence of the GPDs $E_g^{(1)}(X,0,t)$, $E_g^{(2)}(X,0,t)$ and $E_g^{(3)}(X,0,t)$. $E_g^{(1,2,3)}(X,0,0)$ is the topmost curve, shown in blue. As $t$ grows to the value of $-2$ GeV$^2$, it becomes more green in the graph. }
\label{fig:eg_q2}
\end{figure*}

%%%%%%%%%%
%%%%%%%%%% FIGURE 19 GLUONS H and E
\begin{figure}
\includegraphics[scale=0.35]{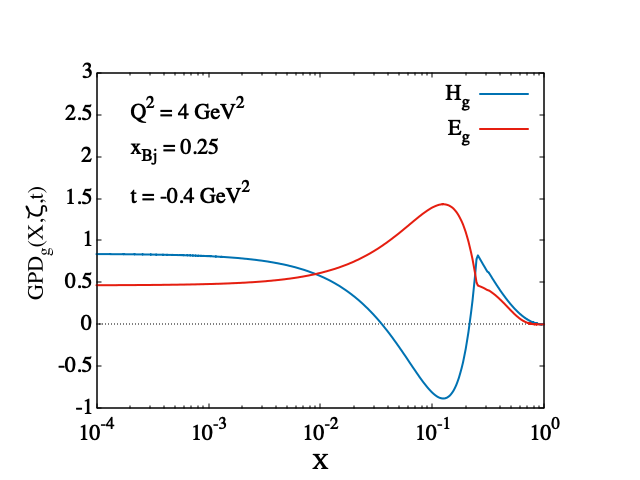}
\caption{Gluon GPDs for parametrization at an initial $Q_{o}^{2} = 0.97$ GeV$^{2}$ in Tables \ref{tab:Hgluon}, \ref{tab:Egluon} for the kinematics $Q^{2} = 4$ GeV$^{2}$, $\zeta = 0.25$, and $t = -0.1$ GeV$^{2}$.}
\label{fig:gluon_H_and_E}
\end{figure}

Results for the gluon GPDs, $H_g$ and $E_g$, are shown in Figures  \ref{fig:gluonH_evol}, \ref{fig:gluon_q2} \ref{fig:HgEg_ff}, \ref{fig:eg_q2}, and \ref{fig:gluon_H_and_E}. 

All parameters were determined similarly to the quark case, {\it i.e.} implementing the procedure described in Section \ref{sec:fittingprocedure} for  either the distributions in $X$, or the Mellin moments as a function of the momentum transfer $t$. To determine the error in the $t$ dependent form factors, we took the size of the RMSE from the GPD envelope in $t$ to be equivalent to the error on the dipole fit of the lattice data moments $A_g(t)$ and $B_{g}(t)$ \cite{Shanahan:2018pib}. As a result, the fit for the GPDs are consistent with a dipole fit on the lattice results.

For the dependence on the initial scale parameter we found two viable sets of values, $Q_o^2=0.58$ GeV$^2$ and $Q_o^2=0.97 $ GeV$^2$. The parameters for the two initial scale values  are presented in Table \ref{tab:Hgluon} for the GPD $H_g$ and in Table \ref{tab:Egluon}, for $E_g$. 
Both values are in a range which is higher than the valence quarks scale, according to the discussion in Section \ref{sec:param_1}, and is also acceptable for perturbative evolution. The value that better validates our physical picture is $Q_o^2=0.58$ GeV$^2$, displayed in Tables \ref{tab:E}, \ref{tab:tilde}, since it is closer to the fitted valence quarks value of $0.1$. Nevertheless, we use both values to study the various trends for the gluon GPDs as $t$, $\zeta$ and $Q^2$ vary. Results from both fitted initial scales generate the same PDF constraint for the GPD $H_g$, shown in Fig.\ref{fig:gluonH_evol} (the curves in the figure reproduce the LO parametrization from Ref.\cite{Alekhin:2002fv}).  

Figure \ref{fig:gluon_q2} shows the effect of pQCD evolution in a typical kinematic bin, $x_{Bj} \approx \zeta =0.25$, $t= - 0.4 $ GeV$^2$, similarly to what shown in Fig.\ref{fig:valGPD} for the valence quarks. Note that because for gluons we are using a logarithmic scale, the ERBL region, $X< 0.25$, is emphasized and the symmetry of the distribution around $X= \zeta/2$ is no longer evident. One can see that, similarly to what described for the valence quarks, the highest valued curve in the DGLAP region ($X>0.25$) corresponds to the lowest value of $Q^2= 1$ GeV$^2$, the effect of evolution moving strength to lower $X$ values. As a consequence, in order to satisfy polynomiality, integrating to the gluon form factor, at $Q^2 = 1$ GeV$^2$, $H_g$ dips to the negative values; as $Q^2$ increases, the dip decreases, until it changes its concavity for the highest considered value of $Q^2=50$ GeV$^2$. 
Notice that in this case the form factors, {\it i.e.} the integrated values are not constant but they are also $Q^2$ dependent, however, this dependence is slower. 

The fit to $A_g(t)$ (Eq.\ref{eq:Diracg}), is shown in Figure \ref{fig:HgEg_ff} (upper panel). Notice that the parametrization fit differs from the lattice result at $t\rightarrow 0$ (although it is consistent within errors), because in our case we impose the constraint, $H_g(X,0,0)=x g(x)$, at $Q^2=4$ GeV$^2$, from the Alekhin parametrization \cite{Alekhin:2002fv}.

%%%%%%%%%%%%%%%%%%%%%%%
%%%%%%%%% SUBSECTION EG
\subsubsection{The GPD $E_g$}
\label{sec:Eg}
The GPD $E_g$ is a more elusive object, being relatively lesser known from phenomenology. The largest obstacle to a clear determination of $E_g$ is that, at variance with $H_g$, it lacks a clear constraint from the forward limit ($t=0$ and $\zeta=0$) . 
However, similarly to $H_g$, we can make use of lattice QCD results to constrain the parametric form's  Mellin moment Eq.(\ref{eq:Paulig}).
%Two alternative fitting methods were used to understand this unique distribution. 
The lattice results are shown in Fig.\ref{fig:HgEg_ff}, lower panel.

The lattice results are consistent with either a dipole form,
\begin{eqnarray}
\label{eq:dipoleEg}
B_{g}(t) &=& \frac{\alpha}{(1- t/\Lambda^{2})^{2}},
\end{eqnarray}
or a constant value, where the parameters $\alpha$ and the dipole mass $\Lambda$ were found to be $\alpha = 0.0978 \pm 0.0466$ and $\Lambda = -2.5578 \pm 2.0849$ GeV.
\[ B_g(t) = 0.075 \pm 0.101 .\]
We denote the dipole GPD parametrization, $E_g^{(1)}$. The GPD fitted to the constant value in $t$, performed starting from the initial scale, $Q_o^2=0.97$ GeV$^2$, is denoted by $E_g^{(2)}$; the one starting at $Q_o^2=0.58$ GeV$^2$, is denoted by $E_g^{(3)}$. 
We perform the fit recursively, that is the parameters determining the behavior of the distribution at $t=0$,  specifically $M_X, \alpha, M_\Lambda$ and $\mathcal{N}$ are fitted first.
Subsequently, the moment  of the distribution is fitted to the lattice values $t \neq 0$. Because the parameters $\alpha'$ and $p$ easily reproduce the dipole behavior (see Section \ref{sec:param_1}), we find that their values for $E_g^{(2,3)}$ are consistent with zero. The fit results  shown in Fig.\ref{fig:HgEg_ff} (lower panel), are for $E_g^{(1)}$ at $Q_o^2= 0.97 $ GeV$^2$.
%and in Fig. \ref{fig:g1} for $E_g^{(2)}$ ($Q_o^2= 0.97 $ GeV$^2$), and $E_g^{(3)}$, ($Q_o^2= 0.58 $ GeV$^2$). 
The values of all parameters with their error is shown in Table \ref{tab:Egluon}. 

An important distinction between the dipole and constant value fits is that the latter supports the presence of a node in $E_g$. All three parameterizations are shown in Fig. \ref{fig:eg_q2},  which illustrates both the evolution in $Q^2$ on the left panels, and the $t$-dependence of the GPD $E_g$, on the {\it r.h.s.} panels. Notice that while  $E_g^{(1)}$ displays a ``valence-like" behavior, $E_g^{(2)}$ and $E_g^{(3)}$ clearly display a node in their $X$ dependence. An important consequence of this behavior is that it impacts the $Q^2$ dependence at low $X$. The GPD $E_g^{(3)}$ is our choice for the the complete parametric form  given in Table \ref{tab:E}, Section \ref{sec:param_1}.  

In Figure \ref{fig:gluon_H_and_E} we juxtapose the GPDs $H_g$ and $E_g$ evolved to $Q^2=4$ GeV$^2$ for atypical Jlab kinematic bin. 

%%%%%%%%%%
%%%%%%%%%%. FOURIER
%\section{Fourier Transforms}
%\label{sec:Fou}
%\hspace{\parindent} Our final task is to compute the Fast Fourier Transform of our GPDs (in the DGLAP region, as we set $\xi = 0$) over $-t$, or $\Delta_{\bot}^2$. For the sake of example, the Fourier Transform for H is given by 
%\begin{eqnarray}
%H(x,\boldsymbol{b}_{\bot}) = \int\frac{H(x,0,-\Delta_{\bot}^2)e^{-\Delta_{\bot}*\boldsymbol{b}_{\bot}}}{(2\pi)^2}d^2\Delta_{\bot}^2
%\end{eqnarray}

%where $\Delta_{\bot}^2$ denotes the longitudinal momentum and $b_{\bot}$ denotes the transverse spatial plane. In the end, the transformed distribution had the following properties:
%\begin{itemize}
%   \item $B_x$ and $B_y$ in units of $GeV^{-1}$, where 1 $GeV^{-1} \simeq .1975$ fm
%    \item $FFTW(H)$ in units of $GeV^2$
%\end{itemize}
%So, a fourier transform of the GPDs give a distribution of mass energy in the transverse spatial plane.

%\section{the notes from kyle, phnew_graphs/ilip, emma, fernanda}

%%%%%%%%% CONCLUSIONS
\section{Conclusions and Outlook}
\label{sec:conclusions}
Measuring GPDs in a wide kinematic range in $x_{Bj}$, $t$ and $Q^2$, will provide a powerful tool allowing a greater insight into the internal structure of the nucleon by uncovering the spatial distribution of its constituent quarks and gluons, and shedding light onto the origin of its mass and spin. 

A quantitative extraction of GPDs from a global analysis can be performed in a consistent QCD-parton framework using experimental data from various deeply virtual exclusive experiments (DVCS, DVMP and theire crossed channels), along with constraints from inclusive deep inelastic scattering, and from the elastic form factors. We provide a first scaffolding of such a framework with a flexible parametrization for all chiral even GPDs in the valence quark $u_v, d_v$, antiquark $\bar{u}, \bar{d}$, and gluon sectors. 
These parametrizations can be readily implemented in theoretical calculations of various derived observables, codes and event generators to evaluate the Compton form factors used in Deeply Virtual Compton Scattering (DVCS) and related experiments. 

A further application that we are currently pursuing includes the computation of Fast Fourier Transforms to obtain quantitative renderings of partonic transverse spatial distributions. A specific goal of the  analysis is to study the sensitivity to the ranges in $t$ that are necessary to obtain a meaningful image in the transverse plane. The availability of a parametrization such as the one presented here is mandatory since it allows us to tune several of the GPD parameters to study the impact of features of the GPD behavior on the Fourier transform. Furthermore, using GPDs in the gluon sector is unprecedented and it provides an alternative approach complementing recent studies of diffractive scattering at EIC kinematics \cite{Mantysaari:2020lhf}.
Our future endeavour will also include a complete analysis at NLO.

Finally, the envelopes of GPDs obtained by appropriately varying different parameters provide an essential background for generating pseudo-data which are a fundamental input in a separate Machine Learning (ML) effort \cite{Grigsby:2020auv}. The latter will ultimately provide the first realistic,  model independent pictures of the proton at the femtometer scale.

\acknowledgements
We thank Gary Goldstein, Osvaldo Gonzalez Hernandez, Jon Poage and Kyle Pressler for joining the initial phases of this project, and Matthias Burkardt and Abha Rajan for many insightful discussions. This work was funded by DOE grant DE-SC0016286 (B.K. and S.L.) and in part by the DOE Topical Collaboration on TMDs (B.K. and S.L.). In particular, the undergraduate students involved in this project, P.V., E.Y, F.Y. worked as a subgroup in the Summer Institute for Wigner Imaging and Femtography (SIWIF) at the University of Virginia funded by SURA grant C2019-FEMT-002-04, and at the (virtual) subsequent effort (Femtonet) at UVA during Summer and Fall 2020, SURA Grant C2020-FEMT-006-05,  
https://pages.shanti.virginia.edu/Femtography/. S.L. also acknowledges partial support from the SURA grants.

\vspace{1cm}
\appendix

%%%%%%%%%%
%%%%%%%%%% APPENDIX A
\section{Kinematics in symmetric and asymmetric frames}
\label{sec:appa}
\begin{subequations}
\begin{eqnarray}
p & \equiv & \left(p^+,\frac{M^2}{2p^+},0 \right) \\
k & \equiv & \left(X p^+,\frac{M^2}{2p^+} - \frac{M_X^2+{\bf k}_\perp^2}{2(1-X) p^+}, {\bf k}_\perp \right) \\
p' & \equiv & \left( (1-\zeta) p^+,\frac{M^2+{\bf \Delta}_\perp^2}{2(1-\zeta) P^+}, {\bf \Delta}_\perp \right) \\
k' & \equiv & \left( (X-\zeta) p^+,\frac{M^2+{\bf \Delta}_\perp^2}{2(1-\zeta) p^+} - \frac{M_X^2+{\bf k}_\perp^2}{2(1-X) p^+}, {\bf k}_\perp - {\bf \Delta}_\perp  \right) \nonumber \\
\end{eqnarray} 
\end{subequations}
In this frame the proton lies on the $z$-axis. One can easily translate into the more commonly used symmetric frame, which uses the (average) sum, $P=(p+p')/2$ and difference, $\Delta = p' - p$, of the proton momenta defines as,
%%%
\begin{subequations}
\label{kin:sym}
\begin{eqnarray}
{P} & \equiv & \left( {P}^+, \frac{M^2}{2{P}^+}, 0 \right)   \\
\Delta & \equiv &  \left( \xi \, (2 {P}^+), \frac{ t+ {\bf \Delta}_T^2}{2 \xi {P}^+},  {\bf \Delta}_T  \right) 
\end{eqnarray}
\end{subequations}
In this case, the vector $P$ lies along the $z$-axis and the coordinates of the initial and final proton and parton are, respectively given by,
\begin{subequations}
\begin{eqnarray}
p & \equiv &   \left((1+\xi) {P}^+,  \frac{M^2+  {\bf \Delta}_T^2/4}{(1+\xi){P}^+},   \frac{{\bf \Delta}_T}{2} \right)   \\
k & \equiv &  \left( (x+\xi){P}^+,  k^-, {\bf k}_T +  \frac{{\bf \Delta}_T}{2} \right),   \\
p' & \equiv &  \left( (1-\xi) P^+,  \frac{M^2+  {\bf \Delta}_T^2/4}{(1-\xi){P}^+},- \frac{{\bf \Delta}_T}{2}  \right) 
\\
k' & \equiv &  \left( (x-\xi) P^+,  k'^-,{\bf k}_T - \frac{{\bf \Delta}_T}{2}   \right)  
\end{eqnarray}
\end{subequations}

%%%%%%%
%%%%%%%

%%%%%%%%%%%%%
%%%%%%%%%%%%%
%%%%%%%%%%%%% APPENDIX B
\section{Helicity Amplitudes Structure of GPDs}
\label{sec:appb}
%%%%%%%
%%%%%%%
GPDs are described in terms of parton-proton helicity amplitudes \cite{Diehl:2003ny}. We describe below the detailed structure of the quark and gluon amplitudes.
\begin{widetext}
\subsection{Quark Amplitudes}
\begin{subequations}
\begin{eqnarray}
%%% H
2 H(X,\zeta,t)
&=& A_{++,++} + A_{+-,+-} + A_{--,--} + A_{-+,-+}  \\
%%% E
-   \frac{\Delta_1 E(X,\zeta,t)}{ M}
&=& A_{++,-+}  + A_{+-,--} - A_{--,+-}  - A_{-+,++}  \\
%-(m+Mx) (1-x) (\Delta_1 + i \Delta_2)
%%% H tilde
2 \widetilde{H}(X,\zeta,t) 
&=& A_{++,++} - A_{+-,+-} + A_{--,--} - A_{-+,-+}  \\
%%% E tilde
\xi \frac{ \Delta_1  \widetilde{E}(X,\zeta,t)}{ M} 
&=& A_{++,-+}  -A_{+-,--} -A_{--,+-}  + A_{-+,++}
\end{eqnarray}  
\end{subequations}
\end{widetext}
Parity relations give,
$A_{--,--} = A^*_{++,++} $, $A_{-+,-+}  = A^*_{+-,+-}$, $A_{--,+-} = - A^*_{++,-+} $, and $A_{+-,--} = - A^*_{-+,++}$.      
In the spectator model one has,
\begin{equation} 
\label{As}
A_{\Lambda^\prime \lambda^\prime, \Lambda\lambda}
  =  \int d^2k_\perp\phi^{q *}_{\lambda^\prime,\Lambda^\prime}(k^\prime,p^\prime) \phi_{\lambda,\Lambda}(k,p),
\end{equation} 
with the following vertex functions (see Fig.\ref{fig:spectator}),
\begin{eqnarray}
\phi_{\Lambda,\lambda}(k,p) & = & \Gamma(k) \frac{\bar{u}(k,\lambda) U(p,\Lambda)}{k^2-m^2} \\
%%%
\phi^{q *}_{\Lambda^\prime \lambda^\prime}(k^\prime,p^\prime)& = &\Gamma(k^\prime) \frac{\overline{U}(p^\prime,\Lambda^\prime) u(k^\prime,\lambda^\prime)}{k^{\prime \,2}-m^2},  
\end{eqnarray}
Notice that we use the same form of coupling for the scalar and axial vector diquark, but distinguish the two by allowing for different mass parameters for the $u$ and $d$ quarks. This ansazt was first introduced in parametric forms in Ref.\cite{Jakob:1997wg}. It is justified in our case because the scalar and axial vector couplings give functional shapes which are similar to one another, while flexibility is provided by allowing for the mass parameters to vary.
The proton-quark-diquark vertex function is given by
\begin{subequations}
\cite{GonzalezHernandez:2012jv},
\begin{equation}
\Gamma = g_s \frac{k^2-m^2}{(k^2- M_\Lambda^2)^2}, 
\quad  \Gamma' = g_s \frac{k'^2-m^2}{(k'^2- M_\Lambda^2)^2}
\label{eq:Gamma}
\end{equation}
leading to,
\begin{eqnarray}
& \phi_{++}(k,p) = & \frac{1}{\sqrt{X}} \frac{(m+M X)(1-X)^2 }{\left[(1-X) {\cal M}^2 - {\bf k}_\perp^2 \right]^2}    \\
& \phi_{++}(k^\prime,p^\prime) =  &   \frac{1}{\sqrt{X'}}  \frac{(m + M X')(1-X')^2 }{\left[ (1-X') {\cal M}'^2 - {\bf \tilde{k}}_\perp^2 \right]^2}   \\
& \phi_{+-}(k,P) = & \frac{1}{\sqrt{X}} \frac{(k_1 - i k_2) (1-X)^2}{\left[(1-X) {\cal M}^2 - {\bf k}_\perp^2 \right]^2}  \\
& \phi_{+-}(k^\prime,p^\prime) = & \frac{1}{\sqrt{X'}} \frac{(\tilde{k}_1 - i \tilde{k}_2)(1-X')^2}{\left[ (1-X') {\cal M}'^2 - {\bf \tilde{k}}_\perp^2 \right]^2} , 
\end{eqnarray}
\end{subequations}
where we used, 
\begin{eqnarray}
\bar{u}(k,\pm) U(p,\pm) &=& \frac{1}{4} {\rm Tr}\{(\slashed{P} +M)(1+\gamma^o)(1 \pm \gamma_5 \gamma^3) (k\not\!+m) \} \nonumber \\ \\
\bar{u}(k,\pm) U(p,\mp) &=& \frac{1}{4} {\rm Tr}\{(P\not\! +M)(1+\gamma^o)(\gamma_1 \pm i \gamma_2) (k\not\!+m) \} , \nonumber \\
\end{eqnarray} 
The denominators from Eq.\eqref{eq:Gamma} are defined as,
\begin{widetext}
\begin{eqnarray}
k^2-m^2 & = & X M^2 - \frac{X}{1-X} M_X^2 - m^2  - \frac{{\bf k}_\perp^2}{1-X}  = {\cal M}^2 - \frac{{\bf k}_\perp^2}{1-X} \\ 
k^{\prime \, 2}-m^2 & = & \frac{X-\zeta}{1-\zeta} M^2 - \frac{X-\zeta}{1-X} M_X^2 - m^2  - \frac{1-\zeta}{1-X} 
 \left( {\bf k}_\perp - \frac{1-X}{1-\zeta} {\bf \Delta}_\perp \right)^2 =  {\cal M}'^2 - \frac{ {\bf \tilde{k}}_\perp^2}{1-X'} \,
.
\end{eqnarray}
\end{widetext}
The final expressions entering Eq.\eqref{fit_form} are, 
%%%%%
%%%%% H and E
%%%%%
\begin{widetext}
\begin{eqnarray}
 H_{M_X,m}^{M_\Lambda} & = &  \displaystyle \mathcal{N} \frac{1-\zeta/2}{1-X}   \int d^2k_\perp 
\frac{% 
 \left[ (m+M X)  \left(m + M X' \right) + {\bf k}_\perp^2 - (1-X') {\bf k}_\perp\cdot {\bf \Delta}_\perp\right](1-X)^2(1-X')^2  }{\left[(1-X) {\cal M}(X)^2 - {\bf k}_\perp^2 \right]^2 \left[ (1-X') {\cal M}(X')^2 - {\bf k}_\perp^2  + 2 (1-X') {\bf k}_\perp\cdot {\bf \Delta}_\perp - (1-X')^2{\bf \Delta}_\perp^2\right]^2 }    \nonumber \\
&   + &  \frac{\zeta^2}{4(1-\zeta)}E_{M_X,m}^{M_\Lambda},
\label{GPDH}
\end{eqnarray} 
\begin{eqnarray}
 E_{M_X,m}^{M_\Lambda}  & = & \displaystyle \mathcal{N} \frac{1-\zeta/2}{1-X} \int  d^2k_\perp 
%%%
 \frac{-\displaystyle \frac{2M (1-\zeta)}{{\bf \Delta}^2_\perp}%
 \left[  ( M (X  - X' ) {\bf k}_\perp \cdot {\bf \Delta}_\perp  - \left(m + M X \right)(1-X'){\bf \Delta}^2_\perp  \right] (1-X)^2(1-X')^2}%
 {\left[(1-X) {\cal M}^2 - {\bf k}_\perp^2 \right]^2 \left[ (1-X') {\cal M}^2 - {\bf k}_\perp^2  + 2 (1-X') {\bf k}_\perp\cdot {\bf \Delta}_\perp - (1-X')^2{\bf \Delta}_\perp^2\right]^2 }
 \label{GPDE} \nonumber \\
%%%
\end{eqnarray}
%where $\mathcal{N}=1-\zeta/2$.
%{\bf is the $(2\pi)^4$ there?}
%%%%%
%%%%% H tilde and and E tilde
%%%%%
\begin{eqnarray}
 \widetilde{H}_{M_X,m}^{M_\Lambda}  & = &  \displaystyle\mathcal{N} \frac{1-\zeta/2}{1-X}   \int d^2k_\perp
\frac{ \left[ (m+M X)  \left(m + M X' \right) - {\bf k}_\perp^2 + (1-X') {\bf k}_\perp\cdot {\bf \Delta}_\perp \right] (1-X)^2(1-X')^2}%
{\left[(1-X) {\cal M}^2 - {\bf k}_\perp^2 \right]^2 \left[ (1-X') {\cal M}^2 - {\bf k}_\perp^2  + 2 (1-X') {\bf k}_\perp\cdot {\bf \Delta}_\perp - (1-X')^2{\bf \Delta}_\perp^2\right]^2}   \nonumber \\
& + &  \frac{\zeta^2}{4(1-\zeta)}\widetilde{E}_{M_X,m}^{M_\Lambda}
\label{GPDHTILDE}
\end{eqnarray} 
\begin{eqnarray}
 \widetilde{E}_{M_X,m}^{M_\Lambda}  & = & \displaystyle\mathcal{N} \frac{1-\zeta/2}{1-X} \int d^2k_\perp
%%%
\frac{- \displaystyle\frac{4M (1-\zeta) }{\zeta {\bf \Delta}^2_\perp}%
 \left[  \left(2 m + M (X+X') \right) {\bf k}_\perp \cdot {\bf \Delta}_\perp  - \left(m + M X \right)(1-X'){\bf \Delta}_\perp^2   \right] (1-X)^2(1-X')^2}{\left[(1-X) {\cal M}^2 - {\bf k}_\perp^2 \right]^2 \left[ (1-X') {\cal M}^2 - {\bf k}_\perp^2  + 2 (1-X') {\bf k}_\perp\cdot {\bf \Delta}_\perp - (1-X')^2{\bf \Delta}_\perp^2\right]^2}
 \label{GPDETILDE} \nonumber \\
%%%
\end{eqnarray}
From the expressions above one clearly sees the dependence of the GTMDs on ${\bf k}_\perp^2$, ${\bf \Delta}_\perp^2$, $({\bf k}_\perp \cdot {\bf \Delta}_\perp)$.

The integration over the angle $\phi$, namely $d^2 {\bf k}_\perp = d k_\perp k_\perp d \phi$, in Eqs.(\ref{GPDH},\ref{GPDE},\ref{GPDHTILDE}), can be carried out analytically thus obtaining the expressions in Section \ref{sec:param}.
\end{widetext}

%Therefore the contributions to the non reggeized unintegrated GPDs (GTMDs) ${\cal N} G_{M_X,m}^{M_\Lambda}(X,k_\perp,\zeta,t)$, can be written as,
%%%%%%%%%
%%%%%%%%% GLUON AMPS
\subsection{Gluon Amplitudes}
We define the gluon-proton helicity amplitudes similarly to the quark-proton amplitudes as, 
%involving light cone quantization, light cone gauge ($A^+=0$), ``light cone helicity" and the noncovariant Hamiltonian formulation (see e.g. Ref.) %\cite{Diehl1, Diehl2}).
\begin{eqnarray}
\label{Ag:eq}
A^g_{\Lambda^{\prime}\lambda_g^{\prime},\Lambda\lambda_g} = \frac{1}{\bar{P}^+} \int \frac{dz^-}{2\pi}e^{ix\bar{P}^+z^-} \langle p^{\prime},\Lambda^{\prime}|{\mathcal O}^g_{\lambda_g^{\prime}\lambda_g}|p,\Lambda \rangle\Big|_{z^+=0} \nonumber , \\
\end{eqnarray}
where the leading twist gluon strength field operators, $O_{\lambda'_g=\pm \lambda_g=\pm}$, are, 
\begin{eqnarray}
{\mathcal O}^g_{++} &=& \frac{1}{2}\Big[G^{+ i}G_{i}{}^{+} - iG^{+\mu}\tilde{G}_{\mu}{}^{+}    \Big]\\
{\mathcal O}^g_{--} &=&  \frac{1}{2}\Big[  G^{+ i}G_{i}{}^{+} + iG^{+\mu}\tilde{G}_{\mu}{}^{+}   \Big] \\
{\mathcal O}^g_{-+} &=&  \frac{1}{2}\Big[G^{+1}G^{1+} - G^{+2}G^{2+} - iG^{+1}G^{2+} - iG^{+2}G^{1+}   \Big] \nonumber  \\ \\
{\mathcal O}^g_{+-} &=&  \frac{1}{2}\Big[  G^{+1}G^{1+} - G^{+2}G^{2+} + iG^{+1}G^{2+} + iG^{+2}G^{1+}   \Big] \nonumber \\
\end{eqnarray}
with $i = 1,2$.  
%By projecting out the transverse gluon fields, one gets the following relations:
%\begin{eqnarray}
%A^g_{\Lambda^{\prime}\lambda_g^{\prime},\Lambda\lambda_g} =\frac{1}{\bar{P}^+} \int \frac{dz^-}{2\pi}e^{ix\bar{P}^+z^-}\langle P^{\prime},\Lambda^{\prime}|\epsilon_i(\lambda_g^{\prime})G^{+i}(-\frac{1}{2}z)G^{+j}(\frac{1}{2}z)\epsilon(\mu)^{\ast}_j|P,\Lambda \rangle \Big|_{z^+=0,\vec{z}_T=0} 
%\end{eqnarray}
%\begin{equation}
%\epsilon(+) = \frac{-1}{\sqrt{2}}\begin{pmatrix} 1\\ i \end{pmatrix}, \hspace{10pt} \epsilon(-) = \frac{1}{\sqrt{2}}\begin{pmatrix} 1\\ -i \end{pmatrix}
%\end{equation}
By using these operators in Eq.(\ref{Ag:eq}), we find 
%The soft amplitudes are related to equations (\ref{eqn:1}),(\ref{eqn:2}) by combining different $\epsilon,\epsilon^*$ helicity combinations.  
%For example 
%{\bf CHECK THIS?}: 
%\begin{equation}
%G^{+1}G^{+1}+G^{+2}G^{+2} = \epsilon_{i,+}G^{+i}G^{+j}\epsilon^{\ast}_{j,+}+
%\epsilon_{i,-}G^{+i}G^{+j}\epsilon^{\ast}_{j,-}
%\end{equation}
for the gluon helicity conserving amplitudes, 
\begin{eqnarray}
\label{eq:g_helamps}
A_{++,++} &=& \sqrt{1-\xi^2}\Big(\frac{H^g+\tilde{H}^g}{2}-\frac{\xi^2}{1-\xi^2}\frac{E^g+\tilde{E}^g}{2}\Big)\nonumber \\
A_{-+,-+} &=& \sqrt{1-\xi^2}\Big(\frac{H^g-\tilde{H}^g}{2}-\frac{\xi^2}{1-\xi^2}\frac{E^g-\tilde{E}^g}{2}\Big)\nonumber \\
A_{++,-+} &=& -e^{-i\phi}\frac{\sqrt{t_0-t}}{2M}\Big(\frac{E^g-\xi\tilde{E}^g}{2}\Big)\nonumber \\
A_{-+,++} &=& e^{i\phi}\frac{\sqrt{t_0-t}}{2M}\Big(\frac{E^g+\xi\tilde{E}^g}{2}\Big),
%\nonumber \\ \nonumber
\end{eqnarray}
The amplitudes observe the following parity relations,  
\begin{equation}
A^g_{-\Lambda^{\prime} -\lambda_g^{\prime}, -\Lambda -\lambda_g} =(-1)^{\Lambda-\lambda_g-\Lambda^{\prime}+\lambda_g^{\prime}}A^{g \, *}_{\Lambda^{\prime} \lambda_g^{\prime}, \Lambda \lambda_g} . %\left[{\rm or} \, A^{g \, *}_{\Lambda^{\prime},\Lambda_{g^\prime};\Lambda,\Lambda_g}\right] .
\end{equation}

The gluon-proton helicity amplitude can be written as, 
%(factors of $2\pi$ ?)
\begin{eqnarray}
A_{\Lambda^{\prime}\lambda_g^{\prime},\Lambda \lambda_g} = \int \frac{d^2k_{\perp}}{1-X} \:
\sum_{\Lambda_X} \, \phi^{\ast \, \Lambda_X}_{ \lambda_g' \Lambda'}(k', p') \, \phi^{\Lambda_X}_{\lambda_g \Lambda}(k, p),  \nonumber \\
%\int dk^-P^+ \left[ \bar{U}(P^{\prime}, \Lambda')\Phi^{\nu}(k^{\prime})
%\frac{i(\slashed{P}_X + M_x)}{P_X^2 - M_X^2}\Phi^{\mu}(k)U(P'\Lambda)
%\epsilon_{\nu}^{\Lambda_g^{\prime}}(k') 
%\epsilon_{\mu}^{* \, \Lambda_g}(k) 
%+ (k \leftrightarrow k^\prime)
%\right]
\end{eqnarray}
where we defined the initial and final LC vertex functions, $\phi^g_{\Lambda_X \lambda_g \Lambda}(k, p)$,  and $\phi^{g \ast}_{\Lambda_X \lambda_g' \Lambda'}(k', p')$, respectively as, 
\begin{eqnarray}
\phi^{\Lambda_X}_{\lambda_g \Lambda}(k,p) &= & %{\sqrt{\frac{k^+}{p^+-k^+}} }\, 
\Gamma(k) \,  \frac{\bar{U}_{\Lambda_X}(p-k)\, U_\Lambda(p)   }{k^2-m_g^2} \, \slashed\epsilon^*_{\lambda_g}(k) 
\nonumber \\
 \\
\phi^{\ast \, \Lambda_X}_{\lambda_g' \Lambda'}(k',p') &=& %\sqrt{\frac{k'^+}{p'^+-k'^+}} \, 
\Gamma(k) \,  
\frac{\bar{U}_{\Lambda'}(p')U_{\Lambda_X}(p'-k')\,}{k'^2-m_g^2} \,\, \slashed\epsilon_{\lambda_g'}(k^{\prime}) \nonumber \\
%\gamma_{\nu} 
\end{eqnarray}
where the gluon mass, $m_g$, {is present because the gluons are off-shell}.
%In the axial gauge $A^+ = 0$, there is the relation that $F^{+\mu}$ simplifies to $\partial^+ A^\mu$.  The ordinary helicity amplitudes $A^0$ are related to the LC helicity amplitudes in sections (1) and (2) by the following conversion factor:\\
%NOTE::: Check this!  Ji's Paper: [arXiv:hep-ph/9801369v2] , and Radyushkin[Nonforward Parton Distributions]\\
%\begin{equation}
%conversion factor \sim \frac{X(X-\zeta)}{(1-\zeta/2)^2}
%\end{equation}
%%
Analogous to the quark case the coupling at the gluon-proton-octet-proton vertex, $\Gamma(k)$ contains a form factor as in the quark-diquark spectator case, 
%NOTE::: See Bacchetta paper.  The propagator is a result of a completeness relation, and an intermediate $<P g|$ state.  
\begin{eqnarray}
\Gamma(k) &\equiv& g \frac{k^2-m_g^2}{(k^2-M_\Lambda^2)^2},\hspace{10pt} 
\end{eqnarray}
so that $k^2-m_g^2$, cancels out. This coupling is used here as an ultraviolet regulator in  ${\bf k}_T$;
$M_\Lambda$ sets the mass scale for the form factor.

By using,$\slashed{P}_X +M_X = \sum\limits_{\Lambda_X} U_{\Lambda_X}(P_X)\bar{U}_{\Lambda_X}(P_X)$, and 
\begin{equation}
\int dk^- P^+ \frac{i}{P_X^2 - M_X^2} f(k^-,\ldots) = \frac{\pi}{(1-X)}f(\ldots)\mid_{P_X^2=M_X^2}
\end{equation}
for the spectator propagator we find, 
\begin{widetext}
\begin{eqnarray}
A_{\Lambda^{\prime}\lambda_g^{\prime},\Lambda\lambda_g}(X, \zeta, t) = \int \frac{d^2k_{\perp}}{1-X} \:
\bar{U}(p',\Lambda') \left[  \gamma_\nu (\not\!{P} + M_X) \gamma_\mu \right]  U(P,\Lambda) \Gamma(k') \Gamma(k) \epsilon_{\lambda_g}^{* \, \nu}(k') \epsilon_{\lambda_g}^\mu(k)
%\int dk^-P^+ \left[ \bar{U}(P^{\prime}, \Lambda')\Phi^{\nu}(k^{\prime})
%\frac{i(\slashed{P}_X + M_x)}{P_X^2 - M_X^2}\Phi^{\mu}(k)U(P'\Lambda)
%\epsilon_{\nu}^{\Lambda_g^{\prime}}(k') 
%\epsilon_{\mu}^{* \, \Lambda_g}(k) 
%+ (k \leftrightarrow k^\prime)
%\right]
\end{eqnarray}
\end{widetext}
The gluon polarizations are defined as,
\begin{subequations}
\begin{eqnarray}
\epsilon_{\lambda_g}^{* \, \nu}(k) & = \frac{1}{\sqrt{2}}(0; -\lambda_g, i,0)& \\
\epsilon_{\lambda_g'}^{* \, \nu}(k') & = \frac{1}{\sqrt{2}}(0; -\lambda_g', i,0)&.
\end{eqnarray}
\end{subequations}
{Notice that the initial $(k,\lambda_g)$, and final $(k',\lambda_g')$ gluon polarizations are both taken with the particles momenta aligned along the $z$-axis, despite the final gluon is rotated the angle $\hat{P\Delta}$. If we rotate the gluon this introduces a higher order in $k_T/P^+$ correction.}
%gluon aligned along the $z$-axis and we rotated the final gluon by the angle $\hat{P\Delta}$.}

\begin{widetext}
The specific helicity combinations read,
\begin{eqnarray}
A_{++, ++} &=& {\mathcal N} \sqrt{1-\zeta} \int d^2k_{\perp} \frac{ \vec{k}_\perp \cdot \vec{\tilde{k}}_\perp + [(1-X)M-M_X] [(1-X^{\prime})M-M_X] X X' }{(k^2-M_{\Lambda}^2)^2(k^{\prime 2} - M_{\Lambda}^2)^2} 
\\
\label{A01s}
A_{-+, -+} &=& {\mathcal N} \sqrt{1-\zeta} \int d^2k_{\perp} \frac{ \vec{k}_\perp \cdot \vec{\tilde{k}}_\perp (1-X)(1-X^{\prime}) }{(k^2-M_{\Lambda}^2)^2(k^{\prime 2}-M_{\Lambda}^2)^2} \\
\label{A02s}
A_{-+, ++} &=& {\mathcal N} \sqrt{1-\zeta} \int d^2 k_{\perp}  \frac{ (1-X^{\prime}) [(1-X)M-M_X] (\tilde{k}_1 + i \tilde{k}_2) }{X'\, (k^2-M_{\Lambda}^2)^2(k^{\prime 2}-M_{\Lambda}^2)^2 } 
\\
\label{A03s}
A_{++, -+} &=&  {\mathcal N} \sqrt{1-\zeta} \int d^2k_{\perp} \frac{(1-X) [(1-X^{\prime})M-M_X] (k_1 - i k_2) }{X \, (k^2-M_{\Lambda}^2)^2(k^{\prime 2}-M_{\Lambda}^2)^2} 
\label{A04s}
\end{eqnarray}
\end{widetext}
where the normalization factor, ${\cal N}$, absorbs all common factors ($\pi,g_s$, constants).
%{\rm with}\,{\mathcal A} &=& {\mathcal N} \frac{1}{\sqrt{1-X}\sqrt{1-X^{\prime}}(1-X)}\frac{1}{(k^2-M_{\Lambda}^2)^2(k^{\prime 2}-M_{\Lambda}^2)^2} = \frac{\sqrt{1-\zeta}}{(1-X)^2}\frac{1}{(k^2-M_{\Lambda}^2)^2(k^{\prime 2}-M_{\Lambda}^2)^2}  
%${\mathcal A} = {\mathcal N} \frac{1}{\sqrt{1-X}\sqrt{1-X^{\prime}}(1-X)}\frac{1}{(k^2-M_{\Lambda}^2)^2(k^{\prime 2}-M_{\Lambda}^2)^2} = \frac{\sqrt{1-\zeta}}{(1-X)^2}\frac{1}{(k^2-M_{\Lambda}^2)^2(k^{\prime 2}-M_{\Lambda}^2)^2}  $
The components of $\vec{k}_\perp$ are defined relative to the direction of $\vec{\Delta}_\perp$, so that the integral over angles can be specified by choosing $\vec{\Delta}_\perp =\Delta_\perp {\hat x}$ or simply ${\hat \Delta}_\perp$.

Inverting Eqs.\eqref{eq:g_helamps} we find the expressions in Section \ref{sec:gluons}.

%%%%%%%%%%%%%%%%%%%%%%%%%%%%%%%%%%%%%%%%%%%%%%%%%%%%%%
\section{Summary of GPD parametrizations}
\label{sec:appc}
We present a summary of the parametrization for the GPDs, $H_{q_v}, H_{\bar{q}}, H_g, E_{q_v}, E_g, \widetilde{H}_{q_v}, \widetilde{E}_{q_v}$, that can be easily implemented in numerical calculations. Note that the GPD $H_q= H_{q_v} + H_{\bar{q}}$ is obtained as the sum of Eq.(\ref{eq:Hqv_final}). Eq.(\ref{eq:Hqbar_final}). The parameters for each component can be read off Tables \ref{tab:H}, \ref{tab:E}, and \ref{tab:tilde}. The given parametric forms need to be perturbatively evolved to the $Q^2$ of the data. 
\begin{widetext}
Parametric form for $H(X,\zeta,t)$:
\begin{equation}
\label{eq:Hqv_final}
H^{q_v}(X,\zeta,t)  =  \left\{ 
\begin{array}{lcc}
H_{M_X,m}^{M_\Lambda}(X,\zeta,t) \,  
R^{\alpha,\alpha^\prime}_{p}(X,t), & {\rm Eq.(\ref{eq:Hqv})} &  \mbox{ $\zeta \leq X \leq 1$} \\ 
& &\\
 a^- X^2 - a^- \, \zeta X + H({\zeta,t}), &  {\rm Eq.(\ref{eq:hqverbl})} & \mbox{$0 \leq X  < \zeta $} \\ 
& & \\
 0 & & \mbox{$-1 + \zeta \leq X  < 0 $}
\end{array} 
\right.
\end{equation}

\begin{equation}
\label{eq:Hqbar_final}
H^{\bar{q}}(X,\zeta,t)  =  \left\{ 
\begin{array}{lcc}
0 &  & \mbox{ $\zeta \leq X \leq 1$}  \\ 
& &\\
  a^+ X^3 - \frac{3}{2} a^+ \zeta X^2 + c X + d  ,  &  {\rm Eq.(\ref{eq:h+})} & \mbox{$0 \leq X  < \zeta $}  \\ 
& & \\
 H_{M_X,m}^{M_\Lambda}(X,\zeta,t) \,  
R^{\alpha,\alpha^\prime}_{p}(X,t), & {\rm Eq.(\ref{eq:Hqbar})}  & \mbox{$-1 + \zeta \leq X  < 0 $}
\end{array} 
\right.
\end{equation}

\begin{equation}
\label{eq:Hg_final}
H^{g}(X,\zeta,t)  =  \left\{ 
\begin{array}{lcc}
H_{M_X,m}^{M_\Lambda}(X,\zeta,t) \,  
R^{\alpha,\alpha^\prime}_{p}(X,t), & {\rm Eq.(\ref{eq:Hg})} &  \mbox{ $\zeta \leq X \leq 1$} \\ 
& &\\
 a_g X^2 - a_g \, \zeta X + H({\zeta,t}), &  {\rm Eq.(\ref{eq:Hgerbl})} & \mbox{$0 \leq X  < \zeta $} \\ 
& & \\
 0 & & \mbox{$-1 + \zeta \leq X  < 0 $}
\end{array} 
\right.
\end{equation}

Parametric form for $E(X,\zeta,t)$:
%%%%%% E %%%%%%
\begin{equation}
\label{eq:Eqv_final}
E^{q_v}(X,\zeta,t)  =  \left\{ 
\begin{array}{lcc}
E_{M_X,m}^{M_\Lambda}(X,\zeta,t) \,  
R^{\alpha,\alpha^\prime}_{p}(X,t), & {\rm Eq.(\ref{eq:Eqv})} &  \mbox{$0 \leq X  < \zeta $} \\ 
& &\\
 a_E X^2 -a_E \, \zeta X + E({\zeta,t}), &  {\rm Eq.(\ref{eq:eqverbl})} & \mbox{ $\zeta \leq X \leq 1$} \\ 
& & \\
 0 & & \mbox{$-1 + \zeta \leq X  < 0 $}
\end{array} 
\right.
\end{equation}

\begin{equation}
\label{eq:Eg_final}
E^{g}(X,\zeta,t)  =  \left\{ 
\begin{array}{lcc}
E_{M_X,m}^{M_\Lambda}(X,\zeta,t) \,  
R^{\alpha,\alpha^\prime}_{p}(X,t), & {\rm Eq.(\ref{eq:Eg})} &  \mbox{$0 \leq X  < \zeta $} \\ 
& &\\
 a_E X^2 -a_E \, \zeta X + E({\zeta,t}), &  {\rm Eq.(\ref{eq:Egerbl})} & \mbox{ $\zeta \leq X \leq 1$} \\ 
& & \\
 0 & & \mbox{$-1 + \zeta \leq X  < 0 $}
\end{array} 
\right.
\end{equation}

Parametric form for $\widetilde{H}(X,\zeta,t)$:

\begin{equation}
\label{eq:Htil_final}
\widetilde{H}^{q_v}(X,\zeta,t)  =  \left\{ 
\begin{array}{lcc}
\widetilde{H}_{M_X,m}^{M_\Lambda}(X,\zeta,t) \,  
R^{\alpha,\alpha^\prime}_{p}(X,t), & {\rm Eq.(\ref{eq:Htilqv})}  & \mbox{ $\zeta \leq X \leq 1$}  \\ 
& &\\
  a^+ X^3 - \frac{3}{2} a^+ \zeta X^2 + c X + d  ,  &  {\rm Eq.(\ref{eq:Htilqbarerbl})} & \mbox{$0 \leq X  < \zeta $}  \\ 
& & \\
 0 &  & \mbox{$-1 + \zeta \leq X  < 0 $}
\end{array} 
\right.
\end{equation}

Parametric form for $\widetilde{E}(X,\zeta,t)$:

\begin{equation}
\label{eq:Etil_final}
\widetilde{E}^{q_v}(X,\zeta,t)  =  \left\{ 
\begin{array}{lcc}
\widetilde{E}_{M_X,m}^{M_\Lambda}(X,\zeta,t) \,  
R^{\alpha,\alpha^\prime}_{p}(X,t), & {\rm Eq.(\ref{eq:Etilqv})}  & \mbox{ $\zeta \leq X \leq 1$}  \\ 
& &\\
  a^+ X^3 - \frac{3}{2} a^+ \zeta X^2 + c X + d  ,  &  {\rm Eq.(\ref{eq:Etilqbarerbl})} & \mbox{$0 \leq X  < \zeta $}  \\ 
& & \\
 0 &  & \mbox{$-1 + \zeta \leq X  < 0 $}
\end{array} 
\right.
\end{equation}

\end{widetext}

\bibliography{DVCS_BH_bib}

\end{document}